\documentclass[letterpaper,aps,preprint,unsortedaddress,showpacs,nofootinbib]{revtex4}

\usepackage{graphicx}
\usepackage{amsfonts}
\usepackage{amsmath}
\usepackage{amsthm}
\usepackage{amssymb}
\usepackage{setspace}
\usepackage{axodraw}
\usepackage{epsfig}

\newcommand{\beq}{\begin{equation}}
\newcommand{\eeq}{\end{equation}}
\newcommand{\met}{\rlap{\,\,/}E_T}
\newcommand{\mB}{m_{\gamma_1}}

\begin{document}

\title{Kaluza-Klein Dark Matter: Direct Detection vis-a-vis LHC}

\author{Sebastian Arrenberg} \affiliation{Department of Physics, University of Z\"urich, Z\"urich, 8057, Switzerland}
\author{Laura Baudis} \affiliation{Department of Physics, University of Z\"urich, Z\"urich, 8057, Switzerland} 
\author{Kyoungchul Kong} \affiliation{Fermi National Accelerator Laboratory, Batavia, IL 60510, USA} 
\author{Konstantin T. Matchev} \affiliation{Institute for Fundamental Theory, Physics Department, University of Florida, Gainesville, FL 32611, USA}
\author{Jonghee Yoo} \affiliation{Fermi National Accelerator Laboratory, Batavia, IL 60510, USA} 
\date{\today}

\begin{abstract}
We explore the phenomenology of Kaluza-Klein (KK) dark matter in 
very general models with universal extra dimensions (UEDs), emphasizing the
complementarity between high-energy colliders and dark matter direct detection experiments.
In models with relatively small mass splittings between the dark matter 
candidate and the rest of the (colored) spectrum, the collider 
sensitivity is diminished, but direct detection rates are enhanced.
UEDs provide a natural framework for such mass degeneracies. 
We consider both 5-dimensional and 6-dimensional non-minimal UED models,
and discuss the detection prospects for various KK dark matter candidates:
the KK photon $\gamma_1$, the KK $Z$-boson $Z_1$,
the KK Higgs boson $H_1$ and the spinless KK photon $\gamma_H$.  
We combine collider limits such as electroweak precision data and expected 
LHC reach, with cosmological constraints from WMAP, and the
sensitivity of current or planned direct detection experiments.
Allowing for general mass splittings, 
we show that neither colliders, nor direct detection 
experiments by themselves can explore all of the relevant 
KK dark matter parameter space. Nevertheless, they probe different 
parameter space regions, and the combination of the two types of
constraints can be quite powerful. For example, in the case of 
$\gamma_1$ in 5D UEDs the relevant parameter space will be almost completely 
covered by the combined LHC and direct detection sensitivities expected in the near future.
\end{abstract}

\preprint{{\small FERMILAB-PUB-08-141-A-T}}
\preprint{{\small UFIFT-HEP-08-09}}

\maketitle

\section{Introduction}\label{sec:intro}

The Standard Model (SM) has been extremely successful in
explaining all available experimental data in particle physics.
However, there are several unsettling features of the SM, which
have motivated a substantial research effort on physics beyond the 
Standard Model (BSM). The two issues continuously attracting 
the most attention are the hierarchy problem and the 
dark matter problem. The anticipated discovery of the Higgs boson 
of the SM at the Large Hadron Collider (LHC) at CERN would pose
a challenging theoretical question: what is the next fundamental 
energy scale? If it is as high as the Planck scale, then what stabilizes 
the hierarchy between the Planck and electroweak scales? Or, if it is 
much lower than the Planck scale, what is the physics associated with it?
The second issue is related to the now established existence of a dark matter
(DM) component of the universe. Since the SM does not accommodate a 
suitable DM particle candidate, the dark matter problem is the most 
pressing phenomenological evidence for physics BSM \cite{Bertone:2004pz}.

\subsection{The Dark Matter Problem and Physics Beyond the Standard Model}

There are different avenues one could follow in extending the SM and 
addressing the dark matter problem. The common theme among them is 
the introduction of new particles, one of which is neutral and 
serves as the dark matter candidate; and a new symmetry, a remnant of which
survives in the low energy effective theory and ensures that the
lifetime of the DM particle is sufficiently long (at the minimum, 
longer than the age of the universe).
In principle, simply postulating a new stable and neutral particle
would be rather ad hoc and unsatisfactory without further corroborating 
evidence. Fortunately, the DM candidates in most BSM models
typically have some kind of non-gravitational interactions,
which are sufficient to keep them in thermal equilibrium in 
the early universe. Thus, their relic abundance can in fact be 
straightforwardly calculated in any given model 
(for details, see Section~\ref{sec:relic} below).
The generic result of this computation is that 
a weakly interacting massive particle (WIMP) with 
a mass near or below the TeV scale has a relic density in 
the right ballpark, and is a suitable candidate for dark matter.
By now there are many examples of WIMPs in BSMs, perhaps
the most popular being the lightest superpartner (LSP) 
in supersymmetry (SUSY) with R-parity conservation \cite{Jungman:1995df},
the lightest Kaluza-Klein partner (LKP) in Universal Extra Dimensions \cite{Hooper:2007qk}, 
the lightest T-parity odd particle in Little Higgs models \cite{Cheng:2003ju,Birkedal:2006fz}, 
the lightest U-parity odd particle in $U(1)^\prime$-extended models \cite{Hur:2007ur,Lee:2008pc}, etc.

The most exciting aspect of the WIMP DM hypothesis is that it is testable
by experiment. Indeed, WIMPs near the TeV scale can be easily within reach 
of both high-energy colliders and dark matter detection experiments. 
Furthermore, the size of the corresponding DM signals can be readily calculated 
within any given BSM, providing some rough expectations for discovery in each case.
In principle, the signals depend on a typically a large number 
of model parameters. However, speaking in a broader sense, the WIMP DM phenomenology 
mostly depends on the answers to the following two questions:
\begin{itemize}
\item Q1: What is the identity of the DM particle candidate?
\item Q2: What is the size of the mass splitting between the DM particle and the 
rest of the (relevant) spectrum?
\end{itemize}
In the following two subsections we shall discuss each one of these questions 
and thus motivate our setup and methodology.

\subsection{The Nature of the Dark Matter Particle}

Within any given BSM, there are typically several potential dark matter candidates 
(i.e. neutral and stable particles) present in the spectrum. 
The answer to the first question (Q1) therefore selects one 
of them as the ``true'' dark matter. For example, in SUSY, 
the dark matter particle could be either a fermion 
(e.g.~gravitino or the lightest neutralino) or a boson (the lightest sneutrino). 
In turn, the lightest neutralino could be the superpartner of a gauge boson
(e.g.~a Bino, a Wino, possibly a $Z'$-ino), the superpartner of a Higgs boson
(e.g.~a Higgsino or a singlino) or some admixture of these \cite{Jungman:1995df}. 
Similarly, the lightest 
sneutrino could carry any one of the three lepton flavors, and in addition, 
could be left-handed \cite{Arina:2007tm}, right-handed \cite{Lee:2007mt}, 
or some mixture of both \cite{Thomas:2007bu}. Since all of these particles have
rather different properties, it is clear that it is impossible to make
any generic predictions about SUSY dark matter without specifying the exact
nature of the LSP, i.e. providing the answer to Q1 above.

On the positive side, the answer to Q1 goes a long way towards the determination 
of the size of the expected dark matter signals. Once the identity of the dark
matter particle is specified, its couplings are fixed and can be used in the
calculation of both direct and indirect detection rates. What is even better,
the answer to question Q1 can be provided in a rather model-independent way, 
without reference to the exact specifics of the model, such as the physics of
the ultraviolet completion, Renormalization Group Equation (RGE) evolution down from high scales, etc.

In this paper, we shall explore the dark matter phenomenology of general
models with flat universal extra dimensions \cite{Appelquist:2000nn}, 
where the usual Standard Model structure is embedded in 5 or 6 space-time dimensions. 
We shall assume the same gauge symmetry and particle content as in the SM.
Similar to the SUSY case just discussed, the models contain several possible
dark matter candidates (electrically-neutral particles which are stable 
due to KK parity conservation)\footnote{For further details on UED models, see Section~\ref{sec:model}.}.
In five dimensional models with minimal particle content, 
they are: the KK graviton ($G_1$), the KK neutrino ($\nu_1$), 
the KK photon ($\gamma_1$), the KK $Z$-boson ($Z_1$) and the KK Higgs boson ($H_1$).
Six dimensional UED models present additional possibilities:
the spinless KK photon ($\gamma_H$) and the spinless KK $Z$-boson ($Z_H$),
which are linear combinations of the gauge boson polarizations
along the two extra dimensions. Just like the case of SUSY, which 
of these particles is the lightest and thus the dark matter candidate,
depends on the model-building details. The issue is even more 
subtle than in SUSY, since all of these KK particles have tree-level
masses of the same order, proportional to the inverse radius $R^{-1}$ 
of the extra dimension. This mass degeneracy is lifted by two main 
sources: radiative corrections due to renormalization and nonuniversality 
in the boundary conditions at the cut-off scale. The former effect is in principle
computable within any given model \cite{Georgi:2000ks,vonGersdorff:2002as,Cheng:2002iz}, 
while the latter is a priori unknown, as its origin lies in the ultraviolet 
physics above the cut-off scale \cite{Cheng:2002iz,Carena:2002me,delAguila:2003bh,delAguila:2006kj}.
A common assumption throughout the existing literature on UED is to
ignore any boundary terms at the cut-off scale. The resulting model has
been dubbed ``Minimal UED'' and is known to accommodate only $\gamma_1$ and $G_1$ LKP
in five dimensions \cite{Cheng:2002iz,Cembranos:2006gt} and 
$\gamma_H$ in six dimensions \cite{Ponton:2005kx,Burdman:2006gy}. 
However, given our complete ignorance of the physics at and above the 
cut-off scale, the other possibilities for the nature of the KK dark 
matter particle should be given serious consideration as well.

One of the goals of this paper is to start filling these gaps in the literature,
by exploring the phenomenology of the alternative dark matter candidates in UED.
Of course, not all of them are on an equal footing. For example,
the KK graviton $G_1$ interacts with the Standard Model particles 
too weakly to be relevant for direct detection searches. 
The KK neutrino $\nu_1$ is already ruled out due to its large 
elastic scattering cross section \cite{Servant:2002aq}.
We shall therefore concentrate on the remaining two possibilities in 5D UEDs:
the KK $Z$-boson ($Z_1$) and the KK Higgs boson ($H_1$).
We shall also review and update the previously published results on $\gamma_1$ and $\gamma_H$,
so that our work would provide a concrete and complete reference on KK dark matter.

\subsection{The Effect of a Mass Degeneracy on Dark Matter Signals}

The second important issue for dark matter phenomenology is the 
answer to question Q2, namely, what is the mass splitting between the 
dark matter particle and the rest of the spectrum. Of course, it is in principle 
possible to have the dark matter particle as {\em the only} new particle 
in the model, in which case Q2 does not apply, and the predictions 
for the dark matter signals are quite robust, once Q1 is addressed.
However, realistic models typically contain a multitude of new particles,
in addition to the dark matter candidate. Their proximity (in mass) to the
dark matter particle therefore becomes an important issue, at least in three,
very different aspects. 

The first is related to the predicted dark matter relic abundance.
A close mass degeneracy can increase the importance of coannihilation 
processes at freeze-out \cite{Griest:1990kh}, and the results for the
relic density are now sensitive not only to the properties of the dark matter 
particle itself, but also to the properties of the coannihilating particles.
The size of the coannihilation effect depends on the particular scenario, 
and there are several known cases in which it can be significant, e.g.
Bino-like neutralinos in supersymmetry. The calculation of the relic density 
in the presence of coannihilations is a bit more involved (due to the
larger number of processes which need to be considered), but nevertheless
pretty straightforward. For UED models, where mass degeneracies are 
generically expected, the complete set of coannihilation processes
which are relevant for the $\gamma_1$ and $Z_1$ LKP case in 5D UED 
have been calculated \cite{Servant:2002aq,Burnell:2005hm,Kong:2005hn}.
We shall make use of them in our analysis below in Section~\ref{sec:relic}.
After reviewing the case of $\gamma_1$ LKP, which has been previously
discussed in the context of minimal UED, we shall also consider 
$Z_1$ LKP and illustrate the effects of
coannihilations with KK quarks on its relic abundance.
Since a calculation of coannihilations in 6D UED models is still lacking,
there we shall consider only one specific example in detail --
the previously discussed case of $\gamma_H$ \cite{Dobrescu:2007ec}. 
The corresponding results for the direct detection rates of $Z_H$ 
can be obtained by a simple scaling of the gauge couplings.

A small mass splitting also has a large impact on the expected 
direct detection signals, whenever the particle degenerate with
the LKP can be exchanged in an $s$-channel. This situation
may in principle arise in supersymmetry, if the squarks are very light,
but this would be viewed by most people as a fortuitous accident.
On the other hand, such a degeneracy occurs much more naturally 
in UED, where the masses of the KK quarks and the LKP necessarily have a 
common origin (the scale of the extra dimension). 
The mass degeneracy may lead to a substantial enhancement of the 
LKP elastic scattering rate \cite{Cheng:2002ej}. In Section~\ref{sec:sigma}
we first review the calculation of the spin-independent and the 
spin-dependent elastic scattering cross sections for the $\gamma_1$ 
LKP case \cite{Cheng:2002ej}. Then we also consider the case of 
$Z_1$, $H_1$ and $\gamma_H$ LKP, paying special attention to the 
enhancement of the cross sections in the limit of small mass splittings.

Finally, the mass splitting between the dark matter candidate and 
the rest of the new physics spectrum is an important parameter
for collider searches as well. The discovery reach for new physics
at colliders is greatly diminished if the mass splittings are small.
This is because the {\em observable} energy in the detector would then be 
rather small as well, in spite of the large amount of energy present
in the events.
Correspondingly, the {\em measured} missing energy (and any related
variable such as $H_T$) is also rather small, which 
makes it more difficult to extract the new physics signal from 
the SM backgrounds. Fortunately, as mentioned above, this is 
precisely the case when direct detection is more promising.
In Sec~\ref{sec:kkdm} we shall explore this complementarity
for various KK DM scenarios, focusing on KK gauge boson dark matter.
From the previous discussion it should be clear that having specified 
the nature of the DM particle, the two most relevant parameters
are the DM particle mass $m_\chi$ and the mass splitting with the
nearest heavier colored particles. In Sec~\ref{sec:kkdm} we shall utilize
this two-dimensional parameter space, and contrast constraints
from different sources: colliders, cosmological observations, and 
current or planned direct detection experiments (the latter are
first extensively reviewed in Sec.~\ref{sec:wimp}).
As expected, we find that colliders and dark matter searches
are highly complementary, while the WMAP constraint is orthogonal 
to them but is somewhat more model-dependent.
Section~\ref{sec:conclusion} is reserved for a summary and conclusions.
In the appendix we write out some technical details of our analysis.

\section{Universal Extra Dimensions and Kaluza-Klein Dark Matter}\label{sec:ued}

\subsection{Review on Universal Extra Dimensions}
\label{sec:model}

Models with universal extra dimensions place all Standard Model particles 
in the bulk of one or more compactified flat extra dimensions. 
In the simplest and most popular version, there is a single extra dimension 
compactified on an interval, $S_1/Z_2$. In UED, each SM particle has a whole tower of
KK modes. The individual modes are labelled by an integer $n$, called KK number,
which is nothing but the number of quantum units of momentum which the SM particle
carries along the extra dimension. A peculiar feature of UED is the conservation of 
Kaluza-Klein number at tree level, which is a simple consequence 
of momentum conservation along the extra dimension. 

However, the fixed points in orbifold compactifications break translation 
invariance along the extra dimension. As a result, KK number is broken by
bulk and brane radiative effects~\cite{Georgi:2000ks,vonGersdorff:2002as,Cheng:2002iz}
down to a discrete conserved quantity, the so called KK parity $(-1)^n$.
The geometrical origin of KK parity in the simplest ($S_1/Z_2$) case is 
the invariance under reflections with respect to the center of the interval. 
Since KK parity is conserved, the lightest KK parity odd particle is a suitable WIMP
candidate \cite{Dienes:1998vg,Cheng:2002iz,Servant:2002aq,Cheng:2002ej}.
KK parity also ensures that the KK-parity odd KK partners (e.g. those at level one)
are always pair-produced in collider experiments. This is reminiscent of 
the case of supersymmetry models with conserved $R$-parity. Therefore,
the limits on UED KK modes from collider searches are relatively weak 
and are rather similar to the limits on superpartners. KK-parity is also 
responsible for weakening the potential indirect limits on UED models from
low-energy precision data. Just like SUSY models with R-parity, the
virtual effects from new physics only appear at the loop level and are 
loop suppressed \cite{Buras:2002ej,Buras:2003mk,Haisch:2007vb}.

Since all KK modes carry momentum along the extra dimension, at tree-level 
their masses receive a dominant contribution $\frac{n}{R}\sim n\ {\rm TeV}$, 
and a subdominant contribution from the corresponding SM particle mass.
All KK modes at a given KK level $n$ are therefore quite degenerate.
The KK modes of the lightest SM particles (photons, leptons, light quarks)
even appear to be absolutely stable {\em at tree level}. However, this conclusion 
is invalidated after accounting for the radiative corrections to 
the KK masses. The latter are proportional to $\frac{n}{R}$
and are sufficient to lift the degeneracy between the lightest
KK modes, leaving only one of them (the true LKP) as absolutely stable \cite{Cheng:2002iz}.

The nature of the LKP, on the other hand, is more model-dependent.
In the minimal 5D UED model, where the boundary terms at the cut-off scale are ignored,
the lightest KK particle is typically the $n=1$ mode $B_1$ of the 
hypercharge gauge boson~\cite{Cheng:2002iz}. Since the 
Weinberg angle for the level one neutral gauge bosons is rather small,
$B_1$ is essentially also a mass eigenstate, the KK ``photon'', and we 
shall therefore denote it as $\gamma_1$. The KK photon $\gamma_1$ is 
an attractive dark matter candidate~\cite{Servant:2002aq,Cheng:2002ej}, 
whose relic abundance is consistent with the observed dark matter density 
for a mass range between 500\,GeV and about 1.5 TeV, as shown by detailed 
computations including coannihilations~\cite{Burnell:2005hm,Kong:2005hn} and 
level-2 resonances~\cite{Kakizaki:2005en,Kakizaki:2005uy,Kakizaki:2006dz}. 
Direct detection of this KK dark matter may be within reach of the next 
generation experiments~\cite{Cheng:2002ej,Servant:2002hb,Majumdar:2002mw,Oikonomou:2006mh}. 
Indirect detection of KK dark matter also has better prospects than 
the case of neutralinos in SUSY~\cite{Cheng:2002ej,Hooper:2002gs,Bertone:2002ms,Hooper:2004xn,Bergstrom:2004cy,Bergstrom:2004nr,Bringmann:2005pp,Birkedal:2005ep,Barrau:2005au}.

In UED the bulk interactions of the KK modes readily follow 
from the Standard Model Lagrangian and contain no unknown parameters other 
than the mass $m_h$ of the Standard Model Higgs boson. In contrast, the boundary 
interactions, which are localized on the orbifold fixed points, are in 
principle arbitrary, and thus correspond to new free parameters in the theory. 
They are in fact renormalized by bulk interactions, and are scale dependent~\cite{Georgi:2000ks}.
Therefore, we need an ansatz for their values at a particular scale. Virtually all existing
studies of UED have been done within the framework of minimal UED (MUED), 
in which the boundary terms are assumed to vanish at the 
cut-off scale $\Lambda$, and are subsequently generated through 
RGE evolution to lower scales (see~\cite{Cheng:2002iz,Cheng:2002ab} 
for 5D and~\cite{Ponton:2005kx,Burdman:2006gy} for 6D). 
In the minimal UED model therefore there are only two input parameters: 
the size of the extra dimension $R$, and the cut-off scale $\Lambda$. 
Of course, there are no compelling reasons for assuming vanishing boundary terms: 
the UED model should be treated only as an effective
theory which is valid up to the high scale $\Lambda$, where
it is matched to some more fundamental theory, which is generically 
expected to induce nonzero boundary terms at the matching scale $\Lambda$. 
As already mentioned in the introduction, nonvanishing boundary terms may 
change both the nature of the LKP, as well as the size of the KK mass splittings. 
The resulting phenomenology may be very different from the minimal case.
This is why in this paper we shall allow for more general scenarios with 
$Z_1$ and $H_1$ LKP\footnote{Other dark matter candidates, 
such as the level-1 KK mode of the graviton or of a right-handed neutrino, 
are also viable for certain ranges of parameters in models with one universal 
extra dimension~\cite{Feng:2003xh,Feng:2003nr,Shah:2006gs,Cembranos:2006gt,Matsumoto:2006bf,Matsumoto:2007dp}.}. 
In each case, we shall take the LKP mass $m_{LKP}$
and the LKP - KK quark mass splitting 
\begin{equation}
\Delta_{q_1} = \frac{m_{q_1}-m_{LKP}}{ m_{LKP}} \, ,
\label{deltaq}
\end{equation}
as free parameters.
We remind the reader that after compactification, the
low energy effective theory contains two massive (Dirac) KK fermions 
for each (Dirac) fermion in the Standard Model. The KK fermions are properly
referred to as $SU(2)_W$-doublet KK fermions or $SU(2)_W$-singlet KK fermions.
However, in the literature they are sometimes called ``left handed'' and ``right handed'',
referring to the chirality of the corresponding Standard Model fermion at 
the zero level of the KK tower. This nomenclature may lead to some confusion, 
since all KK fermions are Dirac and have both chiralities.
In our study, we shall treat the $SU(2)_W$-doublet KK quarks 
(often denoted by $Q_1$) and the $SU(2)_W$-singlet KK quarks (often denoted by $q_1$) equally,
thus avoiding the need for two separate mass splitting parameters (for example,
a separate $\Delta_{Q_1}$ and $\Delta_{q_1}$). The generalization to the case of different 
KK quark masses is rather straightforward.

We shall also explore cases with more than one universal extra dimension.
Theories with two universal extra dimensions also contain a KK parity. 
Under the simplest compactification which leads to chiral zero-mode fermions
(a ``chiral'' square with adjacent sides identified~\cite{Burdman:2005sr,Dobrescu:2004zi}),
the KK parity transformations are reflections with respect to the 
center of the square. Momentum along the two compact dimensions is 
quantized so that any 6-dimensional field propagating on the square
appears as a set of 4-dimensional particles labelled by two positive 
integers, $(n,m)$. These particles are odd under KK parity when
$n+m$ is odd and are even otherwise. In any process, odd particles 
may be produced or annihilated only in pairs. 
The lightest odd particle, which is one of the (1,0) states, is thus stable.
Gauge bosons propagating in six dimensions may be polarized along the 
two extra dimensions. As a result, for each spin-1 KK particle 
associated with a gauge boson, there are two spin-0 KK fields 
transforming in the adjoint representation of the gauge group. 
One linear combination becomes the longitudinal degree of freedom of the
spin-1 KK particle, while the other linear combination remains as a
physical spin-0 particle, called the spinless adjoint\footnote{In 
contrast to the 6D case, KK particles in 5D UED are labelled by only one 
integer and spinless adjoints do not exist since there is only one gauge 
boson degree of freedom polarized along the extra dimension.}.
In the minimal model with vanishing boundary terms, 
the radiative corrections~\cite{Ponton:2005kx,Burdman:2006gy}, 
are such that the lightest (1,0) particle on the chiral 
square~\cite{Burdman:2005sr,Dobrescu:2004zi} 
is always a linear combination of the electrically-neutral 
spinless adjoints of the electroweak gauge group. 
Due to the small mixing angle, this linear combination is essentially a photon
polarized along the extra dimensions.
Similar to its 5D cousin $\gamma_1$, the spinless photon $\gamma_H$ in 6D UED
is also a viable dark matter candidate~\cite{Dobrescu:2007ec}.
See Refs.~\cite{Mohapatra:2002ug,Hsieh:2006fg,Hsieh:2006qe} 
for KK dark matter candidates in UED models with an extended gauge symmetry.

\subsection{Relic Density Calculation with Coannihilations}
\label{sec:relic}
We briefly review the calculation of the relic density including coannihilation processes.
When the relic particle $\chi$ is nearly degenerate with other particles
in the spectrum, its relic abundance is determined not only by its own 
self-annihilation cross section, but also by annihilation processes involving
the heavier particles. The generalization of the relic density calculation including 
this ``coannihilation'' case is straightforward~\cite{Griest:1990kh,Servant:2002aq}.
Assume that the particles $\chi_i$ are labelled according to their masses, 
so that $m_i < m_j$ when $i < j$. 
The number densities $n_i$ of the various species $\chi_i$ obey
a set of Boltzmann equations. It can be shown that
under reasonable assumptions \cite{Griest:1990kh},
the ultimate relic density $n_\chi$ of the lightest species $\chi_1$ 
(after all heavier particles $\chi_i$ have decayed into it)
obeys the following simple Boltzmann equation
\begin{equation}
\frac{d n_\chi}{ d t} = -3 Hn_\chi - \langle \sigma_{eff} v \rangle ( n_\chi^2 - n^2_{eq})\ ,
\end{equation}
where $H$ is the Hubble parameter, $v$ is the relative velocity between the two incoming particles, 
$n_{eq}$ is the equilibrium number density and 
\begin{eqnarray}
\sigma_{eff}(x) &=& \sum_{ij}^N \sigma_{ij} \frac{g_i g_j}{g_{eff}^2} 
                 (1 + \Delta_i)^{3/2} (1 + \Delta_j)^{3/2} \exp(-x ( \Delta_i + \Delta_j ))\ ,
\label{sigmaeff}\\
g_{eff}(x)   &=& \sum_{i=1}^N g_i (1+\Delta_i)^{3/2} \exp(-x \Delta_i)\ , 
\label{geff}\\
\Delta_i     &=& \frac{m_i - m_1}{m_1}\ , \text{~~~~} x = \frac{m_1}{T}.
\end{eqnarray}
Here $\sigma_{ij}\equiv \sigma(\chi_i\chi_j\to SM)$ are the various pair 
annihilation cross sections into final states with SM particles,
$g_i$ is the number of internal degrees of freedom of particle $\chi_i$ and 
$n_\chi \equiv \sum_{i=1}^N n_i$ is the density of $\chi_1$ we want to calculate.

By solving the Boltzmann equation analytically with appropriate approximations
~\cite{Griest:1990kh, Servant:2002aq}, the abundance of the lightest species
$\chi_1$ is given by 
\begin{equation}
\Omega_\chi h^2 \approx \frac{1.04 \times 10^9\ {\rm GeV}^{-1}}{M_{Pl}}
\frac{x_F}{\sqrt{g_\ast(x_F)}} \frac{1}{I_a+3 I_b/x_F }\ ,
\label{oh2coann}
\end{equation}
where the Planck mass scale is $M_{Pl} = 1.22\times 10^{19}$\,GeV and
$g_\ast$ is the total number of effectively massless degrees of freedom at temperature $T$:
\begin{equation}
g_\ast(T) = \sum_{i=bosons} g_i + \frac{7}{8} \sum_{i=fermions}g_i\ .
\label{gstar}
\end{equation}
The functions $I_a$ and $I_b$ are defined as
\begin{eqnarray}
I_a &=& x_F \int_{x_F}^\infty a_{eff}(x) x^{-2} d x \ ,
\label{I_a} \\
I_b &=& 2 x_F^2 \int_{x_F}^\infty b_{eff}(x) x^{-3} d x\ .
\label{I_b}
\end{eqnarray}
The freeze-out temperature, $x_F$, is found iteratively from
\begin{equation}
x_F = \ln \left ( c(c+2) \sqrt{\frac{45}{8}} \frac{g_{eff}(x_F)}{2\pi^3} 
                 \frac{m_1 M_{Pl} (a_{eff}(x_F)+6b_{eff}(x_F)/x_F)}
{\sqrt{g_\ast(x_F) x_F}}  \right )\ ,
\label{xfcoann}
\end{equation}
where the constant $c$ is determined empirically by comparing to numerical solutions of 
the Boltzmann equation and here we take $c=\frac{1}{2}$ as usual. 
$a_{eff}$ and $b_{eff}$ are the first two terms 
in the velocity expansion of $\sigma_{eff}$
\begin{equation}
\sigma_{eff}(x)\,v = a_{eff}(x) + b_{eff}(x)\, v^2 + {\cal O}(v^4)\ .
\label{sigmaeffv}
\end{equation}
Comparing Eqns.~(\ref{sigmaeff}) and (\ref{sigmaeffv}), one gets
\begin{eqnarray}
a_{eff}(x) &=& \sum_{ij}^N a_{ij} \frac{g_i g_j}{g_{eff}^2} 
                 (1 + \Delta_i)^{3/2} (1 + \Delta_j)^{3/2} \exp(-x ( \Delta_i + \Delta_j ))\ , 
\label{aeff}\\
b_{eff}(x) &=& \sum_{ij}^N b_{ij} \frac{g_i g_j}{g_{eff}^2} 
                 (1 + \Delta_i)^{3/2} (1 + \Delta_j)^{3/2} \exp(-x ( \Delta_i + \Delta_j ))\ ,
\label{beff}
\end{eqnarray}
where $a_{ij}$ and $b_{ij}$ are obtained from $\sigma_{ij}v = a_{ij} + b_{ij} v^2 + {\cal O}(v^4)$ and 
$v$ is the relative velocity between the two annihilating particles in the initial state.
Considering relativistic corrections \cite{Srednicki:1988ce}
to the above treatment results in an additional subleading 
term which can be accounted for by the simple replacement 
\beq
b\to b-\frac{1}{4}a \, ,
\label{bcorr}
\eeq
in the above formulas.
For our calculation of the relic density, 
we use the cross sections given in Refs.~\cite{Servant:2002aq,Burnell:2005hm,Kong:2005hn}.

\begin{figure}[t]
\includegraphics[width=0.48\textwidth]{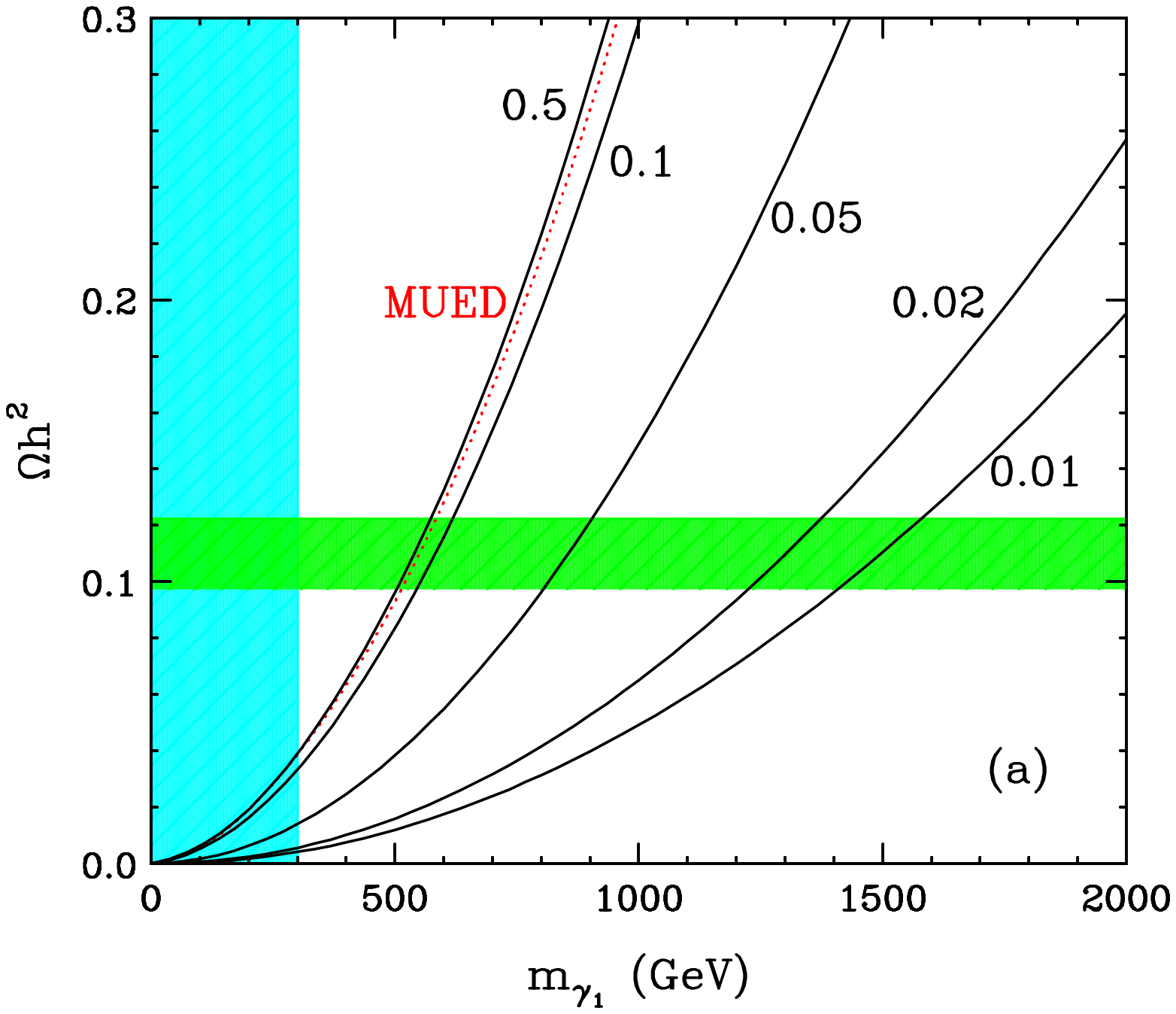}
\hspace{0.1cm}
\includegraphics[width=0.48\textwidth]{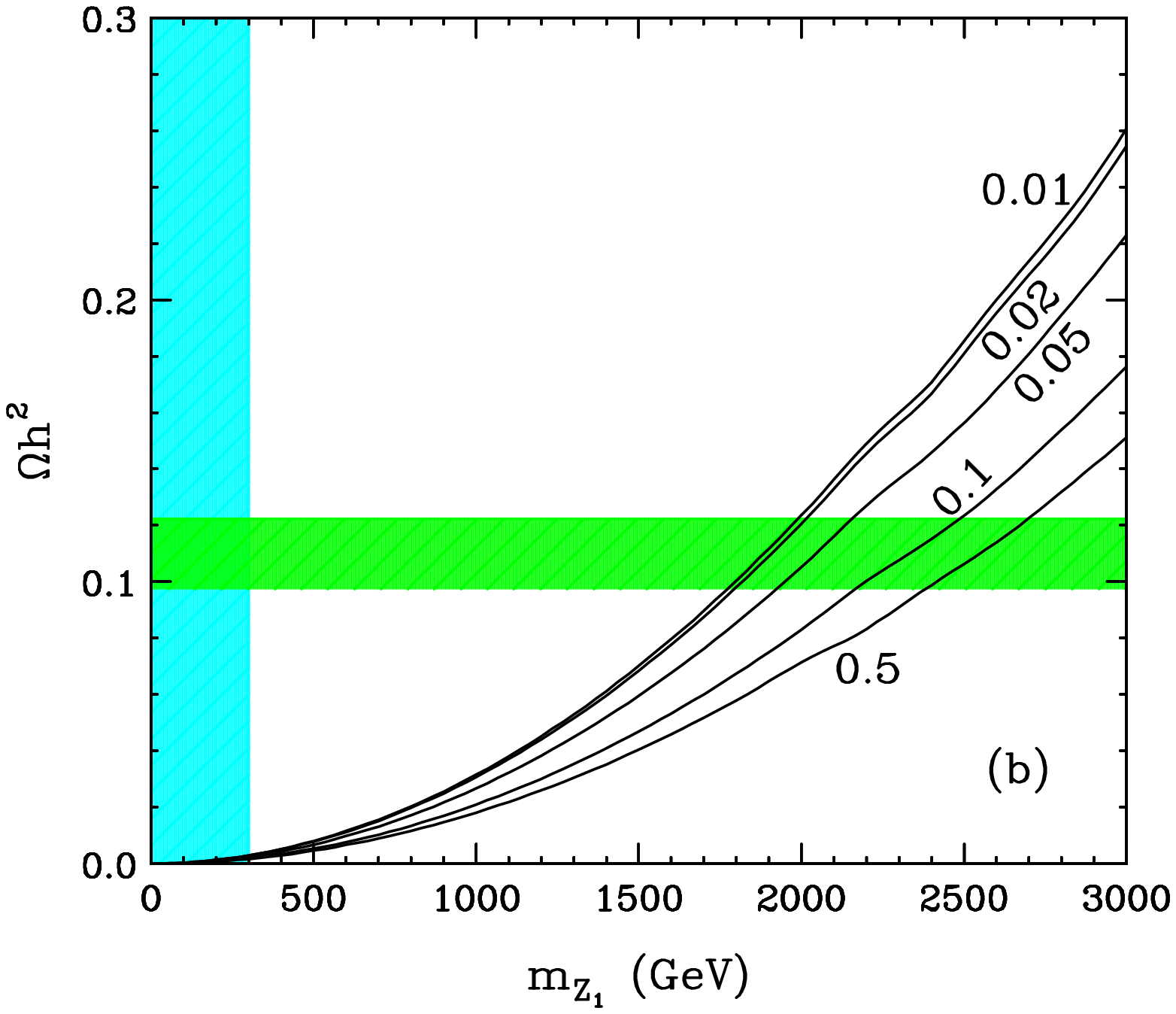}
\caption{\sl Relic density of the LKP ((a) $\gamma_1$ and (b) $Z_1$) 
as a function of the LKP mass. 
The (black) solid lines show the LKP relic density 
for several choices of the mass splitting (\ref{deltaq})
between the LKP and the KK quarks. 
We assume that singlet and doublet KK quarks are degenerate.
The green horizontal band
denotes the preferred WMAP region for the relic density 
$0.1037 < \Omega_{CDM}h^2 < 0.1161$. The cyan vertical band delineates
values of $m_{LKP}$ disfavored by precision data.
(a) We vary the $q_1$ mass by hand, keeping the 
masses of the remaining particles fixed as in MUED.
The solid lines from top to bottom correspond to
$\Delta_{q_1}=0.5, 0.1, 0.05, 0.02, 0.01$.
The (red) dotted line is
the result from the full calculation in MUED, including all coannihilation 
processes, with the proper MUED choice for all masses. 
(b) We assume $Z_1$ and $W_1^\pm$ are degenerate, the gluon is heavier 
than $Z_1$ by 20\%, while all other KK particles are heavier than $Z_1$ by 10\%. 
The solid lines from top to bottom 
correspond to $\Delta_{q_1}=0.01,0.02,0.05,0.1,0.5$.}
\label{fig:Omegah2_B1_Z1}
\end{figure}

\begin{figure}[t]{
\epsfig{file=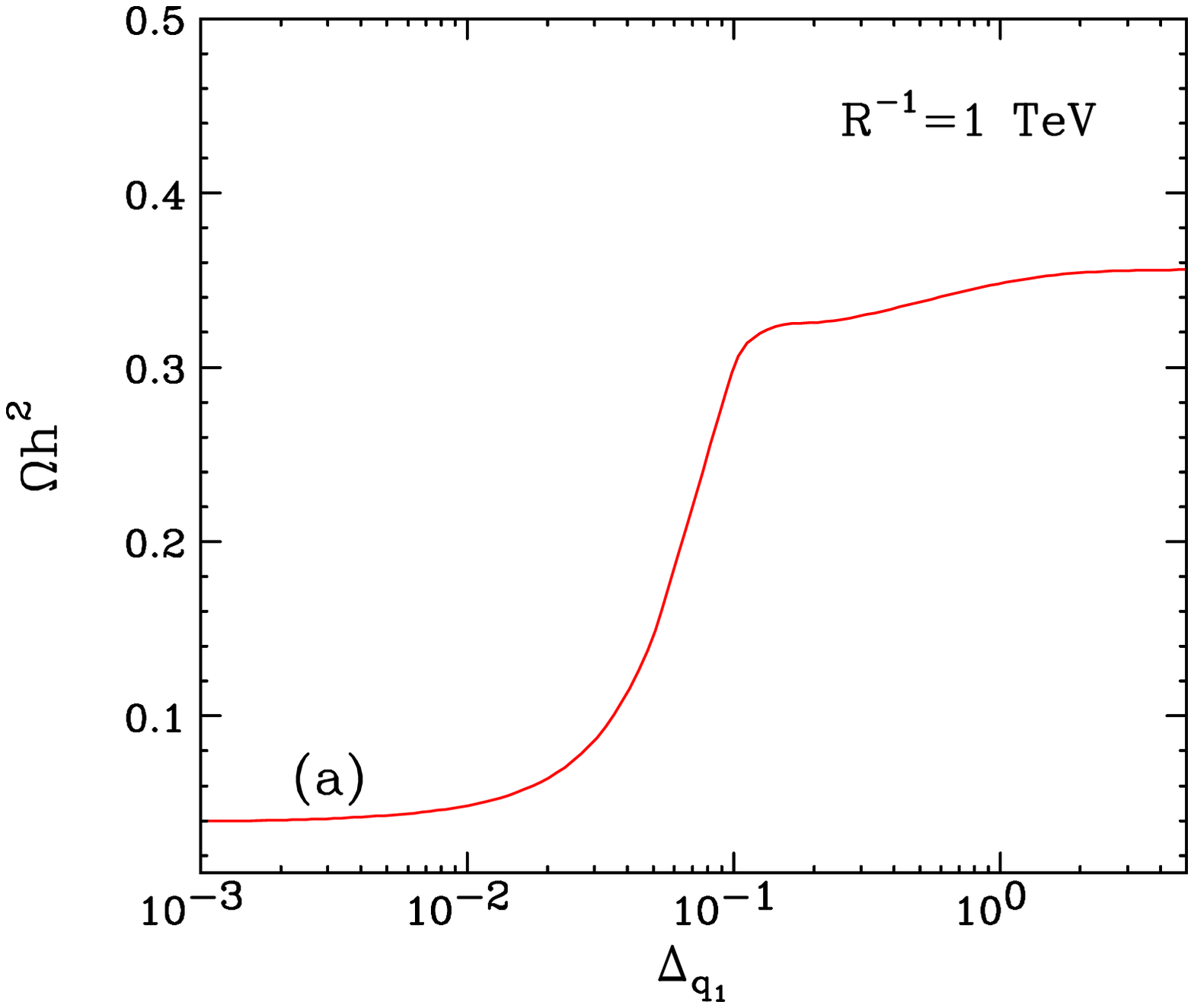,height=6.0cm}\hspace{0.7cm}
\epsfig{file=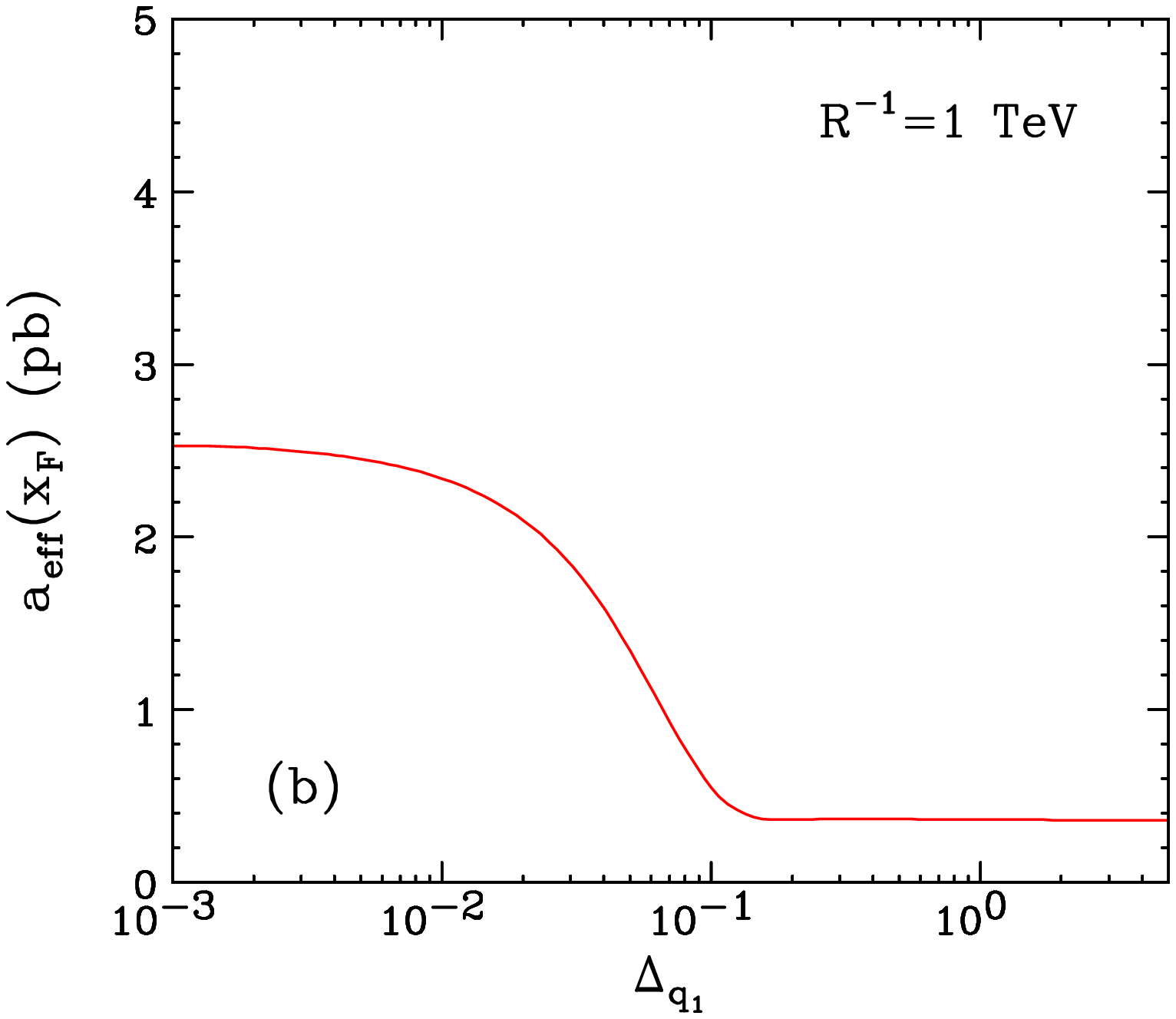,height=6.0cm}\hspace{0.2cm}
\\
\vspace*{0.5cm}\epsfig{file=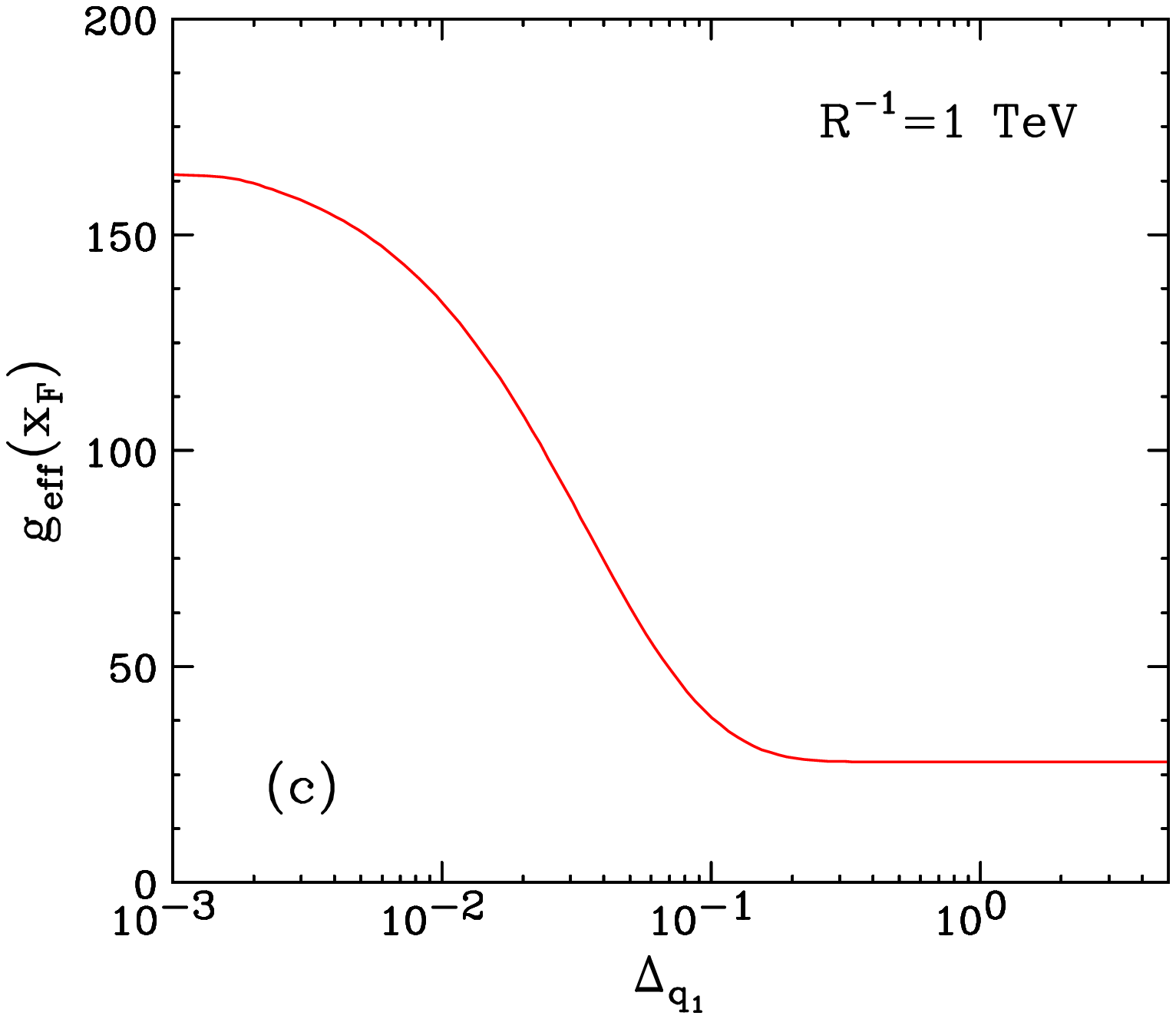,height=6.0cm}\hspace{0.7cm}
\epsfig{file=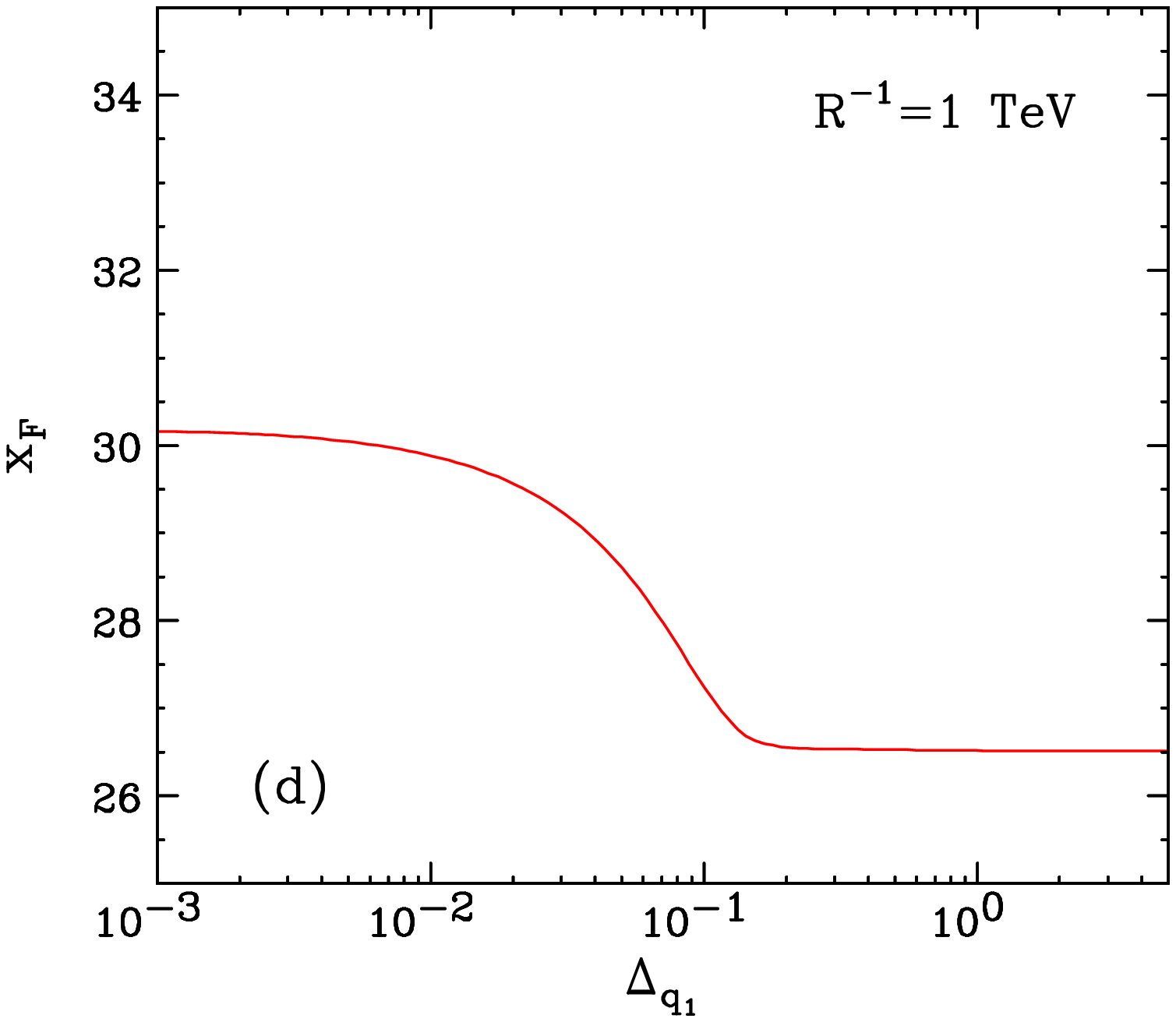,height=6.0cm}
\caption{\sl Plots of various quantities
entering the $\gamma_1$ LKP relic density computation,
as a function of the mass splitting $\Delta_{q_1}$ 
between the LKP and the KK quarks:
(a) relic density, 
(b) $a_{eff}(x_F)$,
(c) $g_{eff}(x_F)$,
and (d) $x_F$. In all four panels, the KK quark masses are varied 
by hand according to the value of $\Delta_{q_1}$, while the 
masses of the $\gamma_1$ and the remaining KK modes are held fixed
at their nominal values predicted in MUED for $R^{-1}=1$ TeV.  }
\label{fig:b1aeff}   
}
\end{figure}

\begin{figure}[t]{
\epsfig{file=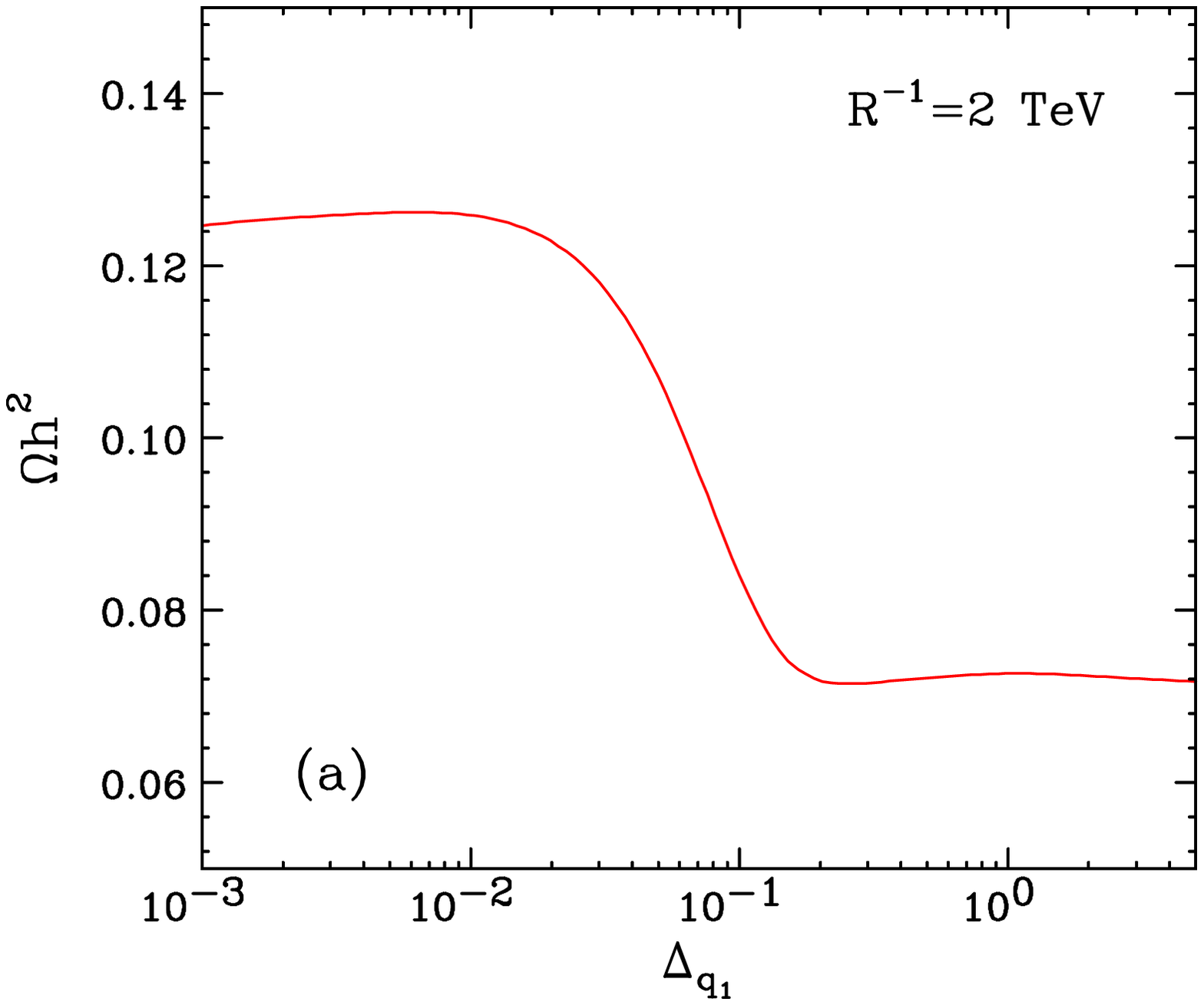,height=6.0cm}\hspace{0.7cm}
\epsfig{file=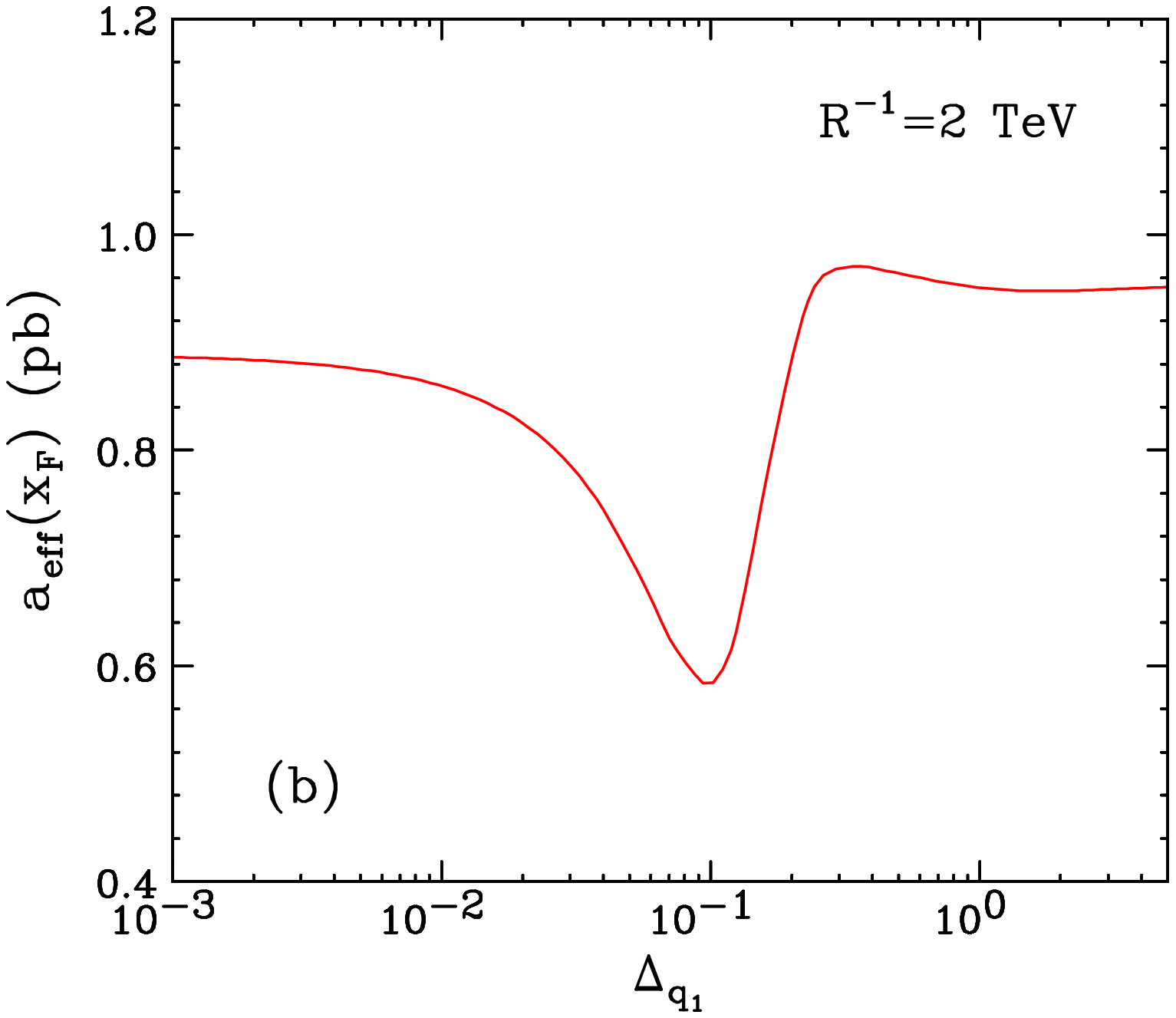,height=6.0cm}\hspace{0.2cm}
\\
\vspace*{0.5cm}\epsfig{file=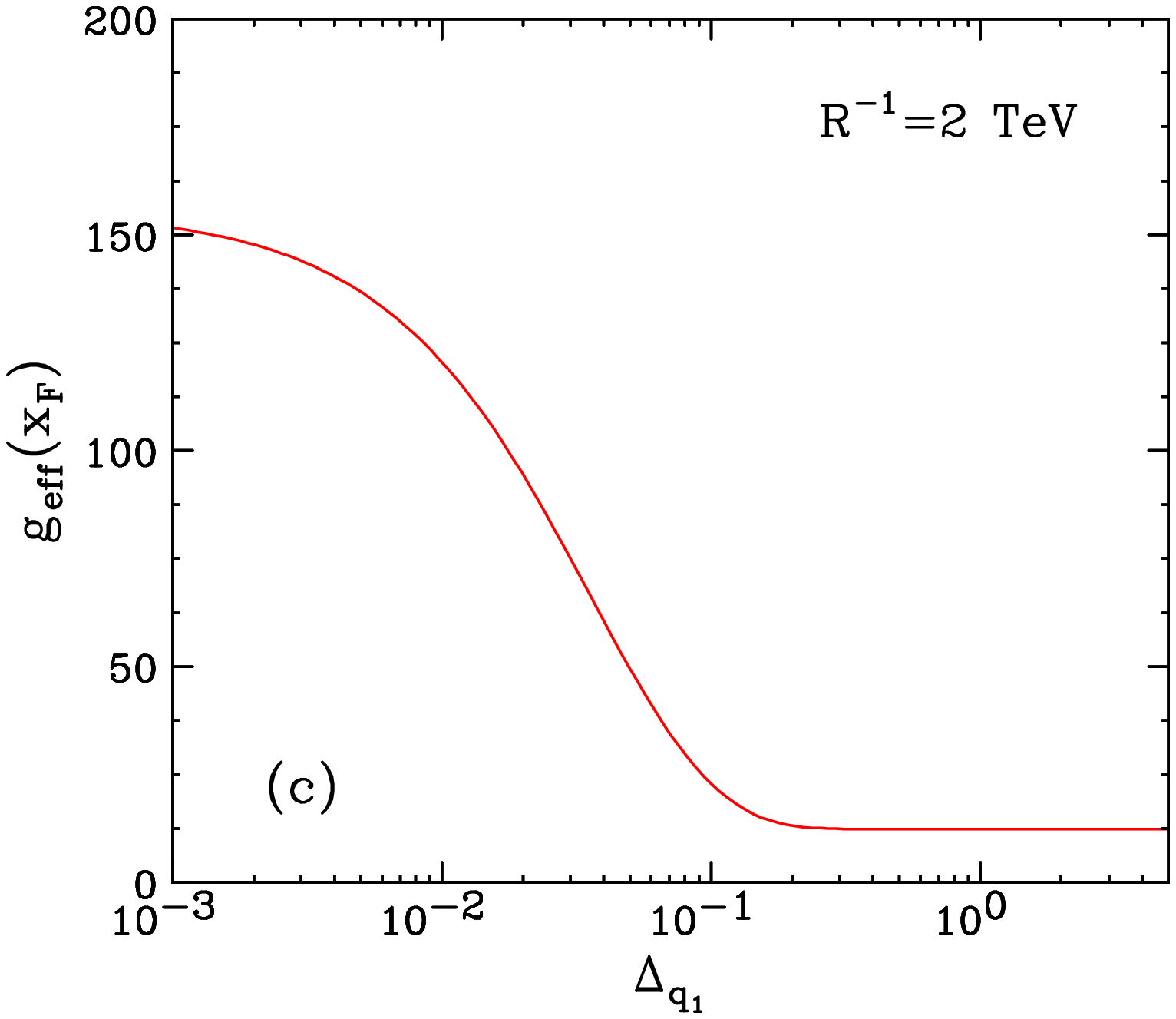,height=6.0cm}\hspace{0.7cm}
\epsfig{file=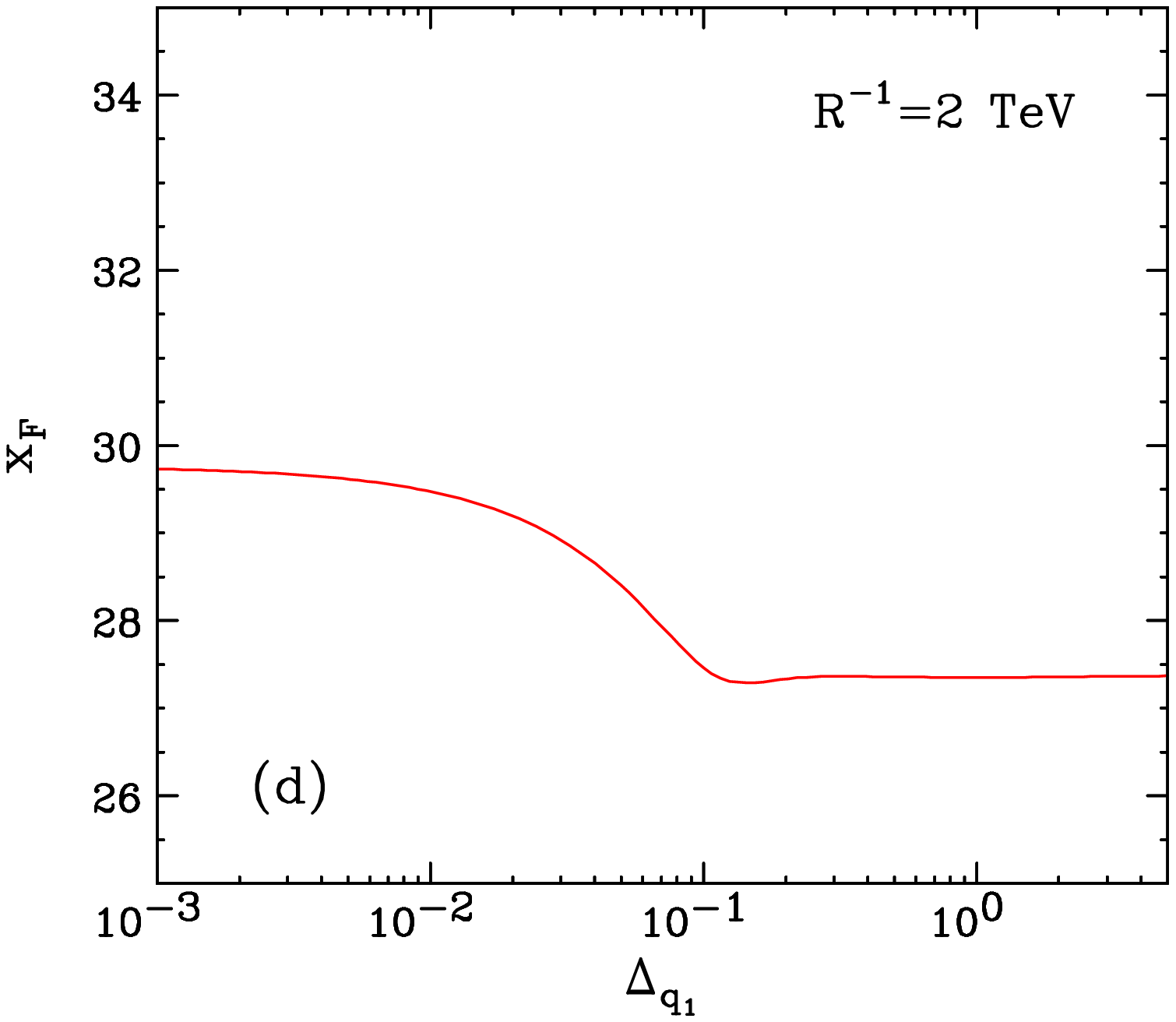,height=6.0cm}
\caption{\sl The same as Fig.~\ref{fig:b1aeff} but for $Z_1$ LKP.
Here the $Z_1$ and $W^\pm_1$ masses are taken to be 2 TeV, 
the KK quark masses are varied in accordance with $\Delta_{q_1}$, 
while the remaining spectrum
is fixed as in Fig.~\ref{fig:Omegah2_B1_Z1}b. }
\label{fig:aeffZ1}
}
\end{figure}

As explained earlier, the assumptions behind MUED can be easily 
relaxed by allowing nonvanishing boundary terms at the scale $\Lambda$ 
\cite{Cheng:2002ab,Carena:2002me,delAguila:2003bh,delAguila:2006kj}. 
This would modify the KK spectrum and correspondingly change the MUED predictions for
the KK relic density. Within the modified KK spectrum, 
any neutral KK particle could be a dark matter candidate.
As an illustration here we shall consider the case of $\gamma_1$ and 
$Z_1$ LKP\footnote{The $Z_1$ is also a good dark matter candidate 
in warped extra dimensions with KK parity~\cite{Agashe:2007jb}.}, 
for which the results for the relevant coannihilation processes 
are available in the literature \cite{Burnell:2005hm,Kong:2005hn}.
In Fig.~\ref{fig:Omegah2_B1_Z1}, we show the relic densities of $\gamma_1$ and $Z_1$ 
as a function of the corresponding LKP mass ($m_{\gamma_1}$ or $m_{Z_1}$) in 5D UED. 
We include coannihilation effects with all $n=1$ KK particles with properly defined masses.
The (black) solid lines show the LKP relic density for several choices of the mass splitting 
(\ref{deltaq}) between the LKP and the KK quarks. 
We assume that singlet and doublet KK quarks are degenerate (i.e., $\Delta_{Q_1}=\Delta_{q_1}$).
The green horizontal band denotes the preferred $2\sigma$-WMAP region for the relic density 
$0.1037 < \Omega_{CDM}h^2 < 0.1161$~\cite{Dunkley:2008ie}. The cyan vertical band delineates
values of $m_{LKP}$ disfavored by precision data~\cite{Gogoladze:2006br, Appelquist:2002wb} \footnote{While 
there have been no studies of non-minimal UED models with $Z_1$ LKP, we anticipate that
the precision bounds in that case will be similar to those in MUED, therefore we display the same 
indirect constraint in Fig.~\ref{fig:Omegah2_B1_Z1}b.}.
In each case of Fig.~\ref{fig:Omegah2_B1_Z1}a, we use the MUED spectrum to fix the masses of the
remaining particles, and then vary the (common) KK-quark mass $m_{q_1}$ by hand.
The solid lines from top to bottom correspond to $\Delta_{q_1}=0.5, 0.1, 0.05, 0.02, 0.01$. 
The (red) dotted line is the result from the full calculation in MUED, 
including all coannihilation processes, with the proper MUED choice for all masses. 
In Fig.~\ref{fig:Omegah2_B1_Z1}b we assumed $Z_1$ and $W_1^\pm$ are degenerate,
the gluon is heavier than $Z_1$ by 20\%, while all other KK particles 
are heavier than $Z_1$ by 10\%. 
The solid lines from top to bottom correspond to $\Delta_{q_1}=0.01,0.02,0.05,0.1,0.5$.
Some individual quantities entering the relic density calculation 
for $\gamma_1$ ($Z_1$) LKP are shown in Fig.~\ref{fig:b1aeff}
(Fig.~\ref{fig:aeffZ1}).

We see that coannihilations in the case of $\gamma_1$ LKP decrease 
the prediction for $\Omega h^2$ and therefore increase the range of 
preferred $m_{\gamma_1}$ values. For $\Delta_{q_1}$
on the order of a few percent, the desired range of 
$m_{\gamma_1}$ is pushed beyond 1 TeV. This poses a challenge for any 
collider searches for UED, since the KK production cross sections
at the LHC become kinematically suppressed for heavier KK modes. 
What is even worse, the small mass splitting $\Delta_{q_1}$ degrades the
quality of the discovery signatures, e.g. the cascade decays of
the KK quarks would yield only (rather soft) jets and no leptons.

On the other hand, Fig.~\ref{fig:Omegah2_B1_Z1}b reveals 
that coannihilations with KK quarks have the opposite effect 
in the case of $Z_1$ LKP\footnote{A similar behavior 
exists in the case of $\gamma_1$ LKP
when coannihilations are caused by the $SU(2)$-singlet KK leptons
$\ell_{R1}$~\cite{Servant:2002aq,Burnell:2005hm,Kong:2005hn}.}.
This time the effect of coannihilations is to increase
the prediction for $\Omega h^2$ and thus lower the preferred
range of values for $m_{Z_1}$. The lesson from Figs.~\ref{fig:Omegah2_B1_Z1}a
and \ref{fig:Omegah2_B1_Z1}b is that while coannihilations 
can be quite important, the sign of the effect cannot be easily 
predicted, since, as will be illustrated in Figs.~\ref{fig:b1aeff} 
and~\ref{fig:aeffZ1}, it depends on the detailed balance 
of several numerical factors entering the computation.
We shall discuss these in some detail in the remainder of 
this subsection. Readers who are not interested in these numerical 
details, are invited to jump to Section~\ref{sec:sigma}.


In Fig.~\ref{fig:b1aeff}a (Fig.~\ref{fig:aeffZ1}a) we plot the relic 
density of the $\gamma_1$ ($Z_1$) LKP, as a function of the mass 
splitting $\Delta_{q_1}$ between the KK quarks and the corresponding LKP.
The rest of the spectrum is held fixed as explained in the figure captions.
Figs.~\ref{fig:b1aeff}a and~\ref{fig:aeffZ1}a demonstrate the importance of
coannihilations at small mass splittings. For $\Delta_{q_1}$ larger 
than about $10-20\%$, coannihilations are turned off, but for KK quarks
within 10\% of the LKP mass, the coannihilation effect is significant.
For $\gamma_1$ LKP, it lowers the prediction for the relic density $\Omega h^2$,
while in the case of $Z_1$ LKP $\Omega h^2$ is enhanced. 
In order to understand this different behavior, it is sufficient 
to investigate the coannihilation effect on the effective cross section, 
and in particular the dominant term $a_{eff}$, which is plotted in 
Figs.~\ref{fig:b1aeff}b and~\ref{fig:aeffZ1}b. As can be seen from
eq.~(\ref{aeff}), every term contributing to $a_{eff}$ is a ratio between
two quantities, each of which has a nontrivial $\Delta_{q_1}$ dependence.
The denominator is common to all terms and is nothing but the
effective number of heavy particle degrees of freedom $g_{eff}$ 
defined in eq.~(\ref{geff}). We show the $g_{eff}$ dependence on 
$\Delta_{q_1}$ in Figs.~\ref{fig:b1aeff}c and~\ref{fig:aeffZ1}c.
As expected, $g_{eff}$ increases significantly after the  
turn-on of coannihilations (below $\Delta_{q_1}\sim0.1$), 
due to the large multiplicity of KK quark states.
At the same time, the numerator of each term contributing to the
$a_{eff}$ sum (\ref{aeff}) is simply the Boltzmann 
suppressed annihilation cross section, which also increases 
with the onset of coannihilations (at small mass splittings $\Delta_{q_1}$). 
The net effect on $a_{eff}$ 
is determined by which of these two quantities increases {\em faster}
at small $\Delta_{q_1}$, relative to the nominal case without coannihilations.
In the case of $\gamma_1$ LKP, the self-annihilation cross sections 
are rather weak, due to the smallness of the hypercharge gauge coupling.
Adding the contributions from the strongly interacting KK quark sector
has therefore a much larger impact than the associated increase 
in the effective number of degrees of freedom $g_{eff}$. As a result, 
$a_{eff}$ increases and $\Omega h^2$ decreases, as shown in
Figs.~\ref{fig:b1aeff}a and \ref{fig:b1aeff}b.
In contrast, in the case of $Z_1$ LKP, the self-annihilation cross sections
by themselves are already larger, due to the larger value of the 
weak gauge coupling. The gain from the addition of the KK quark 
coannihilation processes is more than compensated by the associated  
increase in the effective number of degrees of freedom $g_{eff}$. 
As a result, in this case $a_{eff}$ decreases and $\Omega h^2$ increases, 
as shown in Figs.~\ref{fig:aeffZ1}a and \ref{fig:aeffZ1}b.

In conclusion, we should mention that the KK Higgs boson $H_1$ in principle
can also be a potential dark matter candidate. 
The calculation of its relic density is somewhat more model-dependent 
and we do not consider it here.

\subsection{Elastic Scattering Cross Sections}
\label{sec:sigma}

\begin{figure*}[t]
\unitlength=1.0 pt
\SetScale{1.}
\SetWidth{1.1}      
\begin{center}
\begin{picture}(400,80)(10,20)
\Photon(20,80)( 50,50){3}{4}
\Photon(90,50)( 120,80){3}{4}
\ArrowLine(20,20)(50,50)
\ArrowLine(90,50)(120,20)
\ArrowLine(50,50)(90,50)
\Text(147,50)[r]{{\large +}}
\Photon(180,75)(225,75){3}{4}
\Photon(225,25)(270,25){3}{4}
\ArrowLine(225,25)(225,75)
\ArrowLine(180,25)(225,25)
\ArrowLine(225,75)(270,75)
\Text(297,50)[r]{{\large +}}

\Photon(330,75)(375,75){3}{4}
\ArrowLine(375,25)(420,25)
\ArrowLine(330,25)(375,25)
\DashLine(375,25)(375,75){3}
\Photon(375,75)(420,75){3}{4}

\Text(15,15)[r]{$q$}
\Text(15,85)[r]{$\gamma_1$}
\Text(175,25)[r]{$q$}
\Text(175,75)[r]{$\gamma_1$}
\Text(325,25)[r]{$q$}
\Text(325,75)[r]{$\gamma_1$}
\Text(76,61)[r]{$q_1$}
\Text(130,20)[r]{$q$}
\Text(135,85)[r]{$\gamma_1$}
\Text(285,25)[r]{$\gamma_1$}
\Text(285,78)[r]{$q$}
\Text(435,25)[r]{$q$}
\Text(435,78)[r]{$\gamma_1$}
\Text(242,50)[r]{$q_1$}
\Text(390,50)[r]{$h$}
\end{picture}
\end{center}
%
\caption{\sl Tree-level diagrams for the elastic scattering of $\gamma_1$ LKP with quarks.
The diagrams for the case of $Z_1$ LKP are similar.}
\label{fig:direct}
\end{figure*}
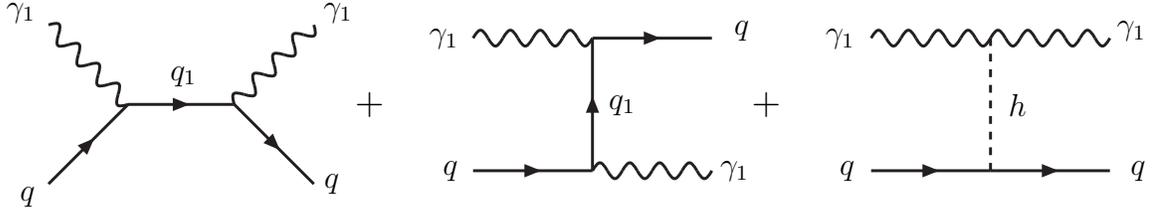

The elastic scattering of the LKP on a nucleon is described by the 
diagrams depicted in Fig.~\ref{fig:direct}. For $\gamma_1$ LKP, the 
corresponding results can be found in~\cite{Cheng:2002ej,Servant:2002hb}.
We follow the computation done in~\cite{Cheng:2002ej}\footnote{The
precise calculation of the heavy quark contribution to the processes
of Fig.~\ref{fig:direct} is rather involved -- the heavy flavors 
contribute only at the loop level, through the gluon content of the nucleon.
In the absence of an exact calculation of these effects in the literature,
we choose to conservatively ignore
the heavy flavor contributions altogether, as was done in \cite{Cheng:2002ej}.}.
The spin-independent cross section is given by 
\begin{equation}\label{scalar}
\sigma_{\text{scalar}} =  \frac{m_T^2}{4\pi\, (\mB + m_T)^2} 
\left[Z f_p +(A-Z) f_n\right]^2 \ ,
\end{equation}
where $m_T$ is the mass of the target nucleus, 
$Z$ and $A$ are respectively the nuclear charge and atomic number, while
\begin{equation}
f_p = \sum_{u, d, s} (\beta_q + \gamma_q) 
\langle p | \bar{q} q | p \rangle 
= \sum_{u, d, s} \frac{\beta_q + \gamma_q}{m_q} m_p 
f^p_{T_q} \ ,
\label{si}
\end{equation}
and similarly for $f_n$. In eq.~(\ref{si}) $m_p$ ($m_n$) stands for the proton (neutron) mass.
For the nucleon matrix elements
we take $f^{p}_{T_u}=0.020\pm 0.004$,
$f^{p}_{T_d}=0.026\pm 0.005$, $f^{n}_{T_u}=0.014\pm 0.003$,
$f^{n}_{T_d}=0.036\pm 0.008$, and $f^{p,n}_{T_s}=0.118\pm
0.062$~\cite{Ellis:2000ds}. The numerical coefficients $\beta_q$ and $\gamma_q$ 
in eq.~(\ref{si}) are defined as\footnote{Ref.~\cite{Servant:2002hb} 
contains a typo in the overall sign of the coefficient $\beta_q$,
which was denoted there as $S_q$.}
\begin{eqnarray} 
\beta_q \! \! &=& \! \!
\label{beta0}\frac{e^2}{\cos^2 \theta_W} \Big [E_q ( Y_{q_L}^2 \cos^2\alpha + Y_{q_R}^2 \sin^2\alpha) 
\frac{m_{q_L^1}^2+ \mB^2}{(m_{q_L^1}^2 - \mB^2)^2}  
+ \frac{  Y_{q_L} Y_{q_R} m_{q_L^1} \, \sin 2 \alpha }{\mB^2 - m_{q_L^1}^2}
+ ( L \to R ) \Big  ]\, \hspace{0.5cm} \\
&\approx& \! \! E_q \frac{e^2}{\cos^2 \theta_W} 
\left[ Y_{q_L}^2 \frac{\mB^2 + m_{q_L^1}^2}{(m_{q_L^1}^2 - \mB^2)^2} 
+ (L \to R) \right] \, \text{~~~for }\alpha = 0,
\label{beta} \\
\gamma_q \! \! &=& \! \! m_q \frac{e^2}{2 \cos^2 \theta_W}
\frac{1}{m_h^2} \label{gamma} \ ,
\end{eqnarray}
where $e$ is the electric charge, $\theta_W$ is the Weinberg angle, 
$m_{q_L^1}$ ($m_{q_R^1}$) is the mass of an $SU(2)_W$-doublet
($SU(2)_W$-singlet) KK quark, and $\alpha$ is the mixing angle in the KK quark mass matrix 
given by $\sin 2\alpha = 2m_q /(m_{q_L^1}+m_{q_R^1})$.
Eq.~\eqref{beta0} includes the mixing effect between two KK quarks and
eq.~\eqref{beta} is obtained in the limit when $\alpha=0$.
This mixing effect gives a minor correction to the cross section (at a few percent level) and 
we do not include it in our figures for 5D.
However it is important to keep it in the 6D case, as shown in Ref.~\cite{Dobrescu:2007ec}.
Our convention for the SM hypercharge
is $Y_i=Q_i-I_{3i}$, where $Q_i$ ($I_{3i}$) is the electric charge
(weak isospin) of particle $i$. $E_q$ in eq.~\eqref{beta} is the energy of a
bound quark and is rather ill-defined. In evaluating eq.~\eqref{si}, we
conservatively replace $E_q$ by the current\footnote{The actual choice of 
the value for $m_q$ is inconsequential since the $m_q$ factor in
eqs.~(\ref{beta}) and (\ref{gamma}) cancels against the $m_q$ factor 
in the denominator of eq.~(\ref{si}).} mass $m_q$.
As alluded to earlier, in eq.~(\ref{beta}) we only sum over light quark flavors, 
thus neglecting couplings to gluons mediated by heavy quark loops. 
Note that the two contributions (\ref{beta}) and (\ref{gamma})
to the scalar interactions interfere constructively:
even with extremely heavy KK quark masses (large $\Delta_{q_1}$), there is an inescapable 
lower bound on the scalar cross section for a given Higgs mass, since 
the Higgs contribution from eq.~(\ref{gamma}) scales with the SM 
Higgs mass $m_h$ and not the KK quark masses.

The analogous results for the case of $Z_1$ LKP can now be obtained 
from the above formulas by simple replacements: $m_{\gamma_1}\to m_{Z_1}$,
$Y_{q_L}\to \frac{1}{2}$ and $Y_{q_R}\to 0$, 
since $Z_1$ is mostly the neutral $SU(2)_W$ gauge boson,
which has no interactions with the $SU(2)_W$-singlet KK quarks
(or equivalently, the right-handed SM quarks). In addition,
one should replace $\frac{e}{\cos\theta_W}\to \frac{e}{\sin\theta_W}$
to account for the different gauge coupling constant. 

\begin{figure}[t]
\includegraphics[width=0.48\textwidth]{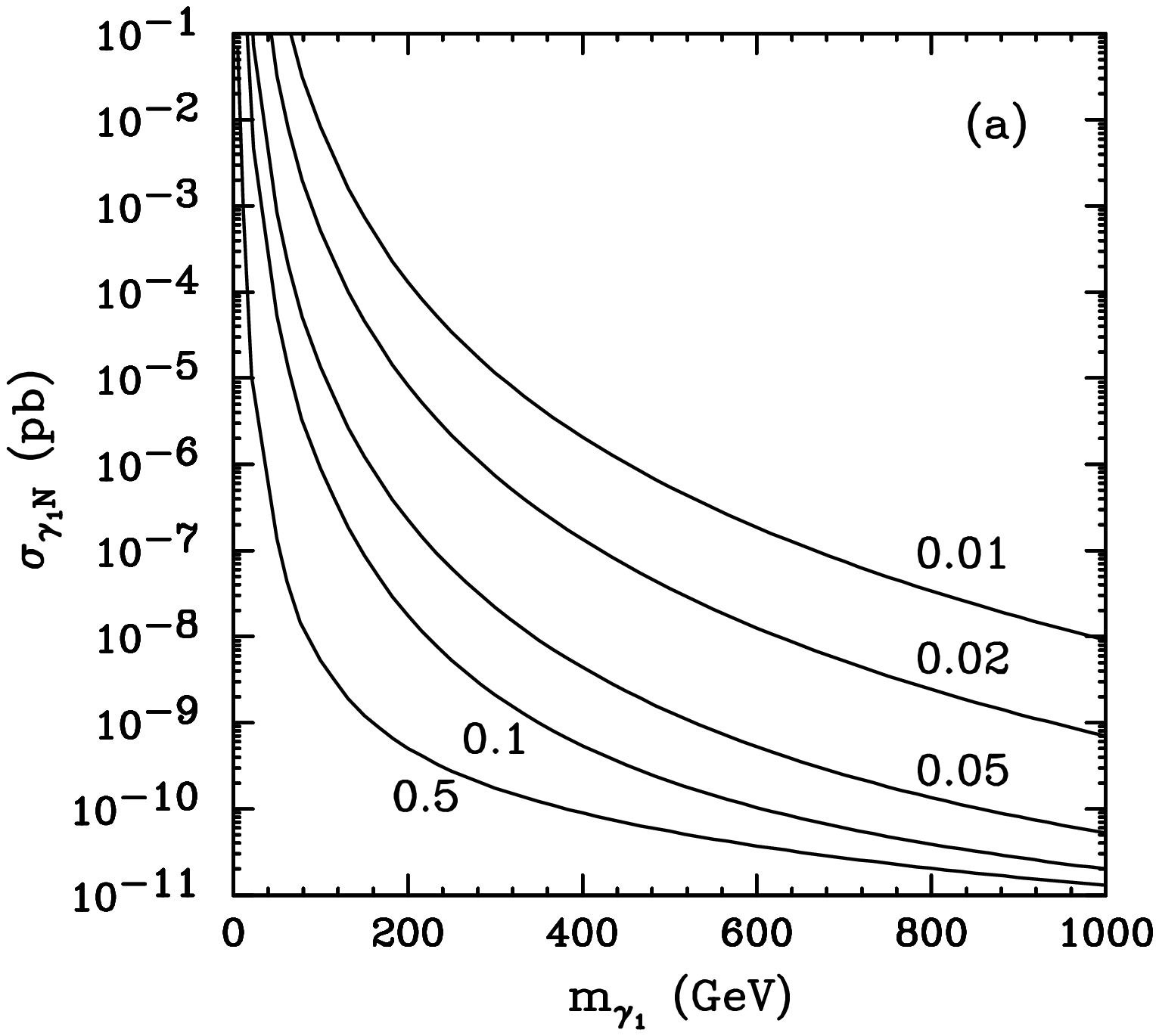}
\hspace{0.1cm}
\includegraphics[width=0.48\textwidth]{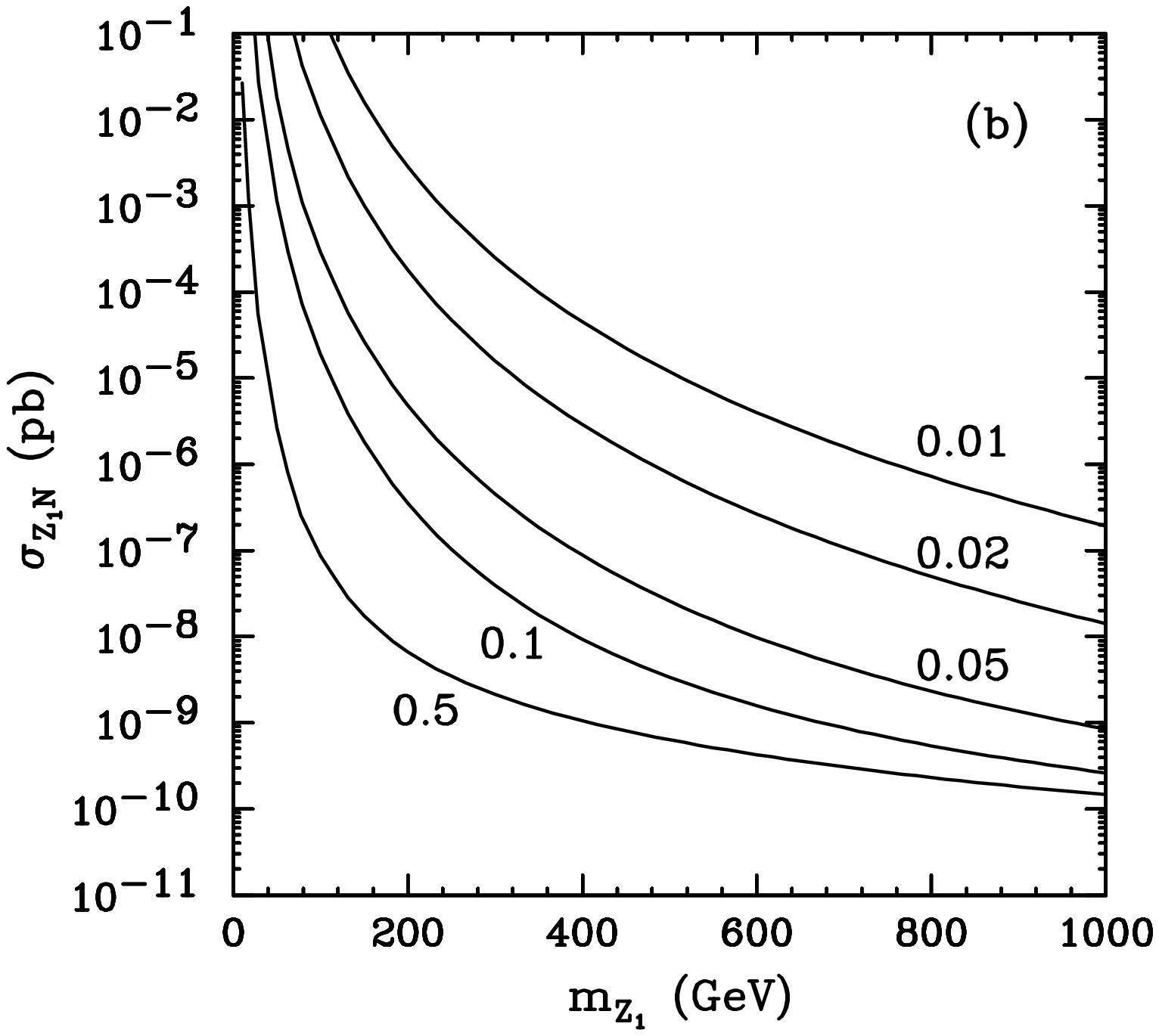}
\caption{\sl Spin-independent elastic scattering cross sections as a function of LKP mass 
for (a) ${\gamma_1}$ and (b) ${Z_1}$. The individual curves are labelled by the value of 
the mass splitting $\Delta_{q_1}$. The SM Higgs mass $m_h$ is fixed to 120\,GeV.}
\label{fig:sigma_scalar_B1_Z1}
\end{figure}

Theoretical predictions for the spin-independent LKP-nucleon
elastic scattering cross sections
are shown in Fig.~\ref{fig:sigma_scalar_B1_Z1} for different fixed values of
the KK quark - LKP mass splitting $\Delta_{q_1}$, and for two different 
LKPs: (a) $\gamma_1$ and (b) $Z_1$. In both cases the cross sections 
decrease as a function of LKP mass. This is due to the inverse scaling 
of the KK quark exchange contributions (\ref{beta}) with the KK mass scale.
Comparing Fig.~\ref{fig:sigma_scalar_B1_Z1}a to \ref{fig:sigma_scalar_B1_Z1}b,
we notice that the scalar cross section for $Z_1$ is more than
one order of magnitude larger than the scalar cross section for 
$\gamma_1$ of the same mass. This is mostly due to the larger 
$SU(2)_W$ gauge coupling. 
Notice that even when the KK quarks are very heavy, there is still 
a reasonable cross section, which is due to the Higgs mediated contribution
(\ref{gamma}). Perhaps the most noteworthy feature of Figs.~\ref{fig:sigma_scalar_B1_Z1}a
and \ref{fig:sigma_scalar_B1_Z1}b is the significant enhancement of 
the direct detection signals at small $\Delta_{q_1}$, often by several 
orders of magnitude. This greatly enhances the prospects for detecting
KK dark matter, if the mass spectrum turns out to be rather degenerate. 

The spin-dependent cross section is given by 
\begin{equation}
\sigma_{\text{spin}} = \frac{1}{6\pi} \frac{m_T^2}{(\mB + m_T)^2}
J_N (J_N+1) \bigg[ \sum_{u,d,s} \alpha_q \lambda_q \bigg]^2 \ ,
\label{sigma_spin}
\end{equation}
where $\alpha_q$ and $\lambda_q$ are 
\begin{eqnarray} 
\alpha_q \! \! &=& \! \! \frac{2 e^2}{\cos^2 \theta_W} \left[ 
\frac{Y_{q_L}^2 \mB}{m_{q_L^1}^2 - \mB^2} +
(L \to R) \right] \, ,  \label{alpha} \\
\lambda_q \! \! &=& \Delta_q^p \langle S_p \rangle/J_N 
+ \Delta_q^n \langle S_n \rangle/J_N \ . \label{lambda}
\end{eqnarray}
Here ${J}_N$ is the nuclear spin operator.
$\Delta_q^{p,n}$ is given by $\langle p,n | {S}^{\mu}_q | p, n
\rangle \equiv \Delta_q^{p,n} {S}^{\mu}_{p,n}$ and is the
fraction of the nucleon spin carried by the quark $q$. We use
$\Delta_u^p = \Delta_d^n = 0.78 \pm 0.02$, $\Delta_d^p =
\Delta_u^n = -0.48 \pm 0.02$, and $\Delta_s^p = \Delta_s^n = -0.15 \pm
0.02$~\cite{Mallot:1999qb}.  $\langle S_{p,n} \rangle / J_N \equiv
\langle N | S_{p,n} | N \rangle / J_N$ is the fraction of the total
nuclear spin $J_N$ that is carried by the spin of protons or neutrons.
For scattering off protons and neutrons, $\lambda_q$ reduces to
$\Delta_q^p$ and $\Delta_q^n$, respectively.

Following \cite{Tovey:2000mm}, we can rewrite eq.~(\ref{sigma_spin})
in the form
\begin{equation}
\sigma_{\text{spin}}=\frac{32}{\pi}\,G_F^2\,\mu^2\,\frac{J_N+1}{J_N}\,
\left( a_p \langle S_p \rangle + a_n \langle S_n \rangle \right)^2\, ,
\label{apaneq}
\end{equation}
where $G_F$ is the Fermi constant and 
\begin{equation}
\mu=\frac{m_T\, m_{\gamma_1}}{m_T+m_{\gamma_1}} \, 
\label{redmass}
\end{equation}
is the reduced mass, 
while the coefficients $a_p$ and $a_n$ are given by
\begin{eqnarray}
a_{p,n}&=&\frac{1}{8\sqrt{3}G_Fm_{\gamma_1}}
\sum_{u,d,s} \alpha_q \Delta_q^{p,n} \nonumber \\
&=&\frac{e^2}{4\sqrt{3}G_F\cos^2\theta_W}
\sum_{u,d,s} \left[ \frac{Y_{q_L}^2}{m_{q_L^1}^2 - \mB^2} + (L \to R) \right] \Delta_q^{p,n}\ .
\end{eqnarray}
The main advantage of introducing the parameters $a_p$ and $a_n$ is that
they encode all the theoretical model-dependence, thus allowing different experiments to
compare their sensitivities in a rather model-independent way.
From eqs.~(\ref{apaneq}-\ref{redmass}) it is clear that for any given target,
the spin-dependent scattering rate depends on only three parameters:
$m_{\gamma_1}$, $a_p$ and $a_n$. Notice that in our setup there
are only two relevant model parameters: $m_{LKP}$ and $\Delta_{q_1}$,
therefore we will have a certain correlation between $a_p$ and $a_n$,
depending on the nature of the LKP\footnote{In introducing the
parameters $a_p$ and $a_n$ we have followed the convention of 
Ref.~\cite{Tovey:2000mm}. We should alert the reader that a different 
convention was used in Ref.~\cite{Servant:2002hb}, where $a_{p,n}$ was 
defined as $a_{p,n}=\sum_{u,d,s} \left[ Y_{q_L}^2 + Y_{q_R}^2 \right] \Delta_q^{p,n}$,
so that $a_p$ and $a_n$ are pure numerical factors, e.g.
$a_n=-0.139167$ and $a_p= 0.280833$ for $\gamma_1$ LKP,
and $a_n=a_p=0.0375$ for $Z_1$ LKP. However, this 
factorization can only be done in the special case of $\Delta_{Q_1}=\Delta_{q_1}$,
and furthermore, the theoretical model dependence (through the KK quark masses) 
creeps back explicitly in the expression for the spin-dependent cross section.}.

\begin{figure}[t]
\includegraphics[width=0.48\textwidth]{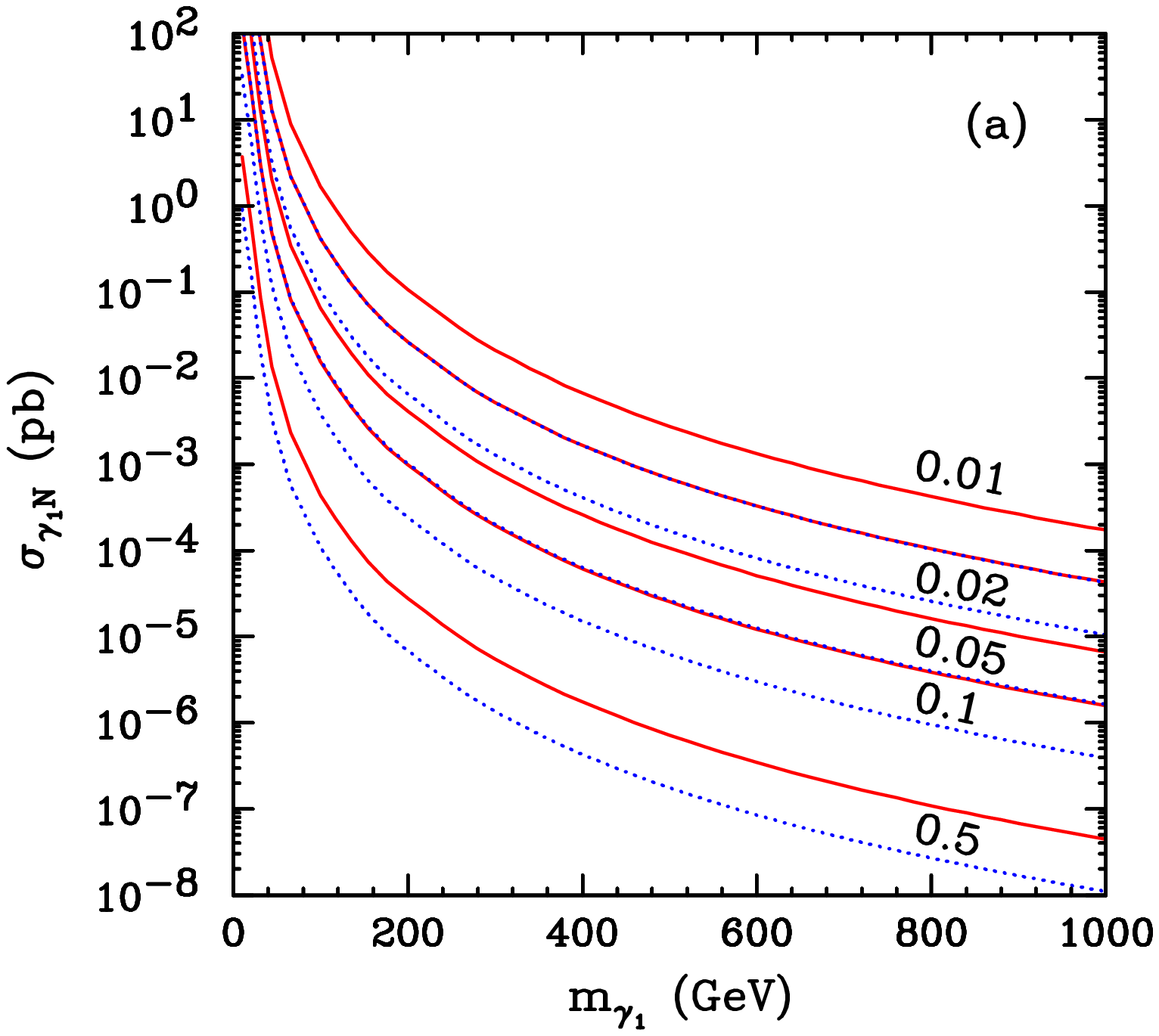}
\hspace{0.1cm}
\includegraphics[width=0.48\textwidth]{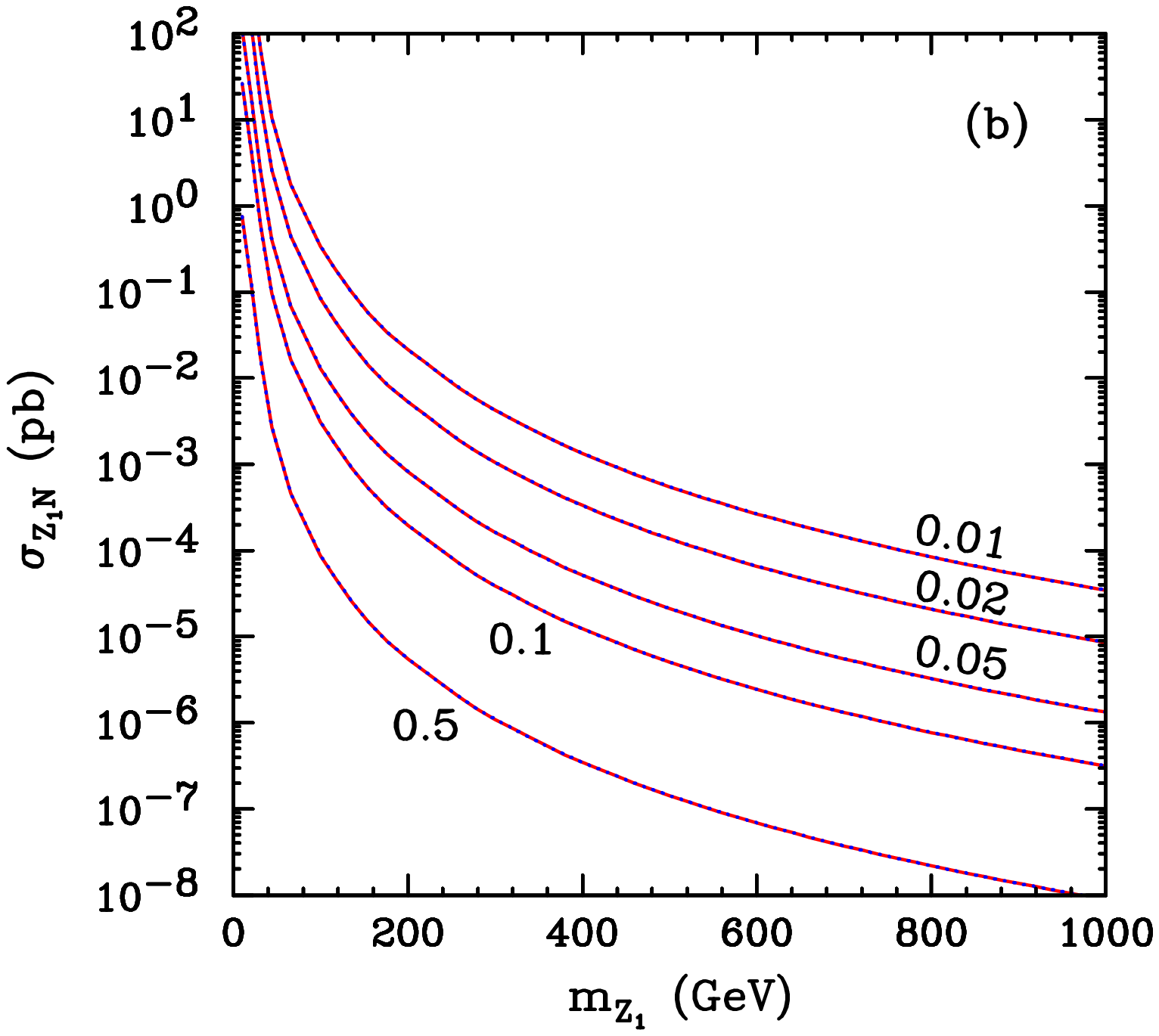}
\caption{\sl The spin-dependent elastic scattering cross sections as a function of LKP mass for 
(a) ${\gamma_1}$ and (b) ${Z_1}$ at various $\Delta_{q_1}$. The red solid curves are 
the LKP-proton cross sections and the blue dotted curves are the LKP-neutron cross sections.
The LKP-proton and LKP-neutron cross sections are identical for $Z_1$.
For $\gamma_1$ the proton cross section is approximately 4 times 
larger than the neutron cross section, for the same values of the LKP mass
$m_{\gamma_1}$ and mass splitting $\Delta_{q_1}$.
}
\label{fig:sigma_spin_B1_Z1}
\end{figure}

In Fig.~\ref{fig:sigma_spin_B1_Z1} we show our result 
for the spin-dependent LKP elastic scattering cross sections off
protons and neutrons for the case of (a) $\gamma_1$ and (b) $Z_1$,
for different mass splittings $\Delta_{q_1}$. The red solid curves are 
the LKP-proton cross sections and the blue dotted curves are the LKP-neutron 
cross sections. All curves exhibit the same general trends as
the corresponding spin-independent results from Fig.~\ref{fig:sigma_scalar_B1_Z1}:
the cross sections decrease with the KK mass scale, and are enhanced for 
small mass splittings $\Delta_{q_1}$. One peculiar feature is that the
proton and neutron spin-dependent cross sections are equal in the case of
$Z_1$, as seen in Fig.~\ref{fig:sigma_spin_B1_Z1}b.
This is an exact statement, which is due to the fact that $Z_1$
does not particularly discriminate between the different quark 
flavors in the nucleon -- it couples with equal strength to both
up- and down-type (left-handed) quarks. 
On the other hand, $\gamma_1$ couples differently to
$u$ and $d$, because of the different hypercharges of the
right-handed quarks. As a result, the cross sections on protons 
and neutrons differ in the case of $\gamma_1$, as seen 
in Fig.~\ref{fig:sigma_spin_B1_Z1}a. Interestingly, for a given
LKP mass $m_{\gamma_1}$ and mass splitting $\Delta_{q_1}$,
the proton cross section in Fig.~\ref{fig:sigma_spin_B1_Z1}a
is larger than the neutron cross section by about a factor of 4, 
which is due to a numerical coincidence involving the 
values of the quark hypercharges and the $\Delta_q^{p,n}$ 
parameters.\footnote{This can be simply understood in terms of the relative scaling of the
$a_p$ and $a_n$ parameters introduced in Eq.~(\ref{apaneq}).
In the case of $\gamma_1$, they differ by a factor of -2,
while for the case of $Z_1$ they are the same.}
Because of this simple scaling, for a given LKP mass $m_{\gamma_1}$,
the proton cross section at a certain $\Delta_{q_1}$ coincides 
with the neutron cross section for half the mass splitting ($\Delta_{q_1}/2$)
since to leading order both the proton and the neutron cross sections are 
proportional to $\left(\Delta_{q_1}\right)^{-2}$.

We shall now review the corresponding results 
for the case of two universal extra dimensions. The sum 
$\beta_q + \gamma_q$ for the spinless photon ($\gamma_H$) LKP 
was computed in~\cite{Dobrescu:2007ec} (note that here we are using 
a different convention for the hypercharges $Y_i$)
\begin{eqnarray}\label{c}
\beta_q + \gamma_q &=& \frac{e^2}{\cos^2 \theta_W} 
\left[
m_q ( Y_{q_L} + Y_{q_R} )^2 \left( \frac{ 1}{m_{q_1}^2 - (m_q-m_{\gamma_H})^2 } +  
                           \frac{1}{m_{q_1}^2 - (m_q+m_{\gamma_H})^2 }  \right) \right.
\nonumber \\ [0.3em]
&+& \left. m_{\gamma_H} ( Y_{q_L}^2 + Y_{q_R}^2 )  \left(  \frac{1}{m_{q_1}^2 - (m_q+m_{\gamma_H})^2 }
                                 - \frac{1}{m_{q_1}^2 - (m_q-m_{\gamma_H})^2 }  \right)
       + \frac{m_q}{2 m_h^2}  \right],
\end{eqnarray}
where $m_{\gamma_H}$ is the mass of the spinless photon,
$m_{q_1}$ is the (common) mass of the $SU(2)_W$-doublet and 
$SU(2)_W$-singlet KK quarks, while $m_q$ is the corresponding 
SM quark mass. 

\begin{figure}[t]
\includegraphics[width=0.49\textwidth]{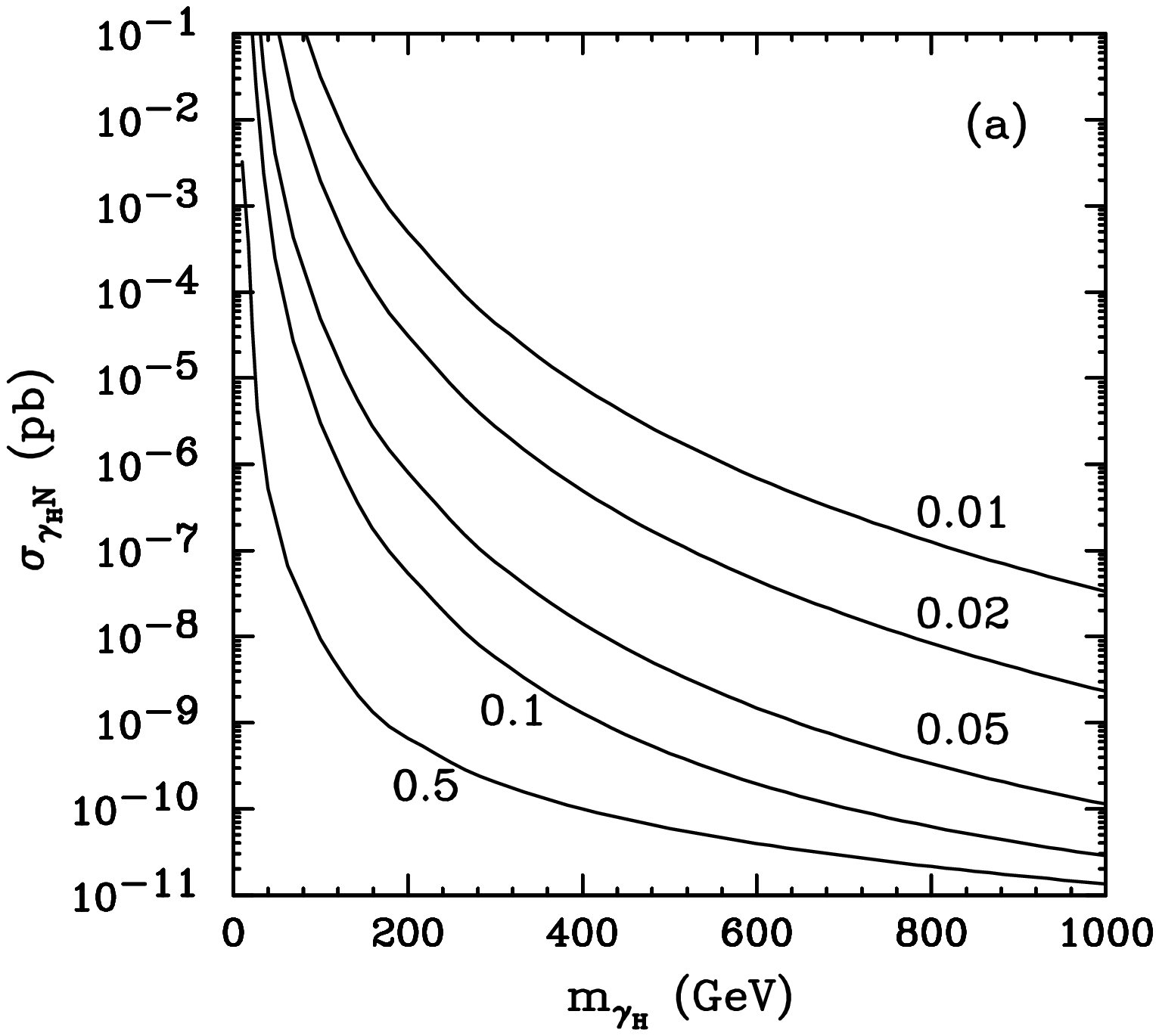}
\hspace{0.1cm}
\includegraphics[width=0.48\textwidth]{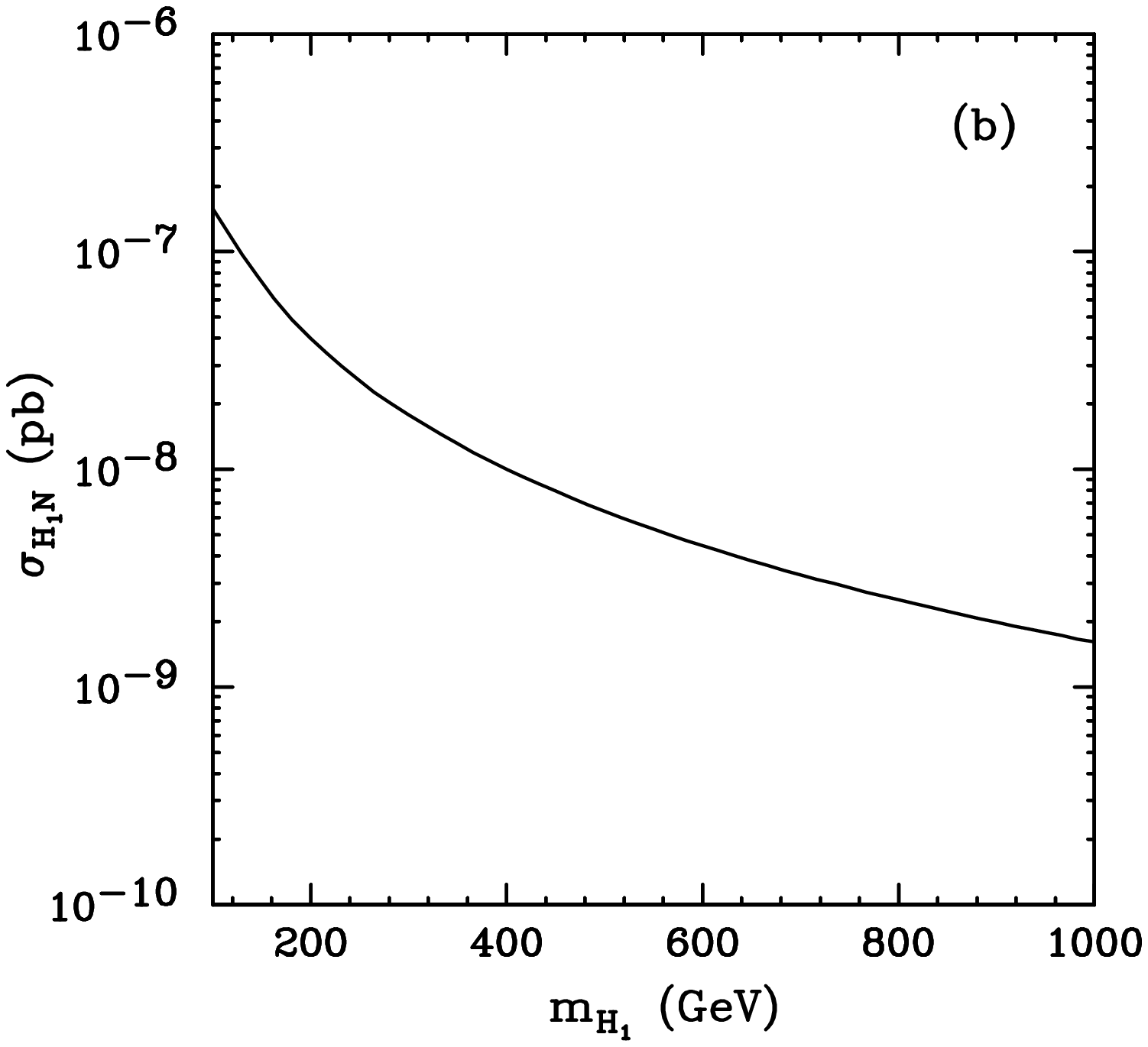}
\caption{\sl The spin-independent elastic scattering cross sections as a function of LKP mass 
for (a) ${\gamma_H}$ in 6D UED and (b) ${H_1}$ in 5D UED. In the case of $\gamma_H$ 
LKP (panel ``a''), we present results for several values of $\Delta_{q_1}$ as shown.
In the case of $H_1$ (panel ``b''), 
we only show the contribution from the SM Higgs exchange, assuming heavy KK quarks.}
\label{fig:sigma_scalar_H1_BH}
\end{figure}

Using Eqn.~\eqref{scalar}, we obtain the spin-independent elastic scattering cross section
for $\gamma_H$ as shown in Fig.~\ref{fig:sigma_scalar_H1_BH}a. The different curves
are labelled by the assumed fixed value of $\Delta_{q_1}$, and are
plotted versus the LKP mass $m_{\gamma_H}$. We see that the size of the
$\gamma_H$ signal is about the same order as the $\gamma_1$ cross sections
from Fig.~\ref{fig:sigma_scalar_B1_Z1}a. On the other hand, the relic density 
constraint would single out somewhat different regions for $m_{\gamma_1}$ and 
$m_{\gamma_H}$. The annihilation cross section for $\gamma_H$ is smaller than that of
$\gamma_1$~\cite{Dobrescu:2007ec}, and correspondingly, lower $\gamma_H$ masses 
would be preferred, with enhanced prospects for direct detection\footnote{One should 
keep in mind that mass splittings $\Delta_{q_1}$ as small as those shown in the plot
mandate the inclusion of coannihilation processes with KK quarks in 6D UED, 
as we did previously in the case of $\gamma_1$ and $Z_1$ in 5D UED
(see Fig.~\ref{fig:Omegah2_B1_Z1}). Unfortunately the computation
of coannihilations in 6D UED does not exist in the literature.
In analogy with the case of $\gamma_1$ LKP in 5D UED, 
we expect the coannihilation effects to
increase the preferred range of $m_{\gamma_H}$.}.
Notice that there is no spin-dependent cross section for $\gamma_H$ since it is a scalar particle.

In conclusion of this section, we shall briefly discuss the scenario of KK Higgs ($H_1$) LKP.
Just like $\gamma_H$, $H_1$ is a scalar and does not have spin-dependent interactions.
Its spin-independent elastic scattering cross section can be readily computed 
following the procedure outlined earlier in this section and in the appendix.
In this case, the KK quark exchange diagrams are also Yukawa suppressed, 
and the dominant among them is the $s$ KK quark contribution.
As in the $\gamma_1$ LKP case, the diagrams with KK quark exchange 
and SM Higgs exchange interfere constructively. Therefore, 
the SM Higgs exchange diagram by itself provides a conservative
lower bound on the elastic scattering cross section, independent 
of the other details of the KK spectrum, and in particular, the KK quark masses.
This absolute minimum of the cross section is plotted in Fig.~\ref{fig:sigma_scalar_H1_BH}b
as a function of the LKP mass $m_{H_1}$. It is worth mentioning that
this result is completely independent of the SM Higgs mass $m_h$.
The contribution corresponding to \eqref{gamma} is given by 
\begin{equation}
\gamma_q = \frac{3}{4} \frac{e^2}{\sin^2\theta_W} \frac{m_q}{m_W^2} \, ,
\end{equation}
where $m_W$ is the mass of the $W^\pm$ boson.
The coupling of the KK Higgs to the SM Higgs boson is the same as 
the triple Higgs coupling of the SM, which is proportional to $m_h^2$.
This $m_h^2$ dependence is exactly cancelled by the $m_h^{-2}$
dependence of the SM Higgs propagator in the non-relativistic limit
(see Eqn.~(\ref{gamma})). Therefore the final cross section is indeed 
independent of the SM Higgs mass, and this fact remains true 
regardless of the values of the KK quark masses.

\section{Direct WIMP Detection and Experiments}\label{sec:wimp}

%
%
The detailed distribution of dark matter in our galaxy, and in particular in the local neighborhood, is not well constrained 
by current observations and high-resolution simulations.  The standard assumption for its distribution is a cored, non-rotating isothermal spherical halo 
with a Maxwell-Boltzmann velocity distribution with a mean of 220\,km/s, and escape velocity from the galactic halo 
of 544\,km/s \cite{Smith:2006ym}. For the local density of dark matter particles we assume $\rho$=0.3\,GeV/cm$^3$ \cite{Gates:1995dw}.  

%
%
The WIMP interaction signature in ultra-low-background terrestrial detectors \cite{Goodman:1984dc} consists of nuclear recoils. 
Direct detection experiments attempt to measure the small ($<$100\,keV) energy deposited when a WIMP scatters from 
a nucleus in the target medium. The recoil energy of the scattered nucleus is transformed into a measurable signal, 
such as charge, scintillation light or lattice excitations, and at least one of the above quantities can be detected. 
Observing two signals simultaneously yields a powerful discrimination against background events, which are mostly interactions with 
electrons as opposed to WIMPs and neutrons, which scatter from nuclei. The WIMP interaction takes place in 
the non-relativistic limit, therefore the total cross section can be expressed as the sum of a spin-independent (SI) 
part (see Eqn.~\eqref{scalar}), a coherent scattering with the whole nucleus, and of a spin-dependent (SD) 
part (see Eqn.~\eqref{sigma_spin}), which describes the coupling to the total nuclear spin~\cite{Lewin:1995rx}.

%
%
Neutrons with energies in the MeV range can elastically scatter from nuclei and mimic a WIMP signal. Two methods are used 
to discriminate against the residual neutron background, which comes from ($\alpha$,n)- and fission-reactions in materials 
and from interactions of cosmic muons with the rock and experimental shields.  First, the  SI WIMP-nucleus cross section is
proportional to the atomic mass-squared of the nucleus, making the expected total WIMP interaction rate material dependent. 
Second, the mean free paths of WIMPs and MeV neutrons are exceedingly different (10$^{10}$\,m versus 8\,cm in a typical WIMP target),  
allowing to  directly constrain the neutron background from the ratio of observed single to multiple interaction events.

The experimental upper bounds of the SI cross section from direct detection experiments are WIMP-type 
independent and thus will not change if we consider different WIMP candidates. 
Similarly, the SD cross section limits can also be reinterpreted for various DM candidates. The only exception
is a spin zero WIMP, such as $\gamma_H$ in 6D UED, which 
does not have an axial-vector coupling with nuclei, hence no SD interaction is expected. 
We will extensively discuss the model dependence of the SD cross section in the next section. 

%
%
In this study, we choose four direct detection experiments which demonstrated best experimental sensitivity to-date in various parts of the 
WIMP search parameter space. The CDMS experiment sets the best SI upper bound above a WIMP mass of 42\,GeV \cite{CDMS:2008},  while XENON10 gives the 
most stringent upper bound on WIMP-neutron SD couplings \cite{XENON10:2008SD} and SI couplings below 42\,GeV \cite{XENON10:2008}. 
The KIMS \cite{KIMS:2007} and COUPP \cite{COUPP:2008} experiments show the best 
sensitivity for SD WIMP-proton couplings. As we shall see in the following section, the combined study of all four experiments 
strongly constrains the SD proton-neutron mixed coupling parameter space (the so called $a_p$-$a_n$ parameter space, 
where $a_p$ and $a_n$ are the dark matter particle's couplings to protons and neutrons, respectively, see eq.~(\ref{apaneq})).

Table~\ref{tab:exp} summarizes the relevant characteristics of the four experiments  such as target material, 
total mass, energy range considered for the WIMP search, and location. In this paper we either calculated the LKP limits based on 
published data (XENON10), or we obtained the data points for the cross section upper bounds  
from the collaboration (CDMS, KIMS and COUPP).

\begin{table}[t!]
  \begin{tabular}{l c c c c c c}
\hline
Experiments & Target & Total mass &  Energy range &  Location & Ref. \\
\hline
\hline
CDMS II  & Ge(Si) & 4.75\,kg (1.1\,kg)&  10\,keV -- 100\,keV & Soudan, USA  & \cite{CDMS:2008}\\
XENON10 & Xe & 15\,kg&4.5\,keV -- 27\,keV  &  Gran Sasso, Italy & \cite{XENON10:2008}\\
KIMS  & CsI &34.8\,kg & 3\,keV -- 11\,keV &  Yangyang, Korea & \cite{KIMS:2007} \\
COUPP & CF$_{3}$I &1.5\,kg & 5\,keV --   &Fermilab, USA  & \cite{COUPP:2008} \\
\hline
\end{tabular}
\caption{Direct WIMP detection experiments considered in this study.}
\label{tab:exp}
\end{table}

%
%

The CDMS experiment~\cite{CDMS:2008} is operated in the Soudan Underground Laboratory, USA. It uses advanced Z(depth)-sensitive Ionization and Phonon (ZIP) detectors, which simultaneously measure the ionization and athermal phonon signals after a particle interacts in the crystal. The ZIP detectors provide excellent event-by-event discrimination of nuclear recoils from the dominant background of electron recoils. The most stringent limits on spin-independent couplings with nucleons above a WIMP mass of 42\,GeV comes from the first two CDMS-II five tower runs with a raw exposure of 397.8\,kg-days in germanium. The null observation of a WIMP signal sets a  WIMP-nucleon cross section upper bound of 6.6$\times10^{-8}$\,pb (for a 60\,GeV WIMP mass) and of 4.6$\times10^{-8}$\,pb when the results are combined with previous CDMS results. 

The SuperCDMS project~\cite{Schnee:2005pj,Brink:2005ej} is a three-phase proposal to utilize CDMS-style detectors with target masses growing from 25\,kg to 150\,kg and up to 1 ton, 
with the aim of reaching a final sensitivity of 3$\times$10$^{-11}$\,pb by mid 2015.  This goal will be realized by developing improved detectors and analysis techniques, and 
concomitantly reducing the intrinsic surface contamination of the crystals.

%
%
The XENON10 collaboration~\cite{XENON10:2008} operated a 15\,kg active mass, dual-phase (liquid and gas) xenon time projection chamber in the Gran Sasso Underground Laboratory (LNGS), in WIMP search mode from August 2006 to February 2007. 
XENON10 uses two arrays of UV-sensitive photomultipliers (PMTs) to detect the prompt and proportional light signals induced by particles interacting in the sensitive 
liquid xenon (LXe) volume. The 3D position sensitivity, the self-shielding of LXe and the prompt versus proportional light ratio are the most important background rejection features. 
The first results, using $\sim$136\,kg-days exposure after cuts, demonstrated that LXe can be used for stable, homogeneous, large scale dark matter detectors, providing 
excellent position resolution and discrimination against the electron recoil background. 
The derived upper bound on SI cross sections on nucleons is 4.5 $\times10^{-8}$\,pb for a WIMP mass of 30\,GeV. Since natural Xe contains $^{129}$Xe (26.4\%) and 
$^{131}$Xe (21.2\%) isotopes,  each of these having an unpaired neutron, the XENON10 results substantially constrain the SD WIMP-nucleon cross section. 
We calculated the XENON10 SD LKP-neutron and LKP-proton upper bounds based on the observation of 10 events, without any background subtraction \cite{XENON10:2008SD}. 
The next phase, XENON100, will operate a total of 170\,kg (70\,kg fiducial) of xenon, viewed by 242 PMTs,  in a dual-phase TPC in an improved XENON10 shield at 
the Gran Sasso Laboratory. While the fiducial mass is increased by more than a factor of 10, the background will be lower by about a factor of 100 (through careful selection 
of ultra-low background materials, the placing of cryogenic devices and high-voltage feed-throughs outside of the shield and by using 100\,kg of active LXe shield) 
compared to XENON10.  XENON100 is currently being commissioned at LNGS, the aim is to start the first science run in fall 2008, probing WIMP-nucleon SI cross sections down to 
$\sim$10$^{-9}$\,pb.

%
%
The Korea Invisible Mass Search (KIMS) experiment~\cite{KIMS:2007} is located at the Yangyang Underground Laboratory, Korea.  
The collaboration has operated four low-background CsI(Tl) crystals, each viewed by two photomultipliers, for a total exposure of 3409\,kg-days.  
Both $^{133}$Cs and $^{127}$I are sensitive to the spin-dependent interaction of WIMPs with nuclei.
KIMS detects the scintillation light after a particle interacts in one of the crystals, kept stably at (0$\pm$0.1)\,$^{\circ}\mathrm{C}$. 
The pulse shape discrimination technique, using the time distribution  of the signal, allows to statistically separate nuclear recoils from 
the electron recoil background.
The KIMS results are consistent with a null observation of a WIMP signal, yielding the best limits on SD WIMP-proton couplings 
for a WIMP mass above 30\,GeV. Specifically, the upper bound for a WIMP mass of 80\,GeV is 1.7$\times$10$^{-1}$\,pb.

%
%
The Chicagoland Observatory for Underground Particle Physics (COUPP) experiment~\cite{COUPP:2008} 
is operated at Fermilab, USA. The experiment has revived the bubble chamber 
technique for direct WIMP searches. The superheated liquid can be tuned such that the detector responds only to keV nuclear recoils, being fully insensitive to 
minimum ionizing particles.  A 1.5\,kg chamber of superheated CF$_3$I has been operated for a total exposure of 250\,kg-days. 
The presence of fluorine and iodine in the target makes COUPP sensitive to both SD and SI WIMP-nucleon couplings. 
The production of bubbles is monitored  optically and via sound emission, reaching a reconstructed 3D spatial resolution of $\sim$1\,mm. 
It allows to reject boundary-events and to identify  multiple neutron interactions. 
The most recent COUPP results set the most sensitive limit on SD WIMP-proton cross sections for a WIMP mass below 30\,GeV. 
As an example, the upper bound on the SD coupling is 2.7$\times$10$^{-1}$\,pb at a WIMP mass of  40\,GeV.

\begin{figure}[t!]
\includegraphics[width=0.6\textwidth]{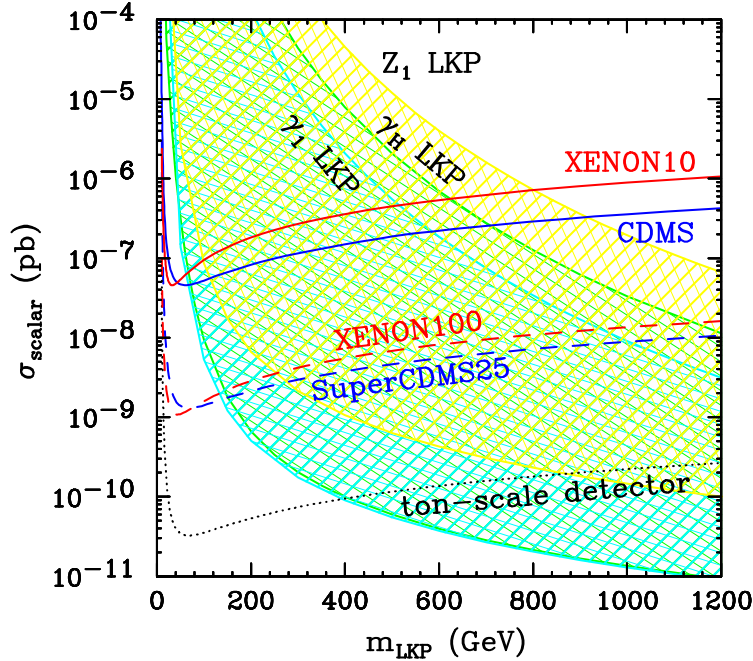}
\caption{\sl Current and projected experimental limits on the SI LKP-nucleon-scattering 
cross section together with the theoretically expected $\gamma_1$ (blue shaded),
$Z_1$ (yellow shaded) and $\gamma_H$ (green-shaded) LKP regions. 
The boundaries of the LKP regions are selected 
for $0.01 < \Delta < 0.5$ while the Higgs mass $m_h$ is fixed to 120\,GeV.
The solid lines are the current experimental upper bounds 
(90\% C.L) from  the CDMS (blue) and XENON10 (red) experiments. 
The dashed lines are expected sensitivities for the SuperCDMS 25\,kg (blue) and XENON100 (red) 
 experiments, which will be operated in the near future. The dotted line  is the expected 
sensitivity for a ton-scale detector. }
\label{fig:SI_CrossSection_Neutron}
\end{figure}

%
%
In Fig.~\ref{fig:SI_CrossSection_Neutron} we show the current CDMS and XENON10 
upper bounds for the SI cross section together with projected sensitivities  
for SuperCDMS 25\,kg,  XENON100 and for a ton-scale detector.  
The LKP boundaries for  $\gamma_1$, $Z_1$ and $\gamma_H$ as dark matter candidates 
are also shown, for a wide range of mass splittings ($0.01 < \Delta_{q_1} < 0.5$)
and a fixed Higgs mass $m_h$ of 120\,GeV.  The small mass splitting regions 
are excluded up to a mass of about 600\,GeV,  900\,GeV and 700\,GeV for $\gamma_1$, 
$Z_1$ and $\gamma_H$, respectively.
For large mass splittings of $\Delta_{q_1} = 0.5$, only masses below about 100\,GeV can be probed. 
Future ton-scale direct detection experiments should cover most of the interesting LKP parameter space. 

\begin{figure}[t!]
\includegraphics[width=0.49\textwidth]{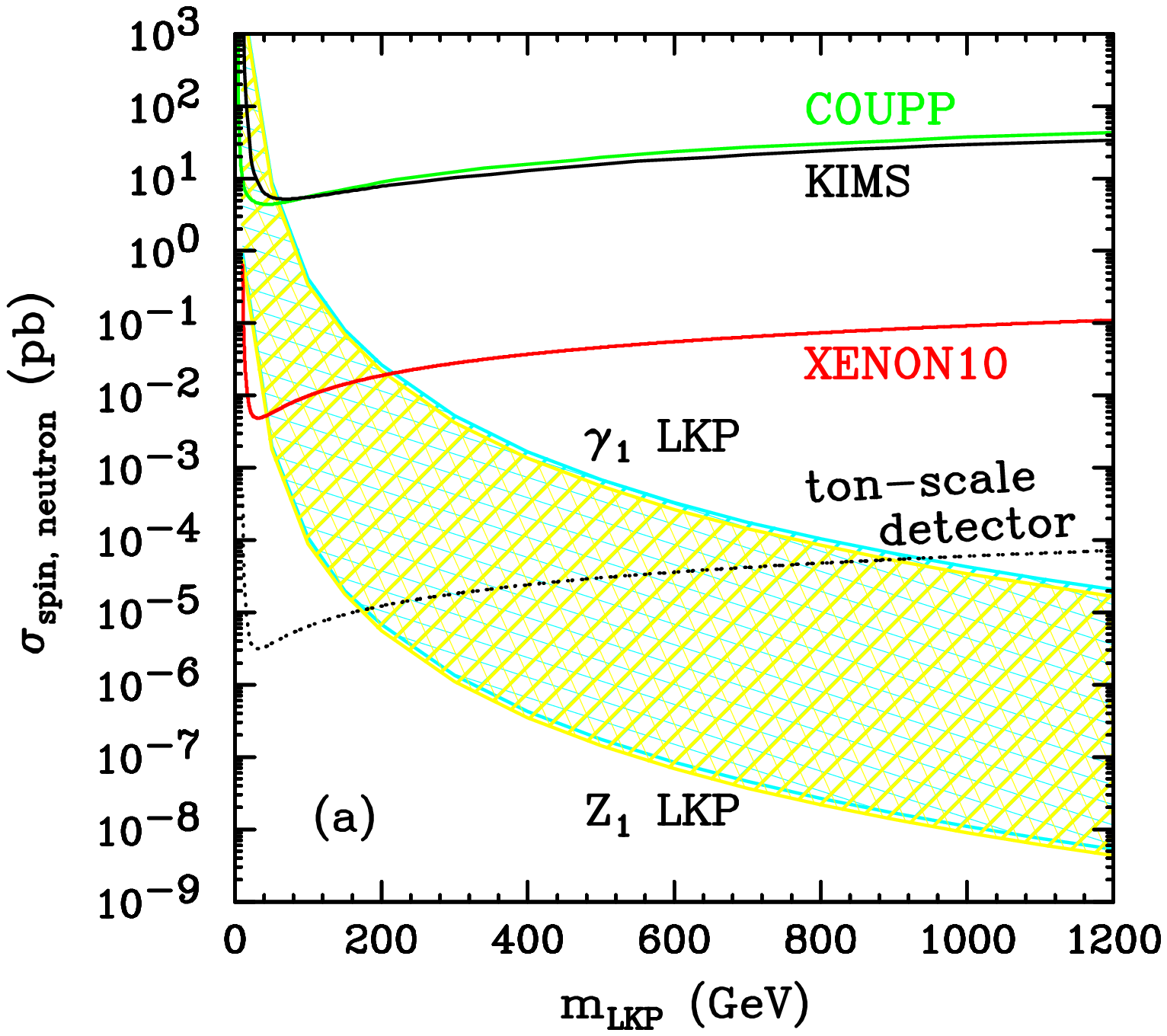}
\includegraphics[width=0.49\textwidth]{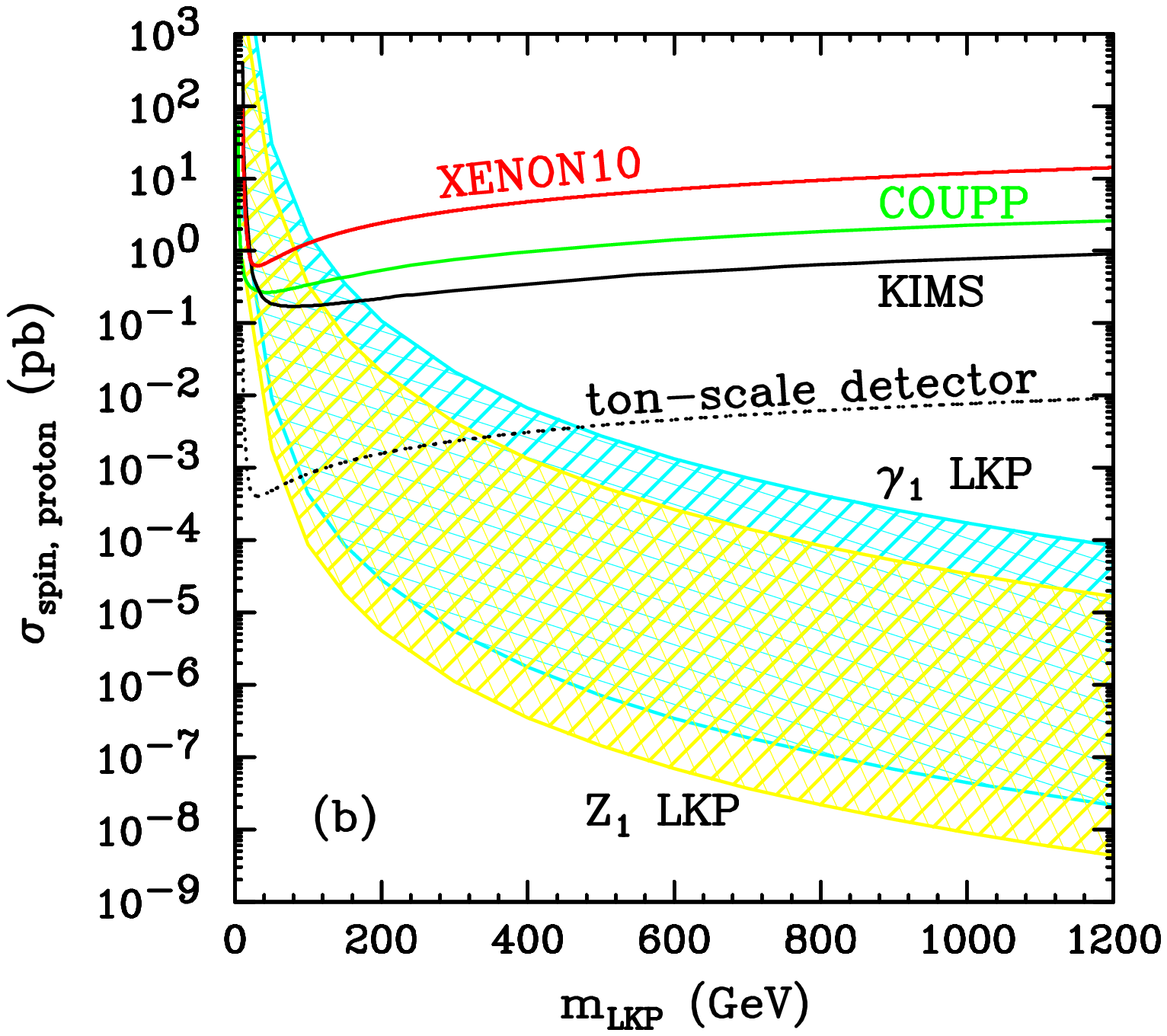}
\caption{\sl Current experimental limits on the SD (a) neutron- and (b) proton-scattering cross section together with the predicted 
SD LKP-neutron (proton) cross sections for $\gamma_1$ (blue shaded) and  $Z_1$ (yellow shaded). 
The solid curves for each plot are the upper bounds (90\% C.L.) from the COUPP (green), 
KIMS (black) and XENON10 (red) experiments. 
The dotted line shows the expected sensitivity for a ton-scale detector 
which is obtained from Fig.~\ref{fig:SI_CrossSection_Neutron} by a proper rescaling considering a Xenon detector.
The $\gamma_1$ and $Z_1$ LKP 
areas are drawn with the same $\Delta_{q_1}$ convention as shown in Fig.~\ref{fig:SI_CrossSection_Neutron}. }
  \label{fig:SD_CrossSection}
\end{figure}
%
%
In Fig.~\ref{fig:SD_CrossSection}, we show the SD cross section limits for both (a) pure neutron 
and (b) pure proton couplings for three experiments together with 
the theoretical predictions for $\gamma_1$ and $Z_1$ for a range of mass splittings ($0.01 < \Delta_{q_1} < 0.5$). 
The most stringent SD pure neutron upper bound is set by the XENON10 experiment, while the best SD cross 
section for pure proton couplings in the region of interesting LKP masses ($>$\,500\,GeV) comes from the KIMS experiment.  
As explained in the previous section, the theoretical $\gamma_1$ and 
$Z_1$ regions are overlapping for  pure neutron couplings, while for pure proton coupling these can be 
distinguished for a given mass splitting $\Delta_{q_1}$. 

In the following section we investigate the details of the LKP specific parameter spaces.

\section{Limits on Kaluza-Klein Dark Matter}\label{sec:kkdm}
\noindent

In the previous sections we introduced the different dark matter
candidates in UED models: KK gauge bosons ($\gamma_1$ and $Z_1$) 
and KK scalars ($\gamma_H$ and $H_1$). On the theoretical side, we discussed 
the calculation of their relic densities and elastic scattering cross sections.
On the experimental side, we described the different types of experiments 
which are sensitive to KK dark matter. We shall now combine our
theoretical predictions with the current/future measurements
discussed earlier. Where applicable, we shall also include constraints 
from high energy collider experiments. 
We shall be particularly interested in the region of small mass splittings $\Delta_{q_1}$, 
which is problematic for collider searches, but promising for direct detection.
We will concentrate on KK gauge boson dark matter (both $\gamma_1$ and $Z_1$),
whose relic density can be reliably calculated, including all
relevant coannihilation processes \cite{Burnell:2005hm,Kong:2005hn}.\footnote{The scalar candidates ($\gamma_H$ and $H_1$) do not have
spin-dependent interactions anyway, while their spin-independent  
scattering rates are similar to the examples we consider, as discussed 
in Sec.~\ref{sec:sigma}.}

\begin{figure}[t]
\includegraphics[width=0.48\textwidth]{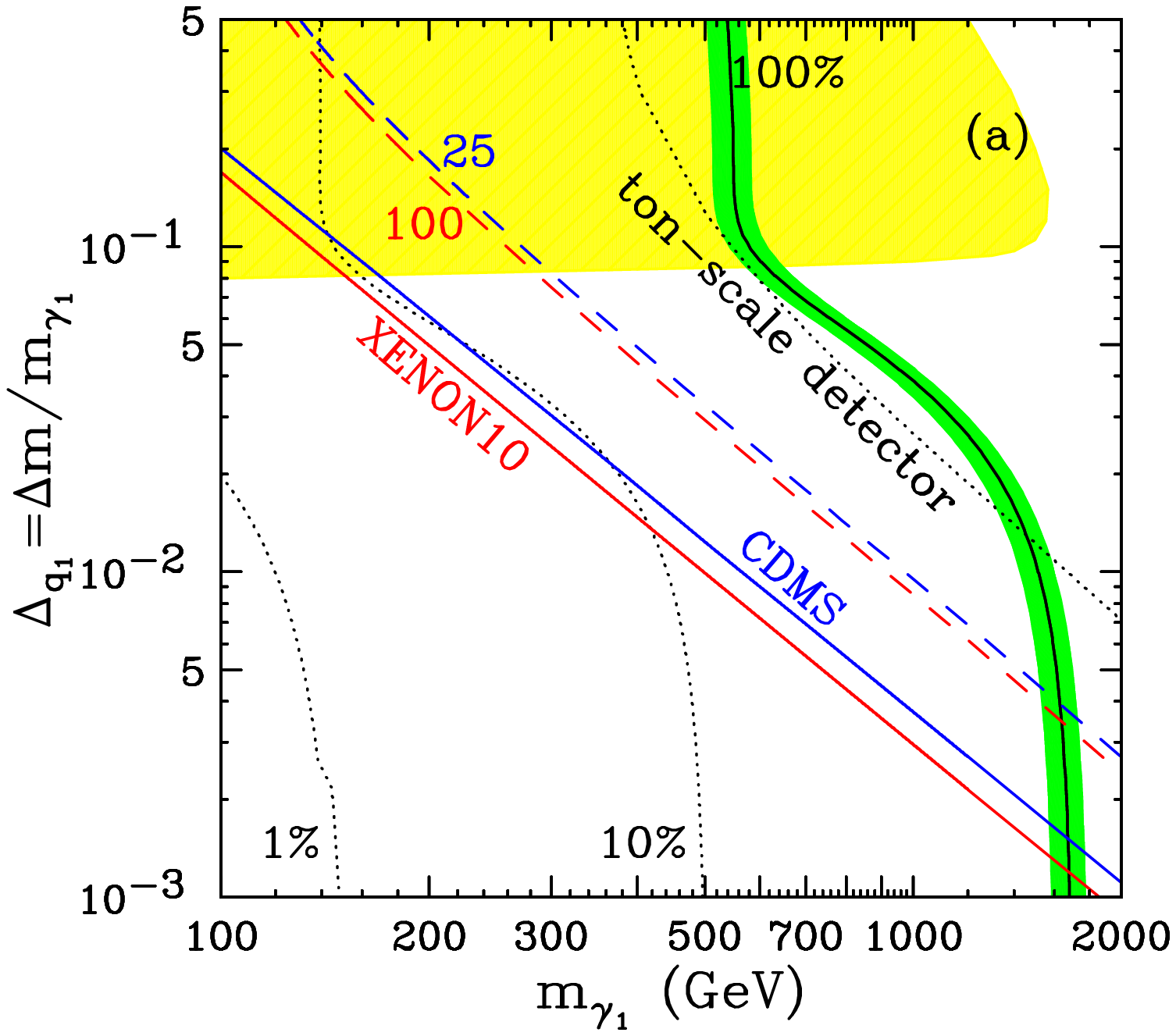}
\hspace{0.1cm}
\includegraphics[width=0.48\textwidth]{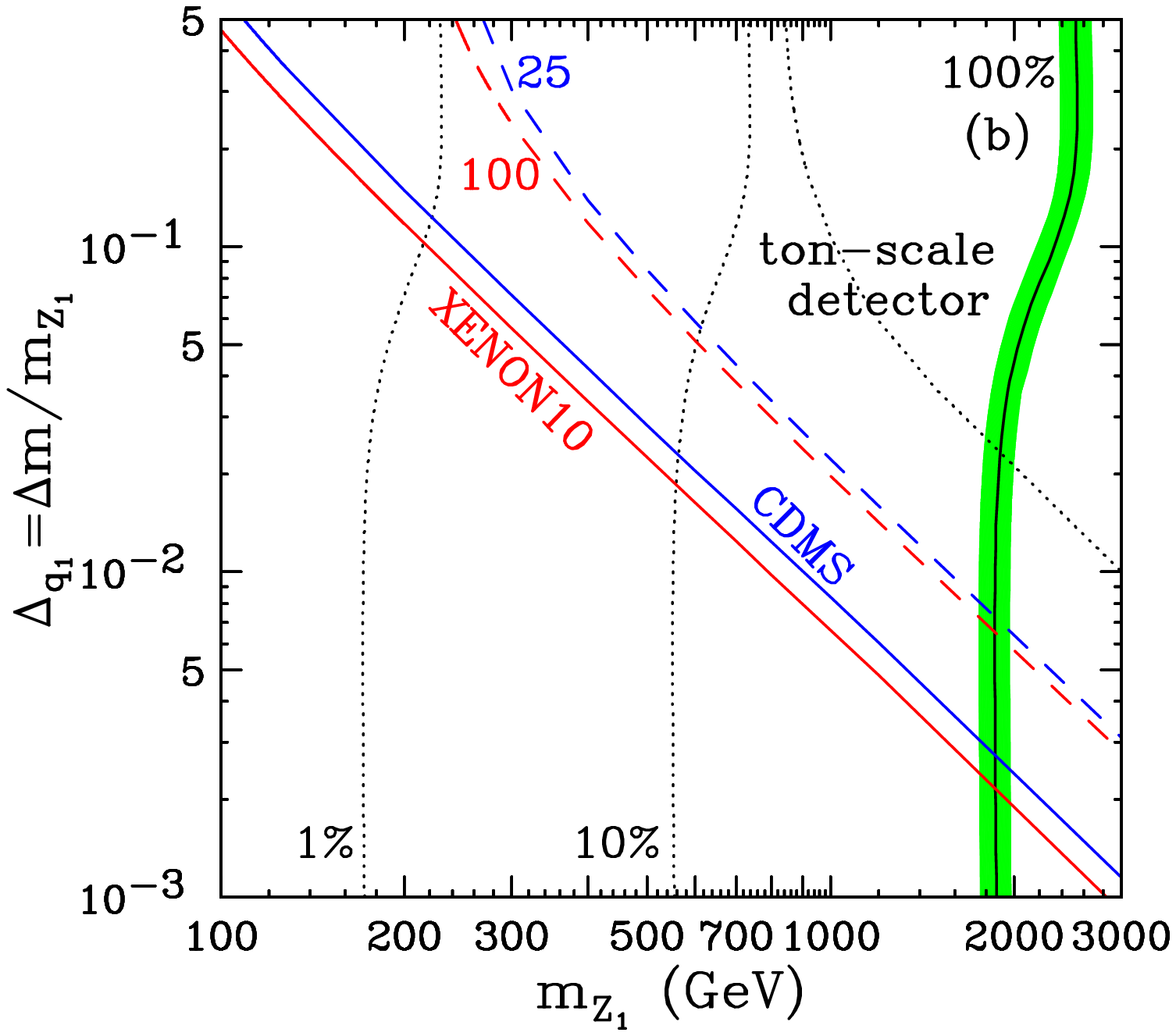}
\caption{\sl
Combined plot of the direct detection limit on the spin-independent cross section, 
the limit from the relic abundance and the LHC reach for (a) $\gamma_1$ and (b) $Z_1$, 
in the parameter plane of the LKP mass and the mass splitting $\Delta_{q_1}$. 
The remaining KK masses have been fixed as in Fig.~\ref{fig:Omegah2_B1_Z1}
and the SM Higgs mass is $m_h=120$\,GeV.
The black solid line accounts for all of the dark matter (100\%) 
and the two black dotted lines show 10\% and 1\%, respectively. 
The green band shows the WMAP range, $0.1037 < \Omega_{CDM}h^2 < 0.1161$.
The blue (red) solid line labelled by CDMS (XENON10) 
shows the current limit of the experiment whereas the dashed and dotted lines 
represent projected limits of future experiments as shown in 
Fig.~\ref{fig:SI_CrossSection_Neutron}. 
In the case of $\gamma_1$ LKP, a ton-scale experiment will rule out 
most of the parameter space while there is little parameter space left in the case of $Z_1$ LKP. 
The yellow region in the case of $\gamma_1$ LKP shows parameter space 
that could be covered by the collider search in the $4\ell+\met$ channel 
at the LHC with a luminosity of 100 fb$^{-1}$ \cite{Cheng:2002ab}. 
}
\label{fig:SI_Delta_Neutron_B_Z}
\end{figure}

In Fig.~\ref{fig:SI_Delta_Neutron_B_Z} we present a combination of results for
the case of (a) $\gamma_1$ and (b) $Z_1$ LKP in 5D UED.
As we emphasized earlier, the two most relevant parameters are the
LKP mass ($m_{\gamma_1}$ or $m_{Z_1}$, correspondingly)
and the mass splitting $\Delta_{q_1}$ between the
LKP and the KK quarks. We therefore take both 
of these parameters as free and do not assume the MUED 
relation among them. For simplicity, we assume that 
the $SU(2)_W$-doublet KK quarks and the $SU(2)_W$-singlet KK quarks 
are degenerate, so that there is a single mass splitting parameter
which we have been calling $\Delta_{q_1}$.
However, this assumption is only made for convenience, and does not
represent a fundamental limitation -- all of our results can be 
readily generalized for different KK quark mass splittings
(i.e. several individual $\Delta$ parameters).
The masses of the remaining KK particles in the spectrum are
fixed as in Fig.~\ref{fig:Omegah2_B1_Z1}: in the case of $\gamma_1$ LKP, 
we use the MUED spectrum, while in the case of $Z_1$ LKP, we 
take the gluon and the remaining particles to be respectively $20\%$ and $10\%$ 
heavier than the $Z_1$. This choice is only made for definiteness, 
and does not carry a big impact on the validity of our results, 
as long as the remaining particles are sufficiently heavy so that
they do not participate in coannihilation processes.

In the so defined parameter plane, in Fig.~\ref{fig:SI_Delta_Neutron_B_Z} we
superimpose the limit on the spin-independent elastic scattering cross section, 
the limit on the relic abundance and the LHC reach in the four 
leptons plus missing energy ($4\ell + \met$)
channel which has been studied in~\cite{Cheng:2002ab}. This signature
results from the pair production (direct or indirect) of $SU(2)_W$-doublet 
KK quarks, which subsequently decay to $Z_1$'s and jets. The leptons (electrons or muons)
arise from the $Z_1\to \ell^+\ell^-\gamma_1$ decay, whose branching fraction 
is approximately $1/3$~\cite{Cheng:2002ab}.
Requiring a 5$\sigma$ excess at a luminosity of 100 fb$^{-1}$, 
the LHC reach extends up to $R^{-1} \approx m_{\gamma_1} \sim 1.5$ TeV, 
which is shown as the right-most boundary of the (yellow) shaded region
in Fig.~\ref{fig:SI_Delta_Neutron_B_Z}a. The slope of that boundary is due to
the fact that as $\Delta_{q_1}$ increases, so do the KK quark masses, and their 
production cross sections are correspondingly getting suppressed, diminishing
the reach. We account for the loss in cross section according to the
results from Ref.~\cite{Datta:2005zs}, assuming also that, as expected, the 
level-2 KK particles are about two times heavier than those at level 1.
Points which are well inside the (yellow) shaded region, of course, 
would be discovered much earlier at the LHC. Notice, however, that the LHC reach 
in this channel completely disappears for $\Delta_{q_1}$ less than about 8\%.
This is where the KK quarks become lighter than the $Z_1$ (recall that 
in Fig.~\ref{fig:SI_Delta_Neutron_B_Z}a $m_{Z_1}$ was fixed according to
the MUED spectrum) and the $q_1\to Z_1$ decays are turned off. 
Instead, the KK quarks all decay directly to the $\gamma_1$ LKP 
and (relatively soft) jets, presenting a monumental challenge for an LHC discovery.
So far there have been no studies of the collider phenomenology of a
$Z_1$ LKP scenario, but it appears to be extremely challenging, especially if the 
KK quarks are light and decay directly to the LKP. This is why
there is no LHC reach shown in Fig.~\ref{fig:SI_Delta_Neutron_B_Z}b.
In conclusion of our discussion of the collider reaches exhibited in
Fig.~\ref{fig:SI_Delta_Neutron_B_Z}, we draw attention once again to the
lack of sensitivity at small $\Delta_{q_1}$: such small mass splittings are 
quite problematic for collider searches (see, for example, 
\cite{Martin:2007gf,Baer:2007uz} for an analogous situation in supersymmetry).

In Fig.~\ref{fig:SI_Delta_Neutron_B_Z} we contrast the LHC reach 
with the relic density constraints and with the sensitivity of 
direct detection experiments. 
To this end we convert our results from Figs.~\ref{fig:Omegah2_B1_Z1} 
and \ref{fig:SI_CrossSection_Neutron} into the $m_{LKP}$-$\Delta_{q_1}$ plane
shown in Fig.~\ref{fig:SI_Delta_Neutron_B_Z}. 
The green shaded region labelled by 100\% represents 2$\sigma$ WMAP band, 
$0.1037 < \Omega_{CDM}h^2 < 0.1161$~\cite{Dunkley:2008ie} and the black solid line inside this 
band is the central value $\Omega_{CDM}h^2 = 0.1099$. 
The region above and to the right of this band is ruled out 
since UED would then predict too much dark matter. 
The green-shaded region is where KK dark matter 
is sufficient to explain all of the dark matter in the universe, while
in the remaining region to the left of the green band 
the LKP can make up only a fraction of the dark matter in the universe.
We have indicated with the black dotted contours the 
parameter region where the LKP would contribute only 10\% and 1\%
to the total dark matter budget. Finally, the
solid (CDMS in blue and XENON10 in red) lines show the 
current direct detection limits, while the dotted and dashed lines 
show projected sensitivities for future experiments (for details, 
refer back to Sec.~\ref{sec:wimp})\footnote{Here and in the rest of the paper, 
when presenting experimental limits in an underdense or an overdense 
parameter space region, we do not rescale the expected 
direct detection rates with the calculated relic density.
The latter is much more model-dependent, e.g. the mismatch with the
WMAP value may be fixed by non-standard cosmological evolution, 
having no effect on the rest of our analysis.}.

Fig.~\ref{fig:SI_Delta_Neutron_B_Z} demonstrates the complementarity between the 
three different types of probes which we are considering. 
First, the parameter space region at very large $m_{LKP}$ is inconsistent
with cosmology -- if the dark matter WIMP is too heavy, 
its relic density is too large. The exact numerical bound on the LKP mass
may vary, depending on the particle nature of the WIMP (compare 
Fig.~\ref{fig:SI_Delta_Neutron_B_Z}a to Fig.~\ref{fig:SI_Delta_Neutron_B_Z}b)  
and the presence or absence of coannihilations (compare the
$m_{LKP}$ bound at small $\Delta_{q_1}$ to the bound at large $\Delta_{q_1}$).
Nevertheless, we can see that, in general, cosmology does provide 
an upper limit on the WIMP mass. On the other hand, colliders are 
sensitive to the region of relatively large mass splittings $\Delta_{q_1}$,
while direct detection experiments are at their best at small
$\Delta_{q_1}$ {\em and} small $m_{LKP}$. 
The relevant parameter space is therefore getting squeezed from
opposite directions and is bound to be covered eventually.
This is already seen in the case of $\gamma_1$ LKP from 
Fig.~\ref{fig:SI_Delta_Neutron_B_Z}a: the future experiments 
push up the current limit almost to the WMAP band. 
Unfortunately in the case of $Z_1$ LKP the available parameter space 
is larger and will not be closed with the currently envisioned experiments 
alone. However, one should keep in mind that detailed LHC studies
for that scenario are still lacking.

\begin{figure}[t]
\includegraphics[width=0.48\textwidth]{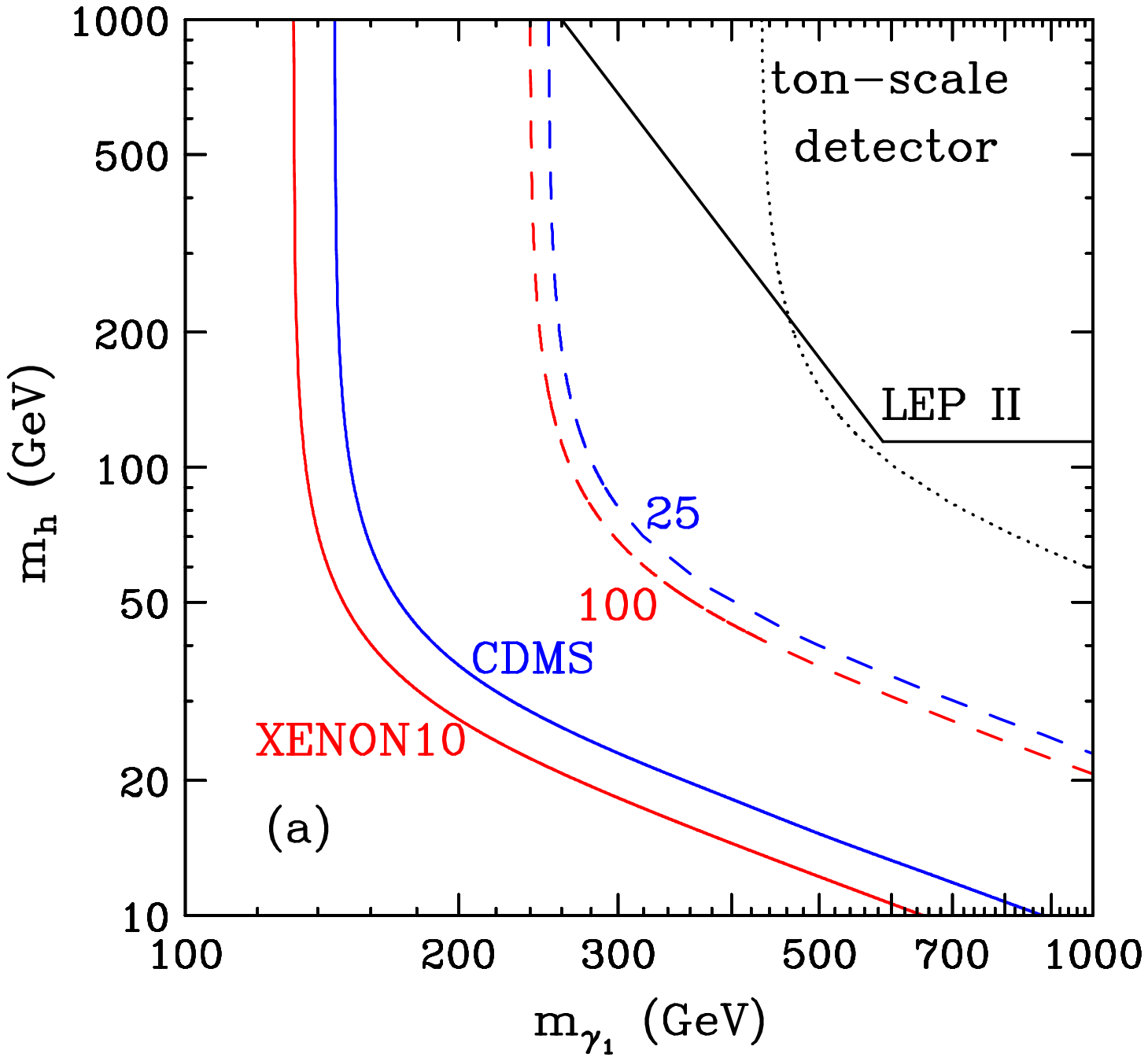}
\hspace{0.1cm}
\includegraphics[width=0.48\textwidth]{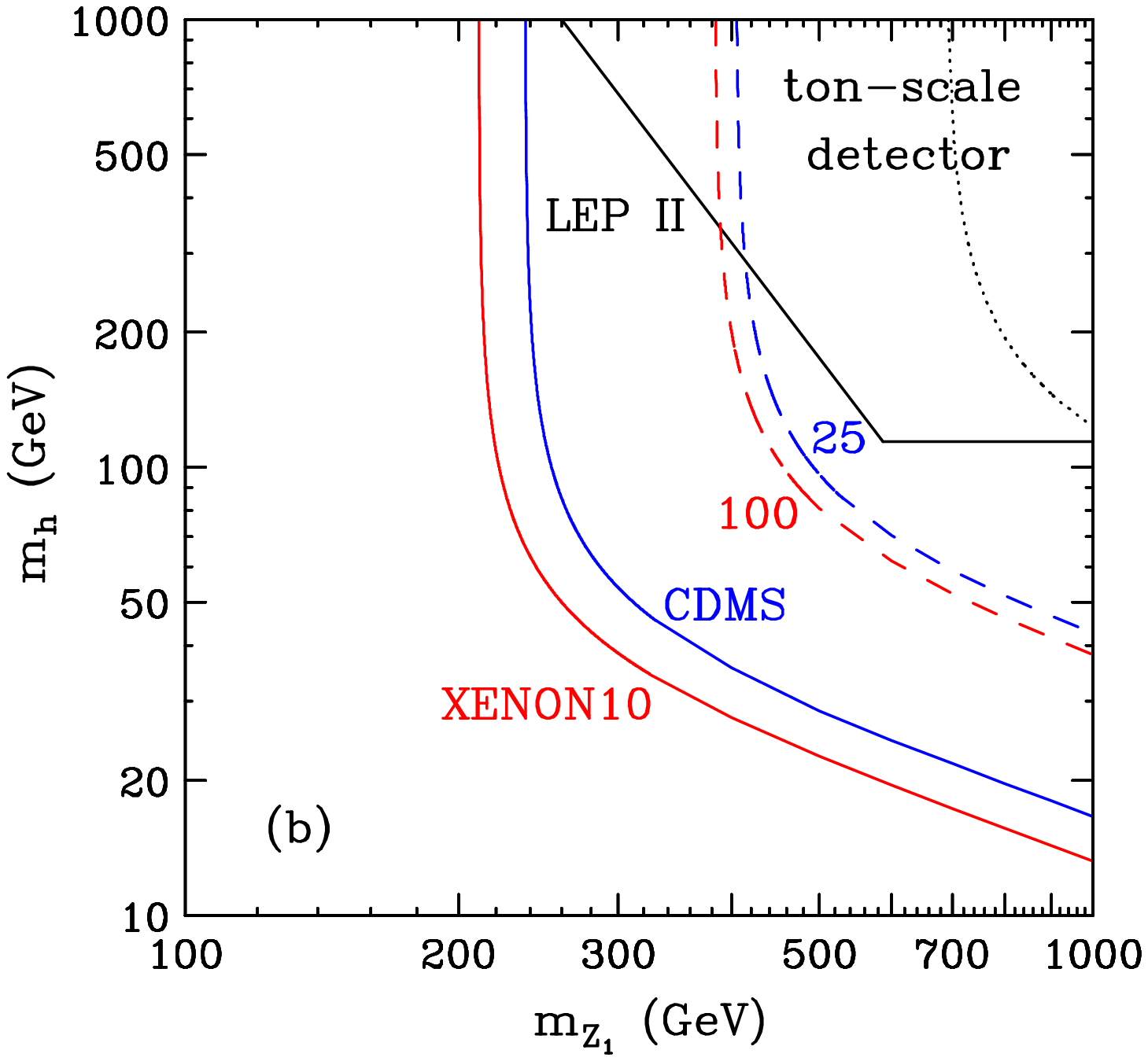}
\caption{\sl Limit on the SM Higgs mass $m_h$ and the LKP mass for (a) $\gamma_1$ and (b) $Z_1$ for a given $\Delta_{q_1}=0.1$. 
The horizontal black solid line is the lower bound on $m_h$ from LEP-II (90\% C.L. at top quark mass 173\,GeV). 
The diagonal black solid line delineates the region disfavored by precision data.
The solid curves are the current (90\% C.L) limits from CDMS (in blue) and XENON10 (in red). 
The dashed curves (SuperCDMS 25\,kg and XENON100) and dotted line (ton-scale detector) 
are the projected sensitivities for the future experiments.}
  \label{fig:SI_HiggsMass_Neutron_B_Z}
\end{figure}

While previously we already argued that $m_{LKP}$ and $\Delta_{q_1}$
are the most relevant parameters for UED dark matter phenomenology,
for completeness we also investigate the dependence on the SM Higgs mass $m_h$,
which is currently still unknown. In Fig.~\ref{fig:SI_HiggsMass_Neutron_B_Z}
we therefore translate the information from Fig.~\ref{fig:SI_CrossSection_Neutron} 
into the $m_{LKP}$-$m_h$ plane, for a given fixed KK mass splitting $\Delta=0.1$
now taking the Higgs mass $m_h$ as a free parameter.
In each panel, the horizontal black solid lines mark the current 
Higgs mass bound of 114\,GeV while the diagonal black solid lines 
show the indirect limit from the oblique corrections in this 
model~\cite{Gogoladze:2006br}.\footnote{One should keep in mind that the 
latter have been calculated only for the case of $\gamma_1$ LKP, and 
only within the framework of minimal UED. The line shown 
in Fig.~\ref{fig:SI_HiggsMass_Neutron_B_Z}b is therefore only for illustration.
Furthermore, the $\gamma_1$ calculation itself may be subject to modifications in
the more general scenarios which we are considering here.} 
For low $m_h$, the limit on the LKP mass (or 
equivalently, the compactification scale) is 
$m_{LKP} \sim R^{-1} \gtrsim 600$\,GeV (for $m_t=173$\,GeV), 
but it gets weaker for larger $m_h$, so that $m_{LKP}$ values 
as low as 300\,GeV are still allowed if the SM Higgs boson is very heavy \cite{Appelquist:2002wb}.
In Fig.~\ref{fig:SI_HiggsMass_Neutron_B_Z} we also show the 
current (solid lines) limits from CDMS (in blue) and XENON10 (in red), 
their projected near-future sensitivities,
SuperCDMS 25\,kg and XENON100 (dashed lines), and the projected sensitivity of  
a ton-scale detector (dotted line).
The shape of these contours is easy to understand.
At large $m_h$, the Higgs exchange diagram in Fig.~\ref{fig:direct} 
decouples, the elastic scattering rate becomes independent of $m_h$
and the direct detection experimental sensitivity is only a 
function of $m_{LKP}$ (since $\Delta_{q_1}$ is held fixed).
In the other extreme, at small $m_h$, the Higgs exchange diagram 
dominates, and the sensitivity now depends on both $m_h$ and 
$m_{LKP}$. Unfortunately, for $\Delta=0.1$
the current direct detection bounds do not extend into the interesting
parameter space region, but future experiments will eventually start
probing the large $m_h$ corner of the allowed parameter space.
On the positive side, one important lesson from Fig.~\ref{fig:SI_HiggsMass_Neutron_B_Z}
is that the $m_h$ dependence starts showing up only at very
low values of $m_h$, which have already been ruled out by the 
Higgs searches at colliders. This observation confirms that when
it comes to interpreting existing and future experimental limits on WIMPs
in terms of model parameters, $m_{LKP}$ and $\Delta_{q_1}$
are indeed the primary parameters, while $m_h$
plays a rather secondary role.

We remind the reader that the LHC will be able to probe all of the 
parameter space shown in Fig.~\ref{fig:SI_HiggsMass_Neutron_B_Z}a
through the $4\ell + \met$ signature, while the discovery of UED in
Fig.~\ref{fig:SI_HiggsMass_Neutron_B_Z}b appears quite problematic. 
Of course, the SM Higgs boson will be discovered in both cases, 
for the full range of $m_h$ masses shown. 

\begin{figure}[t]
\includegraphics[width=0.485\textwidth]{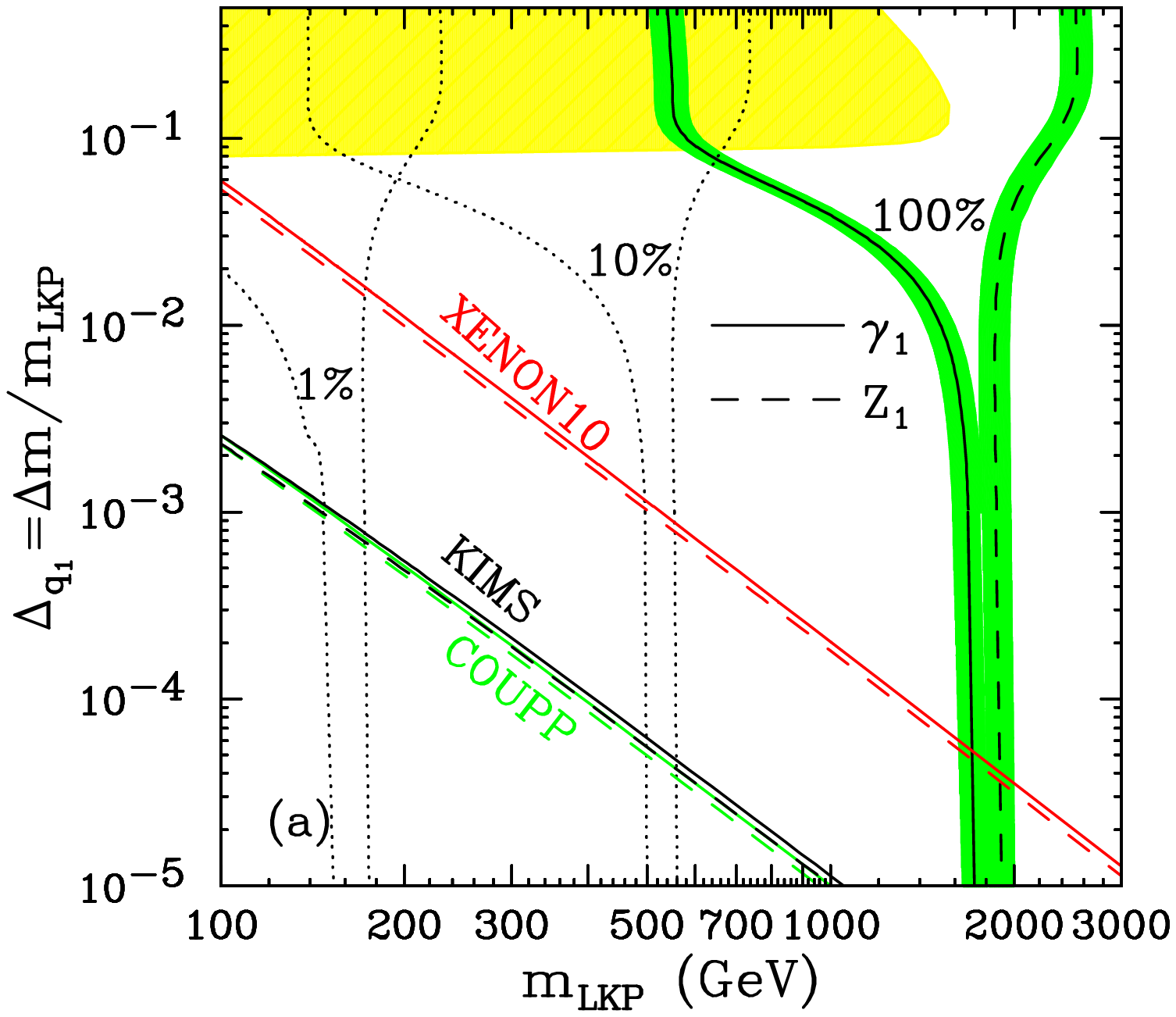}
\hspace{0.1cm}
\includegraphics[width=0.485\textwidth]{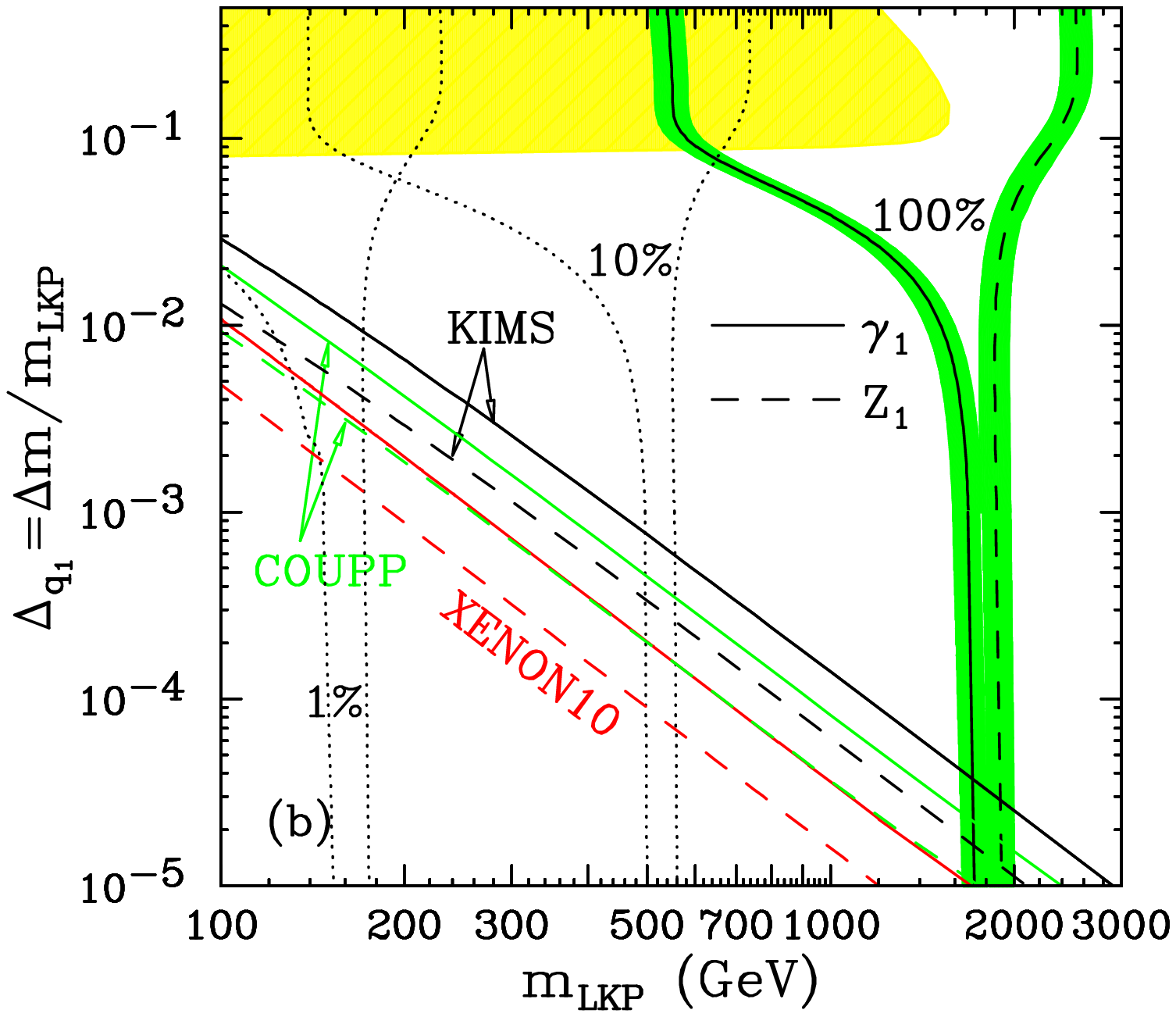}
\caption{\sl Experimental upper bounds (90\% C.L.) on the spin-dependent 
elastic scattering cross sections on (a) neutrons and (b) protons
in the $m_{LKP}$-$\Delta_{q_1}$ plane. 
The solid (dashed) curves are limits on $\gamma_1$ ($Z_1$) for each experiment. Shaded regions and dotted lines are defined in the same way as in Fig.~\ref{fig:SI_Delta_Neutron_B_Z}. The depicted LHC reach (yellow shaded region) applies only to the case of $\gamma_1$ LKP. }
\label{fig:SD_Delta_B}
\end{figure}

We now turn to a discussion of the corresponding
spin-dependent elastic scattering cross sections, which also exhibit 
an enhancement at small $\Delta_{q_1}$, as shown in Fig.~\ref{fig:sigma_spin_B1_Z1}.
Similar to Fig.~\ref{fig:SI_Delta_Neutron_B_Z}, in Fig.~\ref{fig:SD_Delta_B}
we combine existing limits from three different experiments 
(XENON10, KIMS and COUPP) in the $m_{LKP}$-$\Delta_{q_1}$ plane.
Panel (a) (panel (b)) shows the constraints from the 
WIMP-neutron (WIMP-proton) SD cross sections. The rest 
of the KK spectrum has been fixed as in Figs.~\ref{fig:Omegah2_B1_Z1}
and \ref{fig:SI_Delta_Neutron_B_Z}, and $m_h=120$\,GeV. 
The solid (dashed) curves are limits 
on $\gamma_1$ ($Z_1$) for each experiment. The constraints from 
LHC and WMAP on the $m_{LKP}$-$\Delta_{q_1}$ parameter space are the same as in 
Fig.~\ref{fig:SI_Delta_Neutron_B_Z}.

By comparing Figs.~\ref{fig:SI_Delta_Neutron_B_Z} and \ref{fig:SD_Delta_B}
we see that, as expected, the parameter space constraints from SI interactions 
are stronger than those from SD interactions. For example, in perhaps 
the most interesting range of LKP masses from 300\,GeV to 1 TeV, the SI limits 
on $\Delta_{q_1}$ in Fig.~\ref{fig:SI_Delta_Neutron_B_Z} range from a few 
times $10^{-2}$ down to a few times $10^{-3}$. On the other hand, the
SD bounds on $\Delta_{q_1}$ for the same range of $m_{LKP}$ are about an 
order of magnitude smaller (i.e. weaker). We also notice that the constraints 
for $\gamma_1$ LKP are stronger than for $Z_1$ LKP. This can be easily understood by
comparing Fig.~\ref{fig:sigma_spin_B1_Z1}a and Fig.~\ref{fig:sigma_spin_B1_Z1}b:
for the same LKP mass and KK mass splitting, the $\gamma_1$ SD cross sections 
are typically larger.

Fig.~\ref{fig:SD_Delta_B} also reveals that the experiments rank 
differently with respect to their SD limits on protons and neutrons. For example,
KIMS and COUPP are more sensitive to the proton cross section, while
XENON10 is more sensitive to the neutron cross section.
As a result, the current best SD limit on protons comes from KIMS,
but the current best SD limit on neutrons comes from XENON10.
Combining all experimental results can
give a very good constraint on the $a_p$-$a_n$ parameter space.
\begin{figure}[t]
\includegraphics[width=0.48\textwidth]{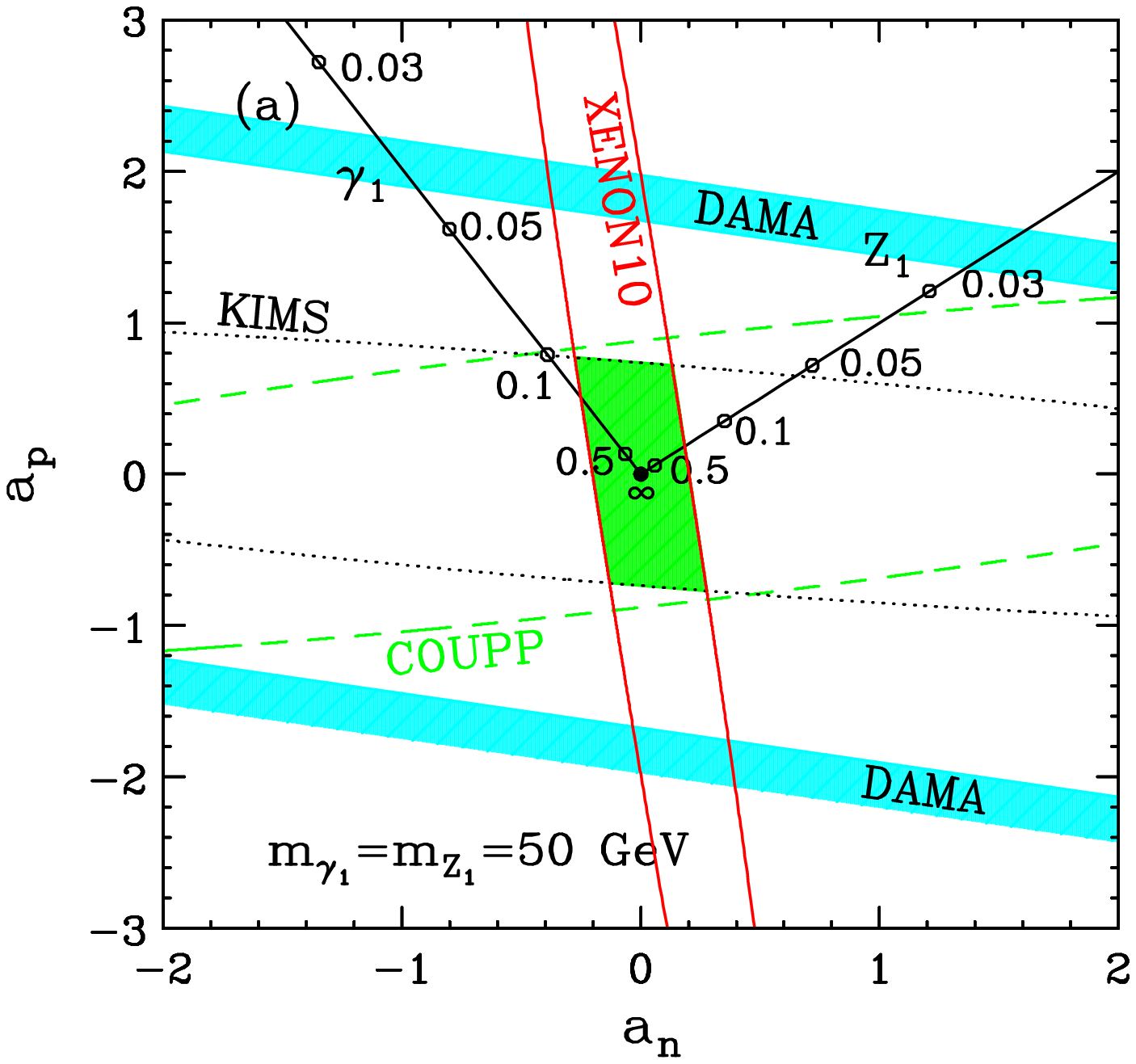}
\hspace{0.1cm}
\includegraphics[width=0.485\textwidth]{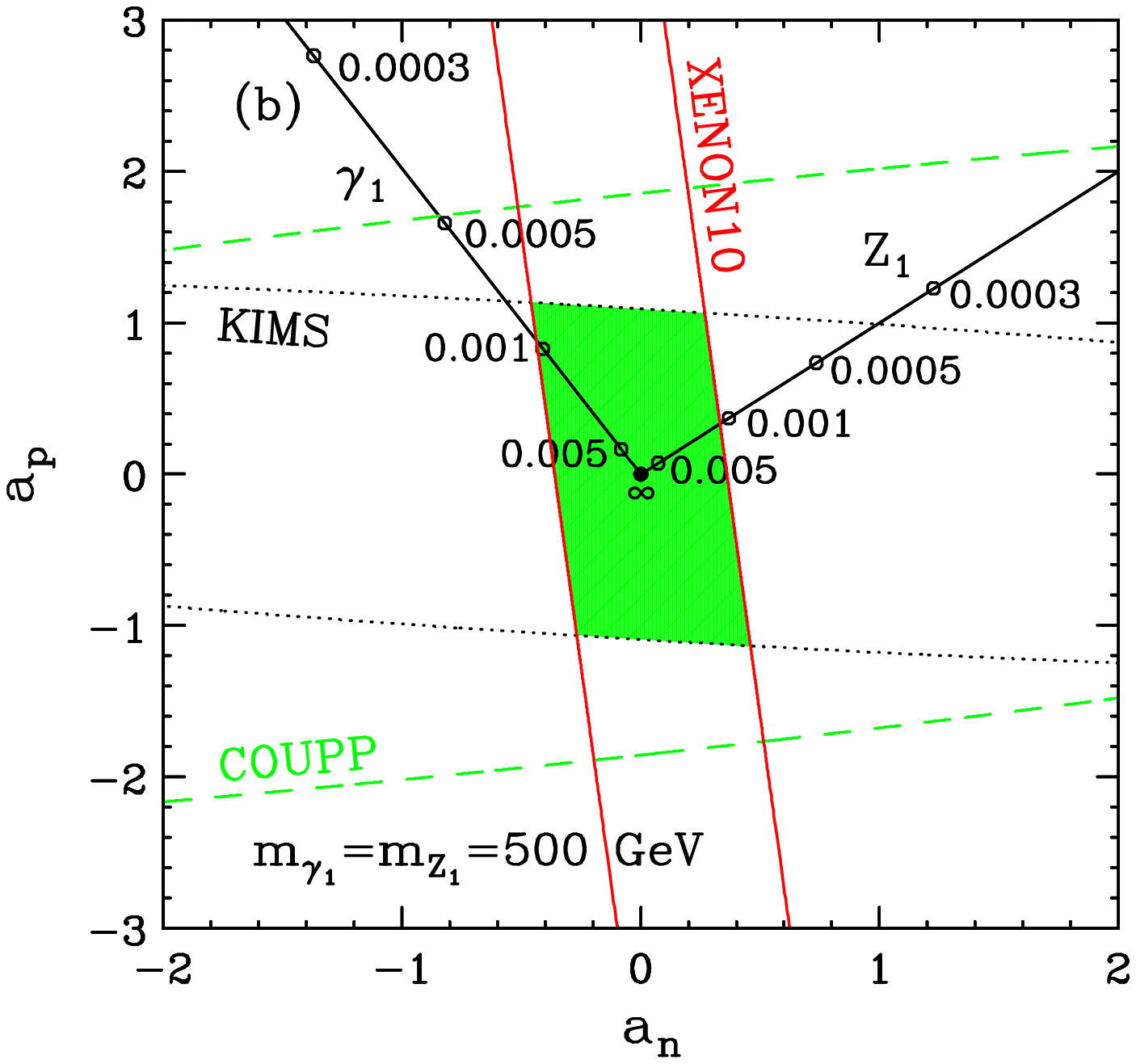}
\caption{\sl Experimental limits on the $a_p$-$a_n$ parameter space 
for (a) $m_{LKP}= 50$\,GeV and (b) $m_{LKP} = 500$\,GeV. 
The contours show limits from XENON10 (red solid line), KIMS (black dotted line) 
and COUPP (green dashed line). The  blue near-horizontal bands show the 
evidence regions allowed by DAMA \cite{Savage:2004fn}, while the green region 
shows the parameter space allowed by all current experimental results. 
The two straight lines originating
from $a_n=a_p=0$ are the theoretical predictions for 
$a_p$ and $a_n$ in the case of $\gamma_1$ or $Z_1$ LKP in 5D UED.
These theory lines are parametrized by the value of $\Delta_{q_1}$
as indicated by a few representative points.}
\label{fig:SD_Ellipses}
\end{figure}
Fig.~\ref{fig:SD_Ellipses}a (Fig.~\ref{fig:SD_Ellipses}b) shows 
combined results for $m_{LKP}= 50$\,GeV ($m_{LKP}= 500$\,GeV)
in the (model-independent) $a_p$-$a_n$ parameter space. 
The contours show limits from XENON10 (red solid line), KIMS (black dotted line) 
and COUPP (green dashed line). 
The  blue near-horizontal bands show the evidence 
regions allowed by DAMA \cite{Savage:2004fn}, while the green region 
shows the parameter space allowed by all current experiments. 
Note that these limits were computed in two different ways. The
results from KIMS and COUPP are based on the method proposed in 
\cite{Tovey:2000mm} whereas those from DAMA and XENON10 are calculated
as advocated in \cite{Savage:2004fn}.
We believe that the latter is more accurate since limits are computed for all
angles in the  $a_p$-$a_n$ plane separately whereas the former solely
relies on the limits calculated considering pure coupling to neutrons and
protons respectively. More details about these calculations can be found in the appendix.
The two straight lines originating
from $a_n=a_p=0$ are the theoretical predictions for 
$a_p$ and $a_n$ in the case of $\gamma_1$ or $Z_1$ LKP in 5D UED.
These theory lines are parametrized by the value of $\Delta_{q_1}$
as indicated by a few representative points.
The feature which is readily apparent in Fig.~\ref{fig:SD_Ellipses} is 
the orthogonality between the regions allowed by the $a_p$-sensitive experiments 
like KIMS and COUPP, on the one side, and the $a_n$-sensitive experiments 
like XENON10, on the other. 
This indicates the complementarity of the two groups of experiments:
the green-shaded region allowed by the combination of all experiments is substantially 
more narrow than the region allowed by each individual experiment.

In conclusion of this section, we shall also consider KK dark matter 
candidates in models with two universal extra dimensions (6D UED).
As mentioned in Sec.~\ref{sec:model} the novel possibility here
compared to 5D UED is the scalar photon ($\gamma_H$) LKP.
As a spin zero particle, it has no spin-dependent interactions 
and can only be detected through its spin-independent elastic scattering.
\begin{figure}[t]
\includegraphics[width=0.485\textwidth]{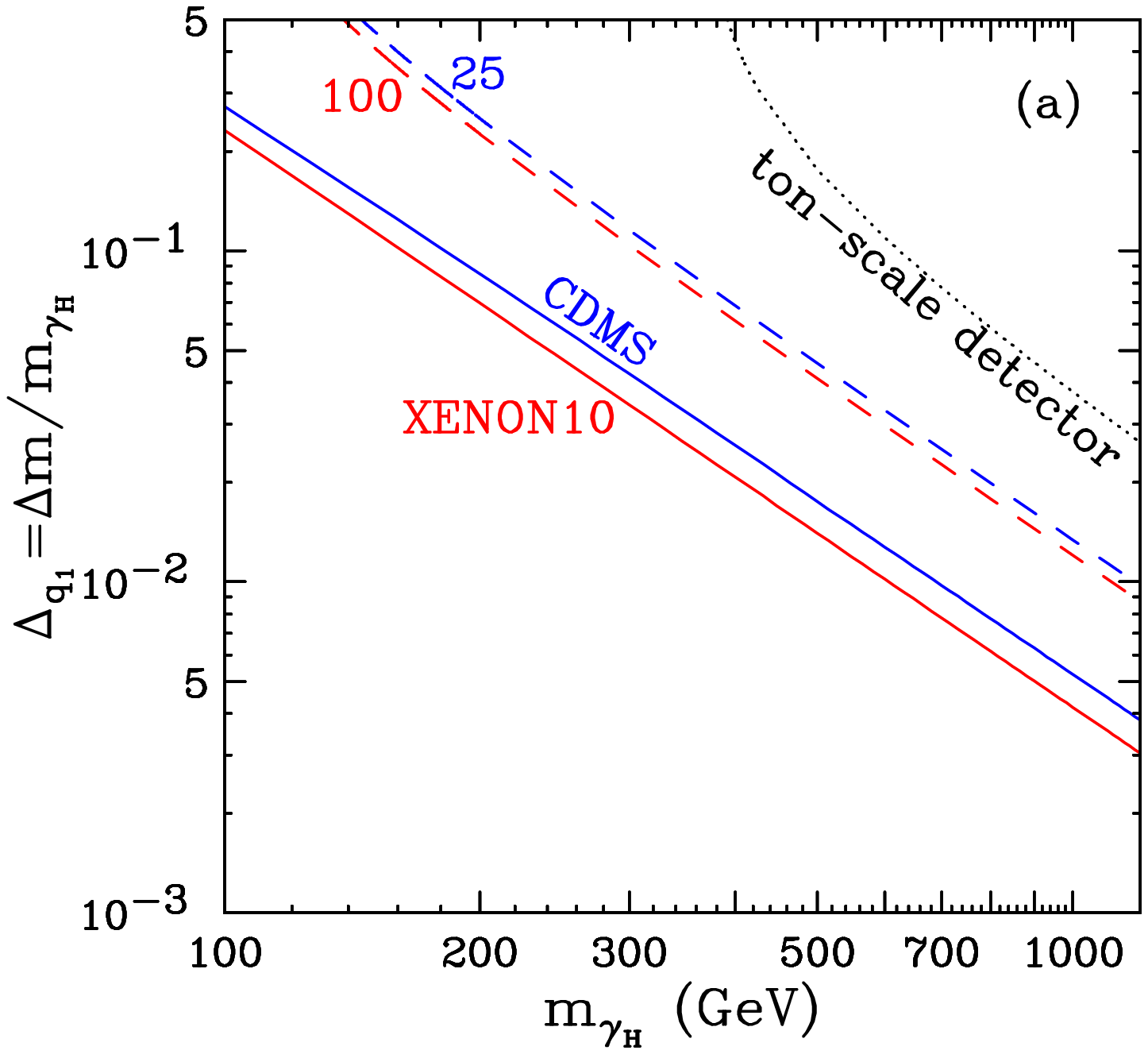}
\hspace{0.1cm}
\includegraphics[width=0.490\textwidth]{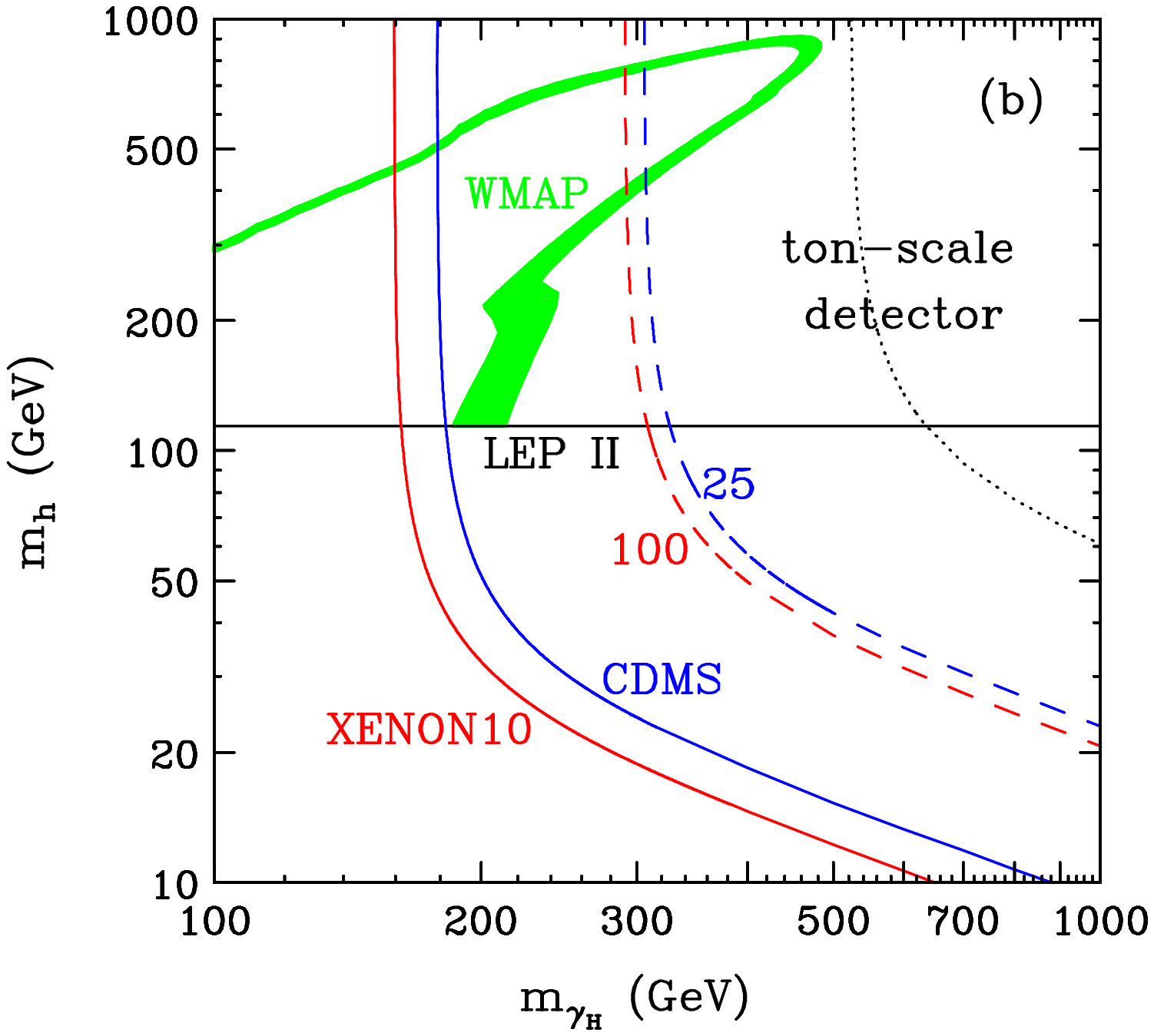}
\caption{\sl Experimental limits (90\% C.L.) on the scalar LKP ($\gamma_H$)
in 6D UED. 
(a) Lower bound of $\Delta_{q_1}$ vs $m_{\gamma_H}$ for $m_h=$120\,GeV. 
The solid lines are the current experimental lower bounds on $\Delta_{q_1}$
for a given $m_{\gamma_H}$ from CDMS (blue) and XENON10 (red). 
SuperCDMS 25\,kg and XENON100 projected sensitivities are
drawn with dashed lines. The dotted line shows the projected sensitivity of 
a ton-scale experiment. (b) Lower bound of the Higgs mass ($m_h$) 
as a function of $m_{\gamma_H}$ for a fixed $\Delta_{q_1} = 0.1$. 
The WMAP allowed range is the green shaded region. 
The LEP II lower bound on $m_h$ is shown as the black solid line. }
\label{fig:SI_Delta_Neutron_BH}
\end{figure}
Fig.~\ref{fig:SI_Delta_Neutron_BH}a (Fig.~\ref{fig:SI_Delta_Neutron_BH}b)
is the analogue of Fig.~\ref{fig:SI_Delta_Neutron_B_Z} 
(Fig.~\ref{fig:SI_HiggsMass_Neutron_B_Z}) for the case of $\gamma_H$ LKP.
In Fig.~\ref{fig:SI_Delta_Neutron_BH}a we show lower bounds on $\Delta_{q_1}$ 
versus the mass $m_{\gamma_H}$ of the scalar photon, for a fixed 
Higgs mass ($m_h=120$\,GeV). The solid lines indicate the current 
experimental limits from CDMS (blue) and XENON10 (red). 
The dashed lines are the projected sensitivities of
SuperCDMS 25\,kg and XENON100 and the dotted line is the projected 
sensitivity of a ton-scale detector.
Since the cosmologically preferred mass range for $\gamma_H$
is much lower ($\sim 200$\,GeV before accounting for coannihilations) 
than for the LKP in 5D UED, 
the constraints are quite powerful -- in particular, 
the future ton-scale experiments are expected to cover most of the 
interesting mass splitting ($\Delta_{q_1}$) region. 

In Fig.~\ref{fig:SI_Delta_Neutron_BH}b we show lower bounds of the Higgs mass 
$m_h$ as a function of $m_{\gamma_H}$ for a fixed $\Delta_{q_1} = 0.1$. 
The WMAP preferred parameter space is marked as the green shaded region, while
the black solid line is the LEP II lower limit on $m_h$. The contours 
resemble in shape those seen earlier in Fig.~\ref{fig:SI_HiggsMass_Neutron_B_Z}.
In particular, we notice that within the LEP II allowed range, 
the Higgs mass does not have a large impact on the direct detection bounds.
However, if the LHC finds a SM Higgs boson with a mass smaller 
than $\sim$300\,GeV, then the WMAP bound would constrain the 
mass of $\gamma_H$ within a relatively narrow mass ranges 
at a given mass splitting ($\Delta_{q_1}$). 
For example in Fig.~\ref{fig:SI_Delta_Neutron_BH}b, 
where the fixed mass splitting is $\Delta_{q_1}=0.1$, 
the corresponding constraint on the mass of $\gamma_H$ 
would be $180\,\text{GeV} < m_{\gamma_H} < 250\,\text{GeV}$.
In fact, this conclusion is rather insensitive to the particular choice of 
$\Delta_{q_1}$. This is due to the fact that $\gamma_H$ self-annihilation 
is helicity-suppressed and gauge boson final states are dominant in the 
WMAP allowed regions. Therefore, Fig.~\ref{fig:SI_Delta_Neutron_BH}b would look qualitatively 
similar, if a different value of $\Delta_{q_1}$ were used.

\section{Conclusions}\label{sec:conclusion}
The dark matter puzzle is among the most intriguing questions in particle physics.
Its origin resides in cosmological observations such as the rotation curves of galaxies, 
cosmic microwave background, gravitational lensing, large scale structure, 
the mass to luminosity ratio and so on.
Interestingly, many scenarios of new physics beyond the Standard Model 
provide a stable neutral particle which, in principle, can be produced and observed at colliders.
In fact, one of the primary motivations for SUSY has always been the fact that it
naturally accommodates a WIMP candidate. 
More recently, we have learned that extra dimensional models 
provide a viable alternative to SUSY dark matter, namely KK dark matter.
Both of these scenarios have been attracting a lot of attention in terms 
of collider and astrophysical aspects. 
In this paper we performed a comprehensive phenomenological analysis of KK dark matter
in universal extra dimensions, extending previous studies by considering 
new LKP candidates ($Z_1$ and $H_1$). We also revisited the cases 
of $\gamma_1$ and $\gamma_H$ LKP, focusing on the possibility of a small mass splitting 
with the KK quarks. All of these features can be realized in non-minimal UED 
scenarios and therefore deserve attention. 

In our analysis we included the relevant theoretical constraints from cosmology
(the relic density of KK dark matter) and particle physics (low energy precision data).
We accounted for all coannihilation processes in our relic density calculation, 
focusing on coannihilations with KK quarks since they play an important role for 
direct detection at small mass splittings.

We then contrasted the sensitivities of the LHC and the different types of 
direct detection experiments, and exhibited their complementarity. 
We demonstrated that the $m_{LKP}-\Delta_{q_1}$ parameter space is both
convenient and sufficient for a simultaneous discussion of collider and 
direct detection searches. Collider experiments like the LHC and possibly ILC
are sensitive to the region of relatively low $m_{LKP}$ and sufficiently large $\Delta_{q_1}$.
On the other hand, direct detection experiments do best at relatively low $m_{LKP}$ and
small $\Delta_{q_1}$. Finally, cosmology rules out the region of very large $m_{LKP}$.
We see that, at least in principle, the combination of all three types of constraints
has the potential to completely cover the relevant parameter space. 
We showed that with the expected sensitivity of the next generation 
direct detection experiments, the coverage is almost complete in the 
case of $\gamma_1$ LKP.

In conclusion, we summarize the main lessons from each of the three main areas 
in our study, and point towards interesting directions for future work. 
\begin{itemize}
\item {\em Direct detection of KK dark matter.} The direct detection prospects are
greatly enhanced when the LKP becomes degenerate with the KK quarks \cite{Cheng:2002ej}.
Therefore, the mass splitting $\Delta_{q_1}$ is a key parameter which
is worth exploiting by the experimental collaborations in presenting their limits.
The conventional approach is to overlay the results of random scans over the full 
parameter space of a given model over the model-independent exclusion curves
from Figs.~\ref{fig:SI_CrossSection_Neutron} and \ref{fig:SD_CrossSection}.
There are several problems with this. First, due to the multitude of parameters 
being scanned, such scans are not sufficiently exhaustive and are 
almost guaranteed to miss the relevant parts of parameter space.
For example, suppose we scan a KK quark mass in the range from 
$m_{LKP}$ up to some maximum value $m_{q_1}^{max}$. In order to obtain 
{\em a single} parameter space point at a given $\Delta_{q_1}$, 
the number of points sampled along the $m_{q_1}$ direction should be at least
$$
N = \frac{m_{q_1}^{max}-m_{LKP}}{m_{q_1}-m_{LKP}} 
= \frac{m_{q_1}^{max}-m_{LKP}}{m_{LKP}}\frac{1}{\Delta_{q_1}}
\sim \frac{1}{\Delta_{q_1}}\ .
$$

In order to probe mass degeneracies of $\Delta_{q_1}\sim 0.01$, one should therefore
generate at least $10^2$ points {\em along each parameter dimension}. 
It is then clear that random scans cannot effectively probe
models which have more than a few input parameters.
For example, typical SUSY scans often utilize a constrained parameter space 
which nevertheless still has on the order of 10-15 parameters. In that case,
seeing the effects of 1\% degeneracies would require $10^{20}-10^{30}$ total 
points -- a number which is obviously impractical. Furthermore, many of the 
scanned parameters often have little effect on the experimental signals being 
discussed. We therefore find it much more efficient and illuminating
to forego such general scans in favor of simple parametrizations 
which would contain only the variables relevant for the experimental search. 
A simple implementation of this idea is the $m_{LKP}-\Delta_{q_1}$ parameter 
space used in Figs.~\ref{fig:SI_Delta_Neutron_B_Z} and \ref{fig:SD_Delta_B}.
Notice that our argument is not limited to UED models --
the mass of the dark matter particle and the mass splitting are 
expected to be the two most important parameters for dark matter searches
in any other theory (such as SUSY, little Higgs, warped extra dimensions, etc.). 
\item {\em Collider searches for KK dark matter.} While beneficial for direct DM
detection, mass degeneracies are generally problematic for collider searches,
since they degrade the quality of the discovery signatures. For example,
at small $\Delta_{q_1}$ the KK quark decay products become softer, and
the leptonic modes may altogether turn off. Detailed LHC studies for
these challenging situations are still lacking and are definitely worth 
undertaking. In particular, the LHC phenomenology of $Z_1$ or $H_1$ LKP
has never been discussed.
\item {\em Theoretical calculations related to KK dark matter.}
In this paper we have reviewed the main ingredients of a complete
analysis of KK dark matter phenomenology. We emphasized the importance of 
including the effect of coannihilations in the case of small mass splittings 
which are relevant for experimental searches. However, there are still 
several missing pieces which are needed to improve the accuracy of
our predictions. While the relic density calculation for $\gamma_1$ and
$Z_1$ LKP is on a relatively firm footing, since the corresponding 
coannihilation processes are known, the complete calculation of
coannihilation effects in case of $H_1$, $\gamma_H$ or $Z_H$
LKP is still lacking. Similarly, one would like to have available 
the precise calculation of the heavy flavor contribution to the 
LKP elastic scattering cross section on nucleons. We are hoping that
it would take less than a WIMP direct detection signal to jumpstart the 
theoretical efforts in this direction. 
\end{itemize}

\begin{acknowledgments}
We would like to thank CDMS collaboration, COUPP collaboration, 
KIMS collaboration and XENON10 collaboration for providing important information, 
and Tim Tait for useful discussion.
SA and LB are supported by the Swiss National Foundation SNF Grant No 20-118119 
and by the Volkswagen Foundation.
KM is supported in part by 
a US Department of Energy Outstanding Junior Investigator 
award under grant DE-FG02-97ER41029. 
Fermilab is operated by Fermi Research Alliance, LLC under
Contract No. DE-AC02-07CH11359 with the U.S. Department of Energy.
\end{acknowledgments}

\section{Appendix:  Event rate calculations for direct detection experiments} 

Given the interaction cross section of LKPs with nucleons the expected event rates for each experiment can be calculated.
A spherical halo model is conventionally used for a consistent comparison of the different experimental results. 
Spin-independent (spin-dependent) interactions with non-zero momentum transfer to the nucleus demand nuclear 
(spin) form factor corrections in the cross section calculations. Furthermore it is a common procedure 
to normalize the cross sections to the scattering from a single nucleon.
In the following we describe the details of our event rate calculations for direct detection experiments
particularly with regard to setting limits on the cross sections and couplings.
Further information can be found in reference~\cite{Lewin:1995rx}.

\subsection{Spherical halo model and differential event rates}
The dark matter halo is assumed to be an isothermal and isotropic sphere of an ideal WIMP gas obeying a Maxwell-Boltzmann velocity distribution 
\begin{equation}
  f(\vec{v}, \vec{v}_{E}) \sim e^{ -\frac{(\vec{v} + \vec{v}_{E})^{2}}{ v_{0}^{2}} } \, ,
  \label{MB_distribution}
\end{equation}
where $\vec{v}$  denotes the velocity of the WIMPs in the rest frame of the earth and $\vec{v}_{E}$ 
the velocity of the earth with respect to the motionless galactic halo. $v_{0}$ (= 220\,km/s) is the characteristic velocity of the distribution which is assumed to be 
equal to the galactic rotation velocity $v_{r}$. There are three contributions to  $\vec{v}_{E}$: the galactic rotation velocity $\vec{v}_{r}$, 
the velocity of the sun with respect to the galactic disc $\vec{v}_{s}$  and the velocity of the earth around the sun $\vec{v}_{orb}$. Its main contribution is given by
\begin{equation}
v_{E}(t) = v_{r} + v_{s} + v_{orb} \cos \beta \cos \left(2 \pi\frac{t-t_{0}}{T}\right) \, ,
\end{equation}
taking the angle $\beta = 59.575 ^{\circ}$ between the earth orbital plane and the galactic plane into
account. The velocities are given by 
$v_{r} = 220 \textrm{ km/s}$, $v_{s} = 12 \textrm{ km/s}$ and  $v_{orb} = 29.79 \textrm{ km/s}$
respectively. $t_0$ is the day in a year corresponding 
to the 2$^{\textrm{nd}}$ June ($t_{0} = 152.5$) and $T$ is the number of days in a year ($T = 365.25$). The modulation of the mean velocity is about $\pm 6.5 \%$.
Since we are not investigating time depending properties of event rates such as annual modulation
effects, we use the mean velocity of the earth $\langle v_{E} \rangle = 232 \text{ km/s}$, averaged over one year. 
The velocity distribution of WIMPs is isotropic in the galactic rest frame. It is limited by the escape velocity of the WIMPs ($v_{esc}$) from the galactic halo 
\begin{equation}
|\vec{v} + \vec{v}_{E}| < v_{esc} \, ,
\end{equation}
which yields a maximum WIMP velocity 
\begin{equation}
  v_{max}(\theta, t) = \sqrt{v_{esc}^{2} - v_{E}^{2}(t) \,(1-\cos^{2}\theta)} - v_{E}(t) \cos \theta \, ,
  \label{esc}
\end{equation}
with  $\theta$ being the scattering angle in the galactic rest frame. In our calculations we take $v_{esc}=544$~km/s \cite{Smith:2006ym}. 
We obtain a mean maximum WIMP velocity given by $\langle v_{max} \rangle = 518.3 \textrm{ km/s}$
by averaging over a year and the scattering angle.
The maximum recoil energy $E_{R_{max}}$ of the nucleus is given by  
\begin{equation}
  E_{R_{max}} = \frac{1}{2} m_{LKP} v_{max}^{2} r \label{max_energy} \, ,
\end{equation}
with the kinematic factor 
\begin{equation}
r = \frac{4 m_{LKP} m_{T}}{(m_{LKP} + m_{T})^{2}} \; \textrm{.}
\end{equation}
The differential event rate is
\begin{eqnarray}\label{dRdE}
\frac{\mathrm{d}R}{\mathrm{d} E_{R}} &=& \frac{R_{0}(\sigma)}{E_{0}r} \frac{k_{0}}{k_{1}}
\Bigg[ \frac{\sqrt{\pi}}{4} \frac{v_{0}}{\langle v_{E} \rangle} \bigg[ \textrm{erf} \left(\frac{v_{min} + \langle v_{E} \rangle}{v_{0}}\right) 
-  \textrm{erf}\left(\frac{v_{min} - \langle v_{E} \rangle}{v_{0}}\right) \bigg]- e^{- \frac{v_{esc}^{2}}{v_{0}^{2}}}\Bigg] \, .\label{das}
\end{eqnarray}
$R_0(\sigma)$ is the total event rate for a given cross section assuming $v_{E}=0$, $v_{esc} = \infty$ and integrated 
over recoil energies from $E_{R} = 0$ to $E_{R} = \infty$. $v_{min}$ is defined as the minimum velocity leading to a certain recoil energy  $E_{R}$ whereas
$E_{0}$ denotes the energy carried by a WIMP with the velocity $v_{0}$. The ratio  $k_{0}/k_{1}$ arises from the normalization of 
the WIMP density distribution. For completeness all relevant formulas are given below:
\begin{eqnarray}
  R_{0}(\sigma) &=& \frac{2}{\sqrt{\pi}} \frac{N_{0}}{A_{u}} \frac{\rho}{m_{LKP}} \, \sigma \, v_{0} \label{nna} \, ,\\
  v_{min} &=& \sqrt{ \frac{E_{R}}{E_{0}r} } \, v_{0} \, , \\
  E_{0}  &=& \frac{1}{2} m_{LKP} v_{0}^{2} \, , \\
  \frac{k_{0}}{k_{1}} &=&  \bigg[  \textrm{erf} \left(\frac{v_{esc}}{v_{0}}\right)  - \frac{2}{\sqrt{\pi}} \frac{v_{esc}}{v_{0}} e^{- \frac{v_{esc}^{2}}{v_{0}^{2}}} \bigg]^{-1} \, .
\end{eqnarray}
$N_{0}$ is the Avogadro constant and $A_{u}$ is the atomic mass unit. The local dark matter density $\rho = 0.3 \textrm{ GeV/cm}^{3}$ is 
taken from \cite{Gates:1995dw}. $\sigma$ denotes either the spin-independent or spin-dependent cross section which for the case of $\gamma_1$ LKP are given in \eqref{scalar} and \eqref{apaneq}
respectively.

\subsection{Expected event rate: spin-independent interactions}
The spin-independent cross-section given in equation \eqref{scalar} normalized to a single nucleon can be written as 
\begin{equation}
  \sigma^{p,n}_{scalar} = \frac{1}{A^2}\frac{\mu^2_{p,n}}{\mu^2}\sigma_{scalar} \, .
\end{equation}
Owing to the finite size of the target nucleus, the LKP-nucleus cross section 
is valid only for the case of zero-momentum transfer ($q=0$). For non-zero momentum transfer, a form factor 
correction needs to be considered:
\begin{equation}
  \sigma_{SI} = \sigma_{\text{scalar}} F_{SI}^{2} \, ,
\end{equation}
where $F_{SI}$ is the nuclear form factor. We used Helm's model \cite{helm} :
\begin{equation}
  F_{SI}(q r_n) = 3 \frac{j_{1}(q r_{n}) }{q r_{n}}e^{- \frac{(qs)^{2}}{2} } \, ,
  \label{fff}
\end{equation}
with the spherical Bessel function $j_{1}$, the effective nuclear radius $r_{n}=\sqrt{c^2+\frac{7}{3} \pi^2 a^2 - 5s^2}$ with $a=0.52 \text{ fm}$ and $c=1.23 \sqrt[3]{A}-0.60 \text{ fm}$, and the nuclear skin thickness $s=1 \text{ fm}$ \cite{Lewin:1995rx}. 

The spin-independent differential event rate for finite-momentum transfer can then be obtained by 
replacing $\sigma$ with $\sigma_{SI}$ in \eqref{nna} and thus in \eqref{dRdE}. If the WIMP target consists of 
more than one element, the respective abundances of each isotope $f_i$ have to be considered. 
The total differential event rate for a specific WIMP target is
\begin{equation}
  \frac{\mathrm{d}R_{SI}}{\mathrm{d} E_{R}} = \sum_{i} \, f_{i} \,
  \frac{\mathrm{d}R^{i}_{SI}}{\mathrm{d} E_{R}} \; \textrm{.}  \label{r_total2}
\end{equation}
The expected number of events for spin-independent interactions for a given experiment 
can then be written as
\begin{equation}\label{n_si}
  N_{SI} = \int_{\textrm{q}_{\textrm{min}}}^{\textrm{min}\big(\textrm{q}_{\textrm{max}} \textrm{,} \langle E_{R_{max}} \rangle\big)} \mathrm{d} E_{R} \, \frac{\mathrm{d}R_{SI}}{\mathrm{d} E_{R}}\,\epsilon(E_R)\,MT \, ,
\end{equation}
where $MT$ denotes the total detector exposure in kg-days and $\epsilon$($E_R$) the WIMP detection efficiency as a function of recoil energy. 
$\textrm{q}_{\textrm{min}}$ and $\textrm{q}_{\textrm{max}}$ denote the lower and upper bound of the WIMP-nucleus recoil energy which is considered in the data analysis. We truncated the integral
at the minimum of the upper analysis limit and the averaged maximum recoil energy 
$\langle E_{R_{max}} \rangle$. Due to its definition there is a fraction of WIMPs which can give rise
to a higher maximum recoil energy especially those which hit the detector in a head-on collision.
However since the differential event rates decrease approximately exponentially the actual value of this cut-off only
effects the results for WIMPs with low masses ($\sim 10 {\rm \,GeV}$). We used the statistical method proposed in \cite{Yellin:2002fn} to obtain the XENON10 limits.

\subsection{Expected event rate: spin-dependent interactions}
Given the spin-dependent cross section \eqref{apaneq} and the differential event rate \eqref{dRdE},
the expected event rate calculation for axial-vector interactions is similar to the spin-independent case. 
We first normalize the cross section to the scattering from a single nucleon. The nucleon spin expectation values are 
$\langle S_{p} \rangle =\frac{1}{2}$ and $  \langle S_{n} \rangle =0$ and $\langle S_{p} \rangle = 0$ and 
$\langle S_{n} \rangle = \frac{1}{2}$, for a proton and a neutron respectively. Using these values the SD cross section for a single nucleon is
\begin{equation}
  \sigma_{\text{spin}}^{p,n} = \frac{24}{\pi} G_F^2 \mu_{p,n}^{2} a_{p,n}^{2} \; \textrm{.}
  \label{sigma_SD_norm}
\end{equation}
Comparing \eqref{sigma_SD_norm} to \eqref{apaneq}, the SD cross section with proper normalization can be written  as 
\begin{equation}
  \sigma_{spin}^{p,n}  = \frac{3}{4} \frac{\mu_{p,n}^{2}}{\mu^{2}} \frac{J_N}{J_N+1} \frac{1}
  {\langle S_{p,n} \rangle^{2}}\sigma_{\text{spin}}\, \textrm{.}
\end{equation}
Similar to the SI case a SD form factor has to be introduced to account for non-zero momentum transfer
\begin{equation}
  \sigma_{SD} = \sigma_{\text{spin}} F_{SD}^{2} \, ,
\end{equation}
where $F_{SD}^{2}$  can be written in the form \cite{Engel:1992bf}
\begin{equation}
  F_{SD}^{2}(q) = \frac{S(q)}{S(0)} \, ,
\label{ooo}
\end{equation}
with the spin structure function $S(q)$. What makes the limit calculations in the SD case involved is the fact that $S(q)$ depends on the WIMP-nucleon couplings.
For the spin-dependent limits for XENON10 we used the form factors obtained
using the Bonn~A potential given in \cite{Ressel:1997fn}.
In the zero-momentum transfer limit, the spin structure function $S(q)$ can be evaluated:
\begin{equation}
  S(0) = \frac{2J_N+1}{\pi}J_N (J_N+1) \Lambda^{2} \, ,
  \label{nice}
\end{equation}
where $\Lambda$ is defined as 
\begin{equation}
  \Lambda = \frac{a_{p} \langle S_{p} \rangle + a_{n}  \langle S_{n} \rangle}{J_N} \, .
  \label{lamb}
\end{equation}
For finite momentum transfer it is a common procedure to 
translate the WIMP-proton and WIMP-neutron couplings  $a_{p}$ and $a_{n}$
into isoscalar and isovector spin couplings $a_{0}$ and $a_{1}$ using
\begin{eqnarray}
a_{0} &=& a_{p}+a_{n} \nonumber \\
a_{1} &=& a_{p}-a_{n}  \, ,\textrm{}
\label{dadada2}
\end{eqnarray}
so that the spin structure function can be written as
\begin{equation}
  S(q) = a_{0}^{2}S_{00}(q)+a_{1}^{2}S_{11}(q)+a_{0} a_{1} S_{01}(q) \, .
  \label{lalal}
\end{equation}
$S_{00}$,  $S_{11}$ and  $S_{01}$ represent an isoscalar, isovector and interference term respectively. The shape of $S(q)$ is determined by 
the ratio $\frac{a_{p}}{a_{n}}$ while its magnitude is proportional to $a_{p}^{2} + a_{n}^{2}$. 
Following ~\cite{Savage:2004fn}  we consider polar coordinates in the ($a_{p}$, $a_{n}$) subspace: 
\begin{eqnarray}
a_{p} & = &a \, \sin \theta \nonumber\\
a_{n} & = &a \, \cos \theta \label{polaris} \, .
\end{eqnarray}
Pure proton and neutron couplings are obtained by setting $\theta = 90^{\circ}$ and $\theta = 0^{\circ}$, respectively. Inserting this ansatz into \eqref{lalal} yields
\begin{eqnarray}
S(q) &=& a^{2} \, \Big(  (\sin \theta + \cos \theta)^{2} \, S_{00}(q) + (\sin \theta - \cos \theta)^{2}\, S_{11}(q) - \cos(2 \theta) \, S_{01}(q) \Big) \, \textrm{.}
\label{lalelu}
\end{eqnarray}
Similar to \eqref{n_si} the number of events is obtained by evaluating the integral 
\begin{eqnarray}
  N_{SD} &=& \sum_{i} f_{i}\,\int_{\textrm{q}_{\textrm{min}}}^{\textrm{min}\big(\textrm{q}_{\textrm{max}} \textrm{,} \langle E_{R_{max}} \rangle\big)} \mathrm{d} E_{R} \frac{\mathrm{d}R^{i}_{SD}}{\mathrm{d} E_{R}} \epsilon(E_R) \,MT \, . 
  \label{n_sd}
\end{eqnarray}
$N_{SD}$ can be re-written in the form 
\begin{equation}
N_{SD} = A \, a_{p}^{2} + 2\, B \, a_{p} a_{n} + C \, a_{n}^{2} \, ,
\end{equation}
with  $A$, $B$ and $C$ being constant for a given WIMP mass. 
Inserting \eqref{polaris} yields
\begin{equation}
N_{SD} = a^{2} \Big( A \, \sin^{2} \theta + 2\, B \, \sin \theta \cos \theta + C \, \cos^{2} \theta \Big) \, \textrm{.}
\label{kleiner}
\end{equation}
In order to calculate the limits on the WIMP-nucleon SD couplings, for any WIMP mass of interest we performed 
a scan over the angle $\theta$ from $0^{\circ}-360^{\circ}$. Thus, as in the spin-independent case, a 
limit on $a^{2}$ can be set. Since \eqref{kleiner} is a quadratic equation the limits are expected to be ellipses in the ($a_{p}$, $a_{n}$) subspace.
The allowed ($a_{p}$, $a_{n}$) parameter space is restricted to the inner region of these ellipses.



\end{document}